%% file: main.tex
\pdfoutput=1
\documentclass[a4paper,USenglish,autoref]{lipics-v2021}

\hideLIPIcs
\nolinenumbers

\usepackage{algorithmicx}
\usepackage{algorithm} 
\usepackage[noend]{algpseudocode}
\usepackage{xifthen}
\usepackage{graphicx}

\input{commands}

\input{sty/defer.tex}

\newdefergroup{appendix}[notdeferred]

\usepackage{xr}
\begin{ifdeferred}{appendix}
\end{ifdeferred}

\title{From Permissioned to Proof-of-Stake Consensus}
\author{Jovan Komatovic}{École Polytechnique Fédérale de Lausanne (EPFL), Lausanne, Switzerland}{jovan.komatovic@epfl.ch}{https://orcid.org/0009-0006-9714-4079}{}   %
\author{Andrew Lewis-Pye}{London School of Economics (LSE), London, United Kingdom}{a.lewis7@lse.ac.uk}{https://orcid.org/0000-0003-0228-2243}{}
\author{Joachim Neu}{a16z Crypto Research, New York, NY, USA}{jneu@a16z.com}{https://orcid.org/0000-0002-9777-6168}{}
\author{Tim Roughgarden}{Columbia University, New York, NY, USA \and a16z Crypto Research, New York, NY, USA}{tim.roughgarden@gmail.com}{https://orcid.org/0000-0002-7163-8306}{}
\author{Ertem Nusret Tas}{Stanford University, Stanford, CA, USA}{nusret@stanford.edu}{https://orcid.org/0000-0001-6061-9700}{}
\authorrunning{J.\ Komatovic, A.\ Lewis-Pye, J.\ Neu, T.\ Roughgarden, and E.\,N.\ Tas}

\Copyright{Jovan Komatovic, Andrew Lewis-Pye, Joachim Neu, Tim Roughgarden, and Ertem Nusret Tas}



\begin{ifdeferred}{appendix}
\relatedversiondetails[cite={fullversionpreprint}]{See full version for appendix}{https://eprint.iacr.org/2025/1139}
\end{ifdeferred}

\funding{%
The work of JK and ENT was conducted in part while at a16z Crypto Research.
The research of TR at Columbia University was supported in part by NSF awards CCF-2006737 and CNS-2212745.
ENT was further supported by the Stanford Center for Blockchain Research and a Research Hub Collaboration agreement between Stanford University and Input Output Global Inc.}

\EventEditors{Zeta Avarikioti and Nicolas Christin}
\EventNoEds{2}
\EventLongTitle{7th Conference on Advances in Financial Technologies (AFT 2025)}
\EventShortTitle{AFT 2025}
\EventAcronym{AFT}
\EventYear{2025}
\EventDate{October 8--10, 2025}
\EventLocation{Pittsburgh, PA, USA}
\EventLogo{}
\SeriesVolume{354}
\ArticleNo{21}

\usepackage{booktabs}
\usepackage{makecell}

\usepackage{pifont}

\definecolor{myParula01Blue}{RGB}{0,114,189}
\definecolor{myParula02Orange}{RGB}{217,83,25}
\definecolor{myParula03Yellow}{RGB}{237,177,32}
\definecolor{myParula04Purple}{RGB}{126,47,142}
\definecolor{myParula05Green}{RGB}{119,172,48}
\definecolor{myParula06LightBlue}{RGB}{77,190,238}
\definecolor{myParula07Red}{RGB}{162,20,47}

\definecolor{jnSUDigitalRed}{HTML}{B1040E}
\definecolor{jnSUDigitalRedLight}{HTML}{E50808}
\definecolor{jnSUDigitalRedDark}{HTML}{820000}
\definecolor{jnSUDigitalBlue}{HTML}{006CB8}
\definecolor{jnSUDigitalBlueLight}{HTML}{6FC3FF}
\definecolor{jnSUDigitalBlueDark}{HTML}{00548f}
\definecolor{jnSUDigitalGreen}{HTML}{008566}
\definecolor{jnSUDigitalGreenLight}{HTML}{1AECBA}
\definecolor{jnSUDigitalGreenDark}{HTML}{006F54}

\definecolor{jnSUAccentIlluminating}{HTML}{FEDD5C}
\definecolor{jnSUAccentIlluminatingLight}{HTML}{FFE781}
\definecolor{jnSUAccentIlluminatingDark}{HTML}{FEC51D}
\definecolor{jnSUAccentPoppy}{HTML}{E98300}
\definecolor{jnSUAccentPoppyLight}{HTML}{F9A44A}
\definecolor{jnSUAccentPoppyDark}{HTML}{D1660F}

\usepackage{tikz}
\usetikzlibrary{calc}
\usetikzlibrary{arrows}
\usetikzlibrary{arrows.meta}
\usetikzlibrary{patterns}
\usetikzlibrary{positioning}
\usetikzlibrary{decorations.pathreplacing}
\usetikzlibrary{calligraphy}
\usetikzlibrary{shapes.misc}
\usetikzlibrary{shapes.symbols}
\usetikzlibrary{shapes.arrows}
\usetikzlibrary{shapes.callouts}
\usetikzlibrary{spy}
\usetikzlibrary{shapes.geometric}
\usetikzlibrary{intersections}
\usetikzlibrary{fit}

\usepackage{pgfplots}
\pgfplotsset{compat=1.17}
\usepgfplotslibrary{fillbetween}

\usepackage{noto-emoji-easy}

\usepackage{import}

\begin{document}
\maketitle
\begin{abstract}
    \import{./sections/}{00_abstract.tex}
\end{abstract}

\import{./sections/}{10_introduction.tex}

\import{./sections/}{20_settings.tex}
\import{./sections/}{30_properties.tex}
\import{./sections/}{40_construction.tex}
\import{./sections/}{50_extensions.tex}

\import{./sections/}{60_relatedwork.tex}
\nocite{fullversionpreprint}
\input{_references_used.tex}
\bibliographystyle{plainurl}
\bibliography{references}

\appendix

\begin{ifdeferred}{appendix}
\section*{Appendix}
See full version for appendix:
\url{https://eprint.iacr.org/2025/1139}~\cite{fullversionpreprint}
\end{ifdeferred}

\begin{ifnotdeferred}{appendix}
\import{./appendix/}{20_modelprelims.tex}
\import{./appendix/}{10_quit_enhanced.tex}
\import{./appendix/}{20_pos_proof.tex}
\import{./appendix/}{60_relatedwork.tex}

\end{ifnotdeferred}

\end{document}

%% file: commands.tex
\usepackage{thmtools}
\usepackage{shortsym}

\usepackage{algorithmicx}
\usepackage{algorithm} 
\usepackage{algpseudocode}

\usepackage{xifthen}
\usepackage{graphicx}
\usepackage{dblfloatfix}

\input{sty/packages.tex}

\input{sty/defs.tex}

\input{sty/code.tex}

\newcommand{\mysubparagraph}[1]{\mysubparagraphNoDot{#1.}}

\newcommand{\mysubparagraphNoDot}[1]{\smallskip\noindent\emph{#1}~~}

\newcommand{\ms}[1]{%
	    \relax\ifmmode
	        \mathord{\mathcode`\-="702D\it #1\mathcode`\-="2200}%
	    \else
	        {\it #1}%
	    \fi
}
\newcommand{\tup}[1]{%
	    \relax\ifmmode
	      \langle #1 \rangle%
	    \else
	        $\langle$ #1 $\rangle$%
	    \fi
}

\newcommand{\ie}{i.e.\xspace}
\newcommand{\eg}{e.g.\xspace}
\newcommand{\cf}{cf.\xspace}
\newcommand{\tx}[0]{\ensuremath{\mathsf{tr}}}
\newcommand{\ed}[0]{\ensuremath{\mathsf{ED}}}
\newcommand{\GST}[0]{\ensuremath{\text{GST}}}
\newcommand{\LOG}[0]{\ensuremath{\mathcal{L}}}
\newcommand{\Pro}[0]{\ensuremath{\mathcal{P}}}

%% file: sty/packages.tex
\usepackage{datetime}
\usepackage{url}
\usepackage{paralist}

\usepackage{hyperref}
\usepackage{url}

\usepackage{amsmath}
\usepackage{amsthm}
\usepackage{amssymb}

\theoremstyle{definition}
\usepackage{graphicx}
\usepackage{arydshln}

\usepackage{multirow}

\usepackage{threeparttable}

\makeatletter
\AddToHook{env/algorithmic/begin}{\def\@currentcounter{ALG@line}}
\makeatother

\usepackage[sort&compress,capitalize,nameinlink]{cleveref}

\AtBeginEnvironment{appendices}{%
    \crefalias{section}{appendix}%
    \crefalias{subsection}{subappendix}%
    \crefalias{subsubsection}{subsubappendix}%
    \crefalias{subsubsubsection}{subsubsubappendix}%
}
\AddToHook{cmd/appendix/after}{%
    \crefalias{section}{appendix}%
    \crefalias{subsection}{subappendix}%
    \crefalias{subsubsection}{subsubappendix}%
    \crefalias{subsubsubsection}{subsubsubappendix}%
}

\hypersetup{hypertexnames=false}

\crefalias{ALG@line}{line}

\crefname{figure}{Fig.}{Figs.}
\Crefname{figure}{Fig.}{Figs.}

\crefname{table}{Tab.}{Tabs.}
\Crefname{table}{Tab.}{Tabs.}

\crefname{section}{Sec.}{Secs.}
\Crefname{section}{Sec.}{Secs.}
\crefname{subsection}{Sec.}{Secs.}
\Crefname{subsection}{Sec.}{Secs.}
\crefname{subsubsection}{Sec.}{Secs.}
\Crefname{subsubsection}{Sec.}{Secs.}
\crefname{subsubsubsection}{Sec.}{Secs.}
\Crefname{subsubsubsection}{Sec.}{Secs.}
\crefname{appendix}{App.}{Apps.}
\Crefname{appendix}{App.}{Apps.}
\crefname{subappendix}{App.}{Apps.}
\Crefname{subappendix}{App.}{Apps.}
\crefname{subsubappendix}{App.}{Apps.}
\Crefname{subsubappendix}{App.}{Apps.}
\crefname{subsubsubappendix}{App.}{Apps.}
\Crefname{subsubsubappendix}{App.}{Apps.}

\crefname{algorithm}{Alg.}{Algs.}
\Crefname{algorithm}{Alg.}{Algs.}
\crefname{line}{ln.}{lns.}
\Crefname{line}{ln.}{lns.}

\crefname{proposition}{Prop.}{Props.}
\Crefname{proposition}{Prop.}{Props.}
\crefname{lemma}{Lem.}{Lems.}
\Crefname{lemma}{Lem.}{Lems.}
\crefname{theorem}{Thm.}{Thms.}
\Crefname{theorem}{Thm.}{Thms.}
\crefname{corollary}{Cor.}{Cors.}
\Crefname{corollary}{Cor.}{Cors.}
\crefname{definition}{Def.}{Defs.}
\Crefname{definition}{Def.}{Defs.}

%% file: sty/defs.tex
\usepackage{xspace}

\usepackage{xcolor}
\usepackage{framed}
\definecolor{lightgray}{gray}{0.90}
\renewenvironment{leftbar}[1][\hsize]
{%
\MakeFramed{\hsize#1\advance\hsize-\width\FrameRestore}%
}
{\endMakeFramed}

\newcommand*\node[1]{\texttt{#1}}

\algsetblockdefx[Upon]{Upon}{EndUpon}{}{}[1]{\textbf{upon} #1:}{}
\algtext*{EndUpon}
\algsetblockdefx[At]{At}{EndAt}{}{}[1]{\textbf{at} #1:}{}
\algtext*{EndAt}
\algsetblockdefx[ForEach]{ForEach}{EndForEach}{}{}[1]{\textbf{for each} #1:}{}
\algtext*{EndForEach}
\algsetblockdefx[GenericBlock]{GenericBlock}{EndGenericBlock}{}{}[1]{#1}{}
\algtext*{EndGenericBlock}
\algsetblockdefx[Function]{Function}{EndFunction}{}{}[3]{#1 #2(#3):}{}
\algtext*{EndFunction}

\algnewcommand{\BlueComment}[1]{\textcolor{gray}{\hfill\(\triangleright\) #1}}
\algnewcommand{\LineComment}[1]{\State \textcolor{gray}{\(\triangleright\) #1}}

\algrenewcommand\alglinenumber[1]{\tiny\textcolor{gray}{#1}}

\newenvironment{proofsketch}{\noindent\textit{Proof sketch.}}{\hfill$\square$}

%% file: sty/code.tex
\crefname{code}{line}{lines}
\Crefname{code}{Line}{Lines}

\usepackage{xcolor}

\definecolor{mygreen}{rgb}{0.254,0.572,0.294}
\definecolor{mygray}{rgb}{0.5,0.5,0.5}
\definecolor{myorange}{rgb}{1,0.35,0}
\definecolor{mymauve}{rgb}{0.58,0,0.82}
\definecolor{myblue}{rgb}{0.2,0.4,0.6}

\definecolor{rakos4orange}{RGB}{255,165,0}
\definecolor{rakos4blue}{RGB}{14,48,173}
\definecolor{rakos4lblue}{RGB}{92,172,238}
\definecolor{rakos4dgray}{RGB}{77,77,77}
\definecolor{plainred}{RGB}{211,63,63}
\definecolor{plainorange}{RGB}{221,105,41}

\usepackage{bold-extra} %

%% file: sections/00_abstract.tex
This paper presents the first generic compiler that transforms any
permissioned consensus protocol into a 
proof-of-stake permissionless consensus protocol.
For
each of the following properties, if the initial permissioned protocol
satisfies that property in the partially synchronous setting, 
the consequent proof-of-stake protocol also satisfies that property in
the partially synchronous and quasi-permissionless setting (with the
same fault-tolerance): 
consistency;  liveness; optimistic responsiveness; every
composable log-specific property; and 
message complexity of a given order.
Moreover, our transformation ensures that the output protocol satisfies accountability (identifying culprits in the event of a consistency violation), whether or not the original permissioned protocol satisfied it.

%% file: sections/10_introduction.tex
\section{Introduction}\label{section:introduction}

\subsection{Permissioned and Permissionless Consensus}

Blockchain protocols are run by large collections of processes that must
stay in sync about the current state of the protocol's virtual
machine.  Keeping a distributed network of processes in agreement in the
face of failures, attacks, and a potentially unreliable communication
network---the problem of {\em consensus}---is a hard problem, but one
for which there is a large body of research developed over the past
four-plus decades. And indeed, many major blockchain protocols achieve
their guarantees by building on protocols and ideas developed in the
1980s and 1990s.

Beyond standing on the shoulders of giants, the rise of Internet-scale
blockchain protocols has pushed the state-of-the-art of consensus
protocols significantly, in two distinct directions.  First,
20th-century consensus protocols were typically designed for the {\em
  permissioned} setting, in which the protocol is to be run by an a
priori known and fixed set of processes (e.g., a bunch of dedicated
servers bought by a corporation for that purpose).  Blockchain
protocols, starting with the Bitcoin protocol, typically aspire to run
in the {\em permissionless} setting in which there is free entry and
exit to the set of processes running the protocol (perhaps after the
acquisition of a costly resource, such as the protocol's native
cryptocurrency). Permissionless consensus is strictly harder than
permissioned consensus, due to a combination of additional challenges,
including sybil-resistance (most commonly addressed through
proof-of-work or proof-of-stake), an ever-changing set of processes
running the protocol, and the possibility of non-faulty processes
periodically going offline.

Second, even after setting aside the challenges posed by the
permissionless setting, the practical deployment of and competition
between Internet-scale state machine replication protocols have fueled
innovation in the design of permissioned consensus protocols.  One
good example is the recent rise of DAG-based consensus protocols,
which differ from previous protocols in, among other things, their use
of simultaneous block proposals by all validators to overcome the
bottlenecks typical of ``leader-based'' protocols in the lineage of
PBFT~\cite{pbft}.\footnote{This feature is not unique to DAG-based protocols; earlier asynchronous consensus protocols---such as~\cite{Cachin2001,DBLP:conf/ccs/MillerXCSS16}---also adopted simultaneous proposals as a fundamental design principle.}  While the formal analysis of DAG-based protocols
has been confined to the permissioned setting thus far~\cite{DBLP:conf/podc/KeidarKNS21,DBLP:conf/ccs/SpiegelmanGSK22,DBLP:journals/corr/abs-2306-03058,DBLP:journals/corr/abs-2405-20488,DBLP:conf/eurosys/DanezisKSS22},
Sui is an example of a permissionless proof-of-stake (PoS) system using a
DAG-based protocol in production (\cf~\cite{DBLP:journals/corr/abs-2310-14821}).

\subsection{The Dream: A Generic Property-Preserving Compiler}

This paper pursues the ``automatic tech transfer'' of innovations for
permissioned consensus protocols to those for permissionless
protocols. The holy grail would be a ``compiler'' that takes as input
an arbitrary permissioned protocol and outputs an ``equally good''
permissionless version. Ideally, such a compiler would obviate the
need for any bespoke work to extend the {\em design} of a permissioned
protocol into a permissionless one, and similarly for its {\em
  analysis}. Given that we make no assumptions about how the initial
permissioned protocol works---i.e., it is given only as a ``black
box''---such a compiler can only interact with the protocol by executing it directly.
Similarly, the analysis of the
compiler's output would be able to rely only on the fact that the
initial permissioned protocol satisfies the desired properties and
not, for example, on any particular design or analysis method by which
those properties might be achieved.

\smallskip
\noindent\textbf{Compromise \#1: Restriction to the quasi-permissionless
  setting.}
The main result of this paper is indeed a ``compiler'' of the above
type, but certain compromises are, or appear to be, unavoidable. 
First,
``sufficiently permissionless'' protocols inherently suffer from provable
limitations that are not shared by permissioned protocols, and for
this reason we focus on the quasi-permissionless setting.  In more
detail, Lewis-Pye and Roughgarden~\cite{DBLP:journals/corr/abs-2304-14701} classify the
``permissionlessness of a protocol''---or more accurately, the maximum
permissionlessness under which a protocol functions as intended---via
a hierarchy with several levels. The permissioned setting is one
extreme point of the hierarchy. The next-most restrictive setting is
the {\em quasi-permissionless (QP)} setting; the assumption here, when
specialized to proof-of-stake protocols, is that correct processes with a
positive amount of locked-up stake are always active. (But unlike the
permissioned setting, the set of such processes can be ever-changing,
and correct processes without locked-up stake can be periodically
inactive.) The next level in the hierarchy is the {\em dynamically
  available (DA)} setting in which, similar to the sleepy setting
of~\cite{DBLP:conf/asiacrypt/PassS17}, correct processes (including those with locked-up stake)
may be periodically offline. Protocols that have even minimal
consistency and liveness guarantees in the DA setting cannot, for
example, be consistent under asynchrony, or accountable (even in
synchrony), or optimistically responsive (even in
synchrony)~\cite{DBLP:conf/fc/NeuTT22,DBLP:journals/corr/abs-2304-14701,DBLP:conf/sp/NeuTT21}.  If the assumptions on the setting
are strengthened from DA to QP, all of the properties are
achievable~\cite{DBLP:journals/corr/abs-2304-14701}. As we are interested in a compiler
that preserves properties like consistency and responsiveness, we
confine our analysis of permissionless protocols to the QP setting.

\smallskip
\noindent\textbf{Compromise \#2: Property-specific analysis.}
The strongest-possible assertion that the compiler's (permissionless)
output protocol shares all the desired properties of its
(permissioned) input protocol would be some type of homomorphism
between executions (in the spirit of e.g.~\cite{crime_punishment}).  Such a
homomorphism would map an execution of the permissionless protocol to one
or more executions of the permissioned protocol showing that,
intuitively, whatever could go wrong in a permissionless execution
could have already gone wrong in a permissioned execution.  There are
several seemingly insurmountable obstacles to achieving this goal. To
spell out one just example, such a homomorphism would presumably
establish a relationship between the units of participation in the
permissionless protocol (e.g., staked coins in a PoS
protocol) and those in the original permissioned protocol (e.g.,
individual processes). But in an execution of the permissionless
protocol, coins might well be transferred back and forth between
correct and Byzantine processes (even if ``locked'' for certain durations of the execution), which would seem to translate to an
execution of the permissioned protocol with a mobile adversary. Such a
homomorphism cannot hope to preserve any property that is achievable
with a static adversary but unachievable with a mobile adversary.

Because of the seeming impossibility of a property-preserving mapping
between executions of the permissioned and permissionless protocols,
we instead prove that our compiler preserves (a long list of) specific
properties using direct, property-specific arguments.

\smallskip
\noindent\textbf{Compromise \#3: Restrictions on the preservable
  properties.}
Because the set of processes in a permissioned protocol is fixed while
that of a permissionless protocol must update periodically (e.g., to
reflect changes in processes' stakes), and because the
permissioned protocol is given only as a closed box, any
implementation of the desired compiler must presumably execute the
permissioned protocol repeatedly, updating the process set
appropriately each time. We refer to each execution of the
permissioned protocol as an {\em epoch}, and this approach to the compiler's design as {\em epoch-based}.
The naive hope would then be that the assumed properties of the
permissioned protocol hold in each epoch of the permissionless
protocol's execution, and thus also for the overall execution of
that protocol. 

Neither part of this hope can be carried out without restrictions on
the properties we aim to preserve. The first part of the hope breaks
down for, for example, properties of the form ``eventually, the
execution satisfies some predicate~$\varphi$.'' The issue here is that
while every (infinite) execution of a permissioned protocol might
eventually satisfy~$\varphi$, there may be no finite epoch length for
which subexecutions of the protocol are guaranteed to satisfy
$\varphi$.  The second part of the hope breaks down for properties
that are not preserved by the ``concatenation'' of the executions of
two consecutive epochs. For a trivial example, the property that ``at
most $n$ distinct identifiers ever vote'' will be true for a
PBFT-style permissioned protocol with a fixed set of~$n$ processes, but
will not generally be true for any corresponding permissionless
protocol (for which there is an unbounded number of identifiers, any
of which may acquire stake and vote at some point in an execution).

The hope, then, is that an epoch-based compiler might still be
capable of preserving all of the specific properties that one would be interested in preserving. 
This brings us to the main result
of the paper.

\subsection{The Main Result}

The primary contribution of this paper is a generic compiler from
permissioned to proof-of-stake permissionless protocols in the
quasi-permissionless and partially synchronous setting. 
The
compilation process is well defined for any permissioned protocol. For
each of the following properties, if the initial permissioned protocol
satisfies that property (in the partially synchronous setting), then
the consequent proof-of-stake protocol also satisfies that property
(in the partially synchronous and quasi-permissionless setting, and
with the same fault-tolerance):
consistency;  liveness with respect to a liveness (latency)
  parameter~$\ell$;\footnote{The liveness parameter is passed as input
  to the compiler and the epoch length of the output protocol will
  depend on its value (\cf \Cref{section:construction}).} optimistic responsiveness; and 
message complexity of a given order.
Furthermore, our construction preserves all composable log-specific safety properties---that is, properties defined solely over logs (the outputs of processes; \cf \Cref{section:model_overview}) and preserved under concatenation (\cf the discussion in the ``Compromise \#3'' paragraph above): roughly speaking, if two consistent logs $L_1$ and $L_2$ each satisfy such a property, then their concatenation $L_1 \Vert L_2$ also satisfies it.
Finally, our compiler guarantees that the output protocol satisfies accountability~\cite{DBLP:journals/corr/abs-2304-14701,CGG19,HKD07,easy-accountability,DBLP:conf/ccs/ShengWNKV21,DBLP:conf/fc/NeuTT22,DBLP:conf/fc/NeuTT24}---that is, in the event of a consistency violation, the responsible parties can be correctly identified---regardless of whether the original permissioned protocol possessed this property.
For example, applying our transformation to existing permissioned DAG-based protocols (e.g.,~\cite{DBLP:conf/podc/KeidarKNS21,DBLP:conf/ccs/SpiegelmanGSK22,DBLP:journals/corr/abs-2306-03058,DBLP:journals/corr/abs-2405-20488,DBLP:conf/eurosys/DanezisKSS22}) yields new quasi-permissionless DAG-based protocols that had not previously been formally analyzed or proven correct.\footnote{To the best of our knowledge, Sui Lutris~\cite{suilutris} is the only DAG-based protocol that has been formally analyzed in the quasi-permissionless setting, explicitly addressing the challenge of validator changes through a reconfiguration mechanism.}

\subsection{Why a Generic Transformation Was Needed}

\noindent\textbf{Limitations of existing practical approaches.}
Several blockchain systems in practice---such as Sui, Aptos, Cosmos, and Celo---already employ partially synchronous, quasi-permissionless consensus protocols.
These systems typically adapt or build upon well-known permissioned protocols like HotStuff~\cite{hotstuff} (in the case of Aptos and Celo), Tendermint~\cite{buchman2018latest} (in the case of Cosmos), or Mysticeti~\cite{DBLP:journals/corr/abs-2310-14821} (in the case of Sui).
While these protocols achieve significant practical performance and are deployed at scale, they rely on \emph{protocol-specific modifications} and often embed \emph{ad-hoc reconfiguration mechanisms} to work in 
PoS/QP settings.
In particular, they are designed and analyzed in a bespoke manner---each new protocol or setting requires revisiting core design choices, reestablishing safety and liveness guarantees, and adapting reconfiguration logic to ensure correctness.
To date, there exists \emph{no general-purpose method} for converting an arbitrary permissioned protocol into a quasi-permissionless PoS protocol with well-defined theoretical guarantees.

\smallskip
\noindent\textbf{Theoretical efforts \& their shortcomings.}
Prior theoretical work has investigated how to adapt specific permissioned consensus protocols to PoS or quasi-permissionless settings.
However, these approaches, like their practical counterparts, are fundamentally \emph{protocol-specific} and rely on tightly coupled mechanisms that are not general.
For example, Lewis-Pye and Roughgarden~\cite{DBLP:journals/corr/abs-2304-14701} construct a quasi-permissionless, PoS version of HotStuff~\cite{hotstuff} by embedding specialized reconfiguration mechanisms directly into the protocol.
Budish \emph{et al.}~\cite{DBLP:journals/corr/abs-2405-09173} similarly transform Tendermint~\cite{buchman2018latest} into a quasi-permissionless, PoS protocol, but again this requires bespoke changes tailored to that specific protocol.
Sui Lutris~\cite{suilutris} is yet another protocol-specific design, featuring a custom hybrid architecture with partial and total ordering, and an ad-hoc epoch-based reconfiguration mechanism.
Other systems, such as Hybrid Consensus~\cite{pass2017hybrid} and PaLa~\cite{pala}, proceed in epochs and incorporate reconfiguration, but are built from the ground up with specific assumptions and analysis for a given protocol architecture.
More broadly, reconfiguration---the task of updating the set of validators---has been widely studied~\cite{dynamicbftfoundations,bchain,DBLP:conf/usenix/OngaroO14,bessani2014state}, but existing solutions embed reconfiguration logic deeply within the protocol's internal machinery.
None of these approaches offer a generic, reusable method for reconfiguration mechanisms or for achieving quasi-permissionless PoS protocols.

\smallskip
\noindent\textbf{Why a generic transformation is needed.}
This disconnect---between \emph{theory}, which mainly focuses on permissioned consensus, and \emph{practice}, which needs permissionless consensus---is precisely what motivates our work. 
We present the first \emph{generic, closed-box transformation} that converts any partially synchronous permissioned protocol into a quasi-permissionless PoS protocol, while preserving a rich set of properties: consistency, liveness, optimistic responsiveness, accountability, and all composable log-specific safety properties.
Our transformation separates the consensus core from reconfiguration logic, enabling a clean and composable analysis. This modularity provides two key benefits:
\begin{compactitem}
    \item For theorists, it allows continued focus on designing and analyzing permissioned protocols, while ensuring their results can be lifted automatically to the permissionless world.
    \item For practitioners, it offers a principled and reusable path to deploying proven permissioned protocols in permissionless environments, without redesigning reconfiguration or re-establishing correctness from scratch.
\end{compactitem}
Our transformation acts as a ``bridge lemma'' between permissioned and permissionless worlds, making decades of theoretical work directly applicable to modern PoS systems.

\smallskip
\noindent\textbf{Why designing this transformation was challenging.}
Naively running an arbitrary permissioned protocol as a subroutine within a PoS protocol---restarting it periodically with a new set of processes reflecting the latest stake amounts---fails to preserve even the most basic properties. 
Next, we highlight key technical challenges in designing and analyzing our generic compiler, along with the ideas that contribute to our solution.

\mysubparagraphNoDot{Quit-enhanced permissioned protocols.}
At a high level, our compiler follows an epoch-based approach, executing the given permissioned protocol in each epoch for a finite duration.
Therefore, the first step in our approach is to analyze, in a general manner, the behavior of permissioned protocols when processes are allowed to quit executing the protocol (which corresponds to the end of an epoch).
The challenge here is the fact that we know nothing about
how the given permissioned protocol might work, and the worry is that
interfering with its execution (e.g., making one process inactive so
that a different process can take its place) could affect its
properties in unpredictable ways (e.g., with the newly inactive process
now viewed as Byzantine by the protocol, violating its assumed
fault-tolerance).  
For instance, consider adapting the Tendermint~\cite{buchman2018latest} permissioned consensus protocol to a quasi-permissionless setting using our epoch-based approach.
Recall that Tendermint, being a permissioned protocol, assumes that all correct processes participate in the protocol forever, i.e., they never quit executing the protocol.
As a result, all of its proven guarantees are built on this crucial assumption.
However, when we attempt to transfer Tendermint into a quasi-permissionless setting using our epoch-based approach, this assumption no longer holds.
Specifically, if the transition from epoch $e$ to epoch $e + 1$ takes place before the network has stabilized (i.e., before GST; \cf \Cref{section:model_overview} for the definition of GST), it is possible that all correct processes except one stop participating in the Tendermint instance associated with epoch $e$.
Therefore, the remaining correct process finds itself isolated, surrounded only by adversarial processes, with no support from other correct participants.
This leads to a critical question: can consistency still be preserved in epoch $e$ under such circumstances?
In particular, is there a risk that the abandoned correct process could be misled by adversarial processes into violating consistency?
Notably, this case is \emph{not} covered by the original Tendermint analysis, as it falls outside the boundaries of the permissioned model assumed in~\cite{buchman2018latest}, where such an abandoned correct process is never considered.
Thus, analyzing the consistency (and other properties) of the resulting quasi-permissionless protocol requires going beyond the guarantees provided by the original Tendermint permissioned protocol, as those guarantees no longer directly apply in this new setting.

To address this challenge, we proceed as follows: (1) we extend the interface of Tendermint---though the idea applies to any permissioned protocol---to allow correct processes to stop participating, and (2) we analyze the guarantees provided by this extended protocol.
We refer to these enriched versions of permissioned protocols as \emph{quit-enhanced} permissioned protocols. 
Given any standard permissioned protocol, in which correct processes are expected to participate forever, the quit-enhanced version behaves identically in terms of internal logic but explicitly permits correct processes to stop executing the protocol.
Thus, quit-enhancing a standard permissioned protocol constitutes a closed-box transformation: only the interface is extended, while the internal mechanisms remain unchanged (and untouched).
Importantly, quit-enhanced permissioned protocols integrate naturally with our epoch-based structure: the properties of quit-enhanced permissioned protocols extend directly to our setting. 
With this in mind, we prove that quit-enhanced versions of arbitrary permissioned protocols preserve the key properties of interest (such as consistency and liveness) originally established in the standard permissioned model, and are thus suitable for use within an epoch-based compiler.

\mysubparagraphNoDot{Preserving consistency.}
Even the basic property of consistency will not be preserved in an
epoch-based approach to a generic compiler without carefully designing
how one epoch transitions into the next. 
For example, consider a permissioned protocol whose algorithm first tentatively confirms a block---such as upon receiving an initial quorum certificate---and only later finalizes it, either through a follow-up quorum certificate or because the block is extended by blocks that become finalized.
(It is important to emphasize that this tentative confirmation is an internal step within the algorithm's operation and does not represent the block's external status.)
If an
epoch-ending block $B$ is only tentatively confirmed, who is then allowed
to extend that block? 
Which processes should constitute the new validator set?
We cannot allow the validator set for the next epoch to be determined based on the transactions in block $B$ and its ancestors, with new validators immediately proposing descendants of $B$, because $B$ has not been finalized.
Since another conflicting block $B'$ might also be tentatively confirmed, this could result in differing views on which validator set should be chosen for the next epoch. 
If we simply wait for the first directly finalized block and treat it as the genesis block of the next epoch, a key question remains:
How can we unambiguously identify which block was directly finalized first?

To overcome this challenge, we introduce the notion of an \emph{epoch-ending block}
(akin to epoch transition in~\cite{pass2017hybrid}).
When the validators of an epoch $e$ determine that the time has come to end the epoch (the conditions for this are discussed in the ``Preserving liveness'' paragraph below), they issue special epoch-ending transactions.
They then wait for the first finalized block that, along with its ancestors, finalizes epoch-ending transactions from at least a quorum of the epoch's validators.
This uniquely determined block becomes the epoch-ending block and serves as the genesis block for the next epoch.
(Any data required to prove finality, such as blocks produced after the epoch-ending one, is retained in the protocol's history but does not contribute to the transactions finalized in that epoch.)

\mysubparagraphNoDot{Preserving liveness.}
Preserving liveness (and optimistic responsiveness, a strengthening of
liveness) presents an orthogonal set of challenges.
For example, in the partially synchronous model, epoch lengths are
most sensibly denominated in blocks (rather than time). The assumed liveness parameter $\ell$ of the underlying permissioned protocol---stating that newly issued transactions are finalized within $\ell$ time after the network stabilizes---is, however, defined in terms of time steps.
Naturally, this is a desirable property that we aim to preserve.
The issue is then that, if epochs
complete in fewer than $\ell$ timesteps (due to an unexpectedly fast
network) and every epoch has a fresh set of new validators,
there is no guarantee that a given transaction will ever be
finalized. 

Our compiler addresses this challenge by ensuring that
all correct validators overlap in each epoch (after the network stabilizes) for at least
$\ell$ time, which allows for all new transactions to be finalized.
To enforce this, each validator of an epoch $e$ locally measures a period of $\ell + \Delta$ time, where $\Delta$ denotes the known bound on message delays after the network stabilizes (i.e., after GST; \cf \Cref{section:model_overview}).
Given that messages propagate within $\Delta$ time, this local timing ensures that correct validators overlap for at least $\ell$ time during the epoch $e$ (assuming the epoch takes place after stabilization).
After this $\ell + \Delta$ interval elapses, a validator knows the epoch has run long enough and issues an epoch-ending transaction (as previously explained in the ``Preserving consistency'' paragraph).
Because each epoch concludes with the finalization of an epoch-ending block---one that includes epoch-ending transactions from a quorum of validators, and thus from at least one correct validator---it is guaranteed that every epoch following network stabilization runs long enough to ensure finalization of new transactions, in accordance with the~$\ell$-liveness property of the underlying permissioned protocol.

\mysubparagraphNoDot{Preserving composable log-specific safety properties.}
As discussed earlier, the epoch-based approach of our compiler necessitates focusing on properties whose satisfaction in every finite prefix of an (infinite) execution guarantees satisfaction in the entire execution---these are known as \emph{safety} properties~\cite{DBLP:journals/dc/AlpernS87}.
Consistency and optimistic responsiveness are examples of such safety properties.
In contrast, eventual liveness is not a safety property, though liveness with respect to a fixed time bound $\ell$ on time-to-finality is.

This raises the question: how broad is the class of safety properties we can hope to preserve?
Given the epoch-based approach, we must limit ourselves not only to safety properties but to those that are ``closed under concatenation''.
To formalize this, we focus on \emph{log-specific} properties---predicates that depend solely on the validators' running logs of finalized transactions (and not on, say, the precise sequence of messages that led to the creation of those logs).
For log-specific properties, ``safety'' then means that a violation of
the property must be evident from a finite-length prefix of
(possibly unbounded-length) logs, and ``composable'' means that
the property is preserved under unions of sets of logs.
A canonical example of a composable log-specific safety property is an
external validity property, such as ``every
finalized transaction is accompanied by appropriate signatures''.
We prove that our transformation preserves, simultaneously, every
composable log-specific safety property.

%% file: sections/20_settings.tex
\section{System Model: Overview} \label{section:model_overview}

We now provide an overview of the system model. 
A detailed description is in \Cref{section:preliminaries}.

\smallskip
\noindent\textbf{Processes, identifiers \& adversary.}
We consider a (potentially infinite) set of processes denoted by $\Pi$.
Each process $p \in \Pi$ is assigned a non-empty and potentially infinite set of \emph{identifiers}, denoted by $\mathsf{id}(p)$.
Intuitively, $\mathsf{id}(p)$ determines the set of public keys for which process $p$ knows the corresponding private key.
We denote by $\mathsf{IDs}$ the set of all identifiers.
Moreover, each process may or may not be \emph{active} at each timeslot.
A \emph{process allocation} is a function specifying, for each process $p \in \Pi$, the timeslots at which process $p$ is active.
To accommodate for clock drifts, our model permits processes to be idle even at timeslots at which they are active.
Concretely, at each timeslot at which a process is active, the process can either be \emph{waiting} or \emph{not waiting}.
Whether a process is waiting or not at a specific timeslot is also determined by the process allocation function.
In this work, we focus on protocols assuming a public key infrastructure (PKI) that allows processes to sign their messages and verify messages received from other processes.\footnote{We discuss how to extend our results in \Cref{section:extensions}.}
Finally, we assume a static adversary that corrupts a fraction of all processes at the beginning of each execution.
A corrupted process is said to be \emph{faulty}; a non-faulty process is said to be \emph{correct}.

\smallskip
\noindent\textbf{Environment.}
There exists an \emph{environment} that sends \emph{transactions} to active and non-waiting processes.
If the environment sends a transaction to an active and non-waiting process $p$ at a timeslot $\tau$, then $p$ receives the transaction at the timeslot $\tau$ (along with messages sent by other processes).

\smallskip
\noindent\textbf{Communication.}
We assume a message-passing model in which processes communicate by exchanging messages.  
In particular, any process can \emph{send} a message directly to another process via point-to-point communication.  
Additionally, we assume the existence of a \emph{gossip} primitive that enables a process to broadcast a message to all processes in the system.  
This gossip mechanism is ``global''---it is not restricted to a specific subset of processes, and messages are intended for the entire network.
An important assumption is that messages---whether sent directly or gossiped---are eventually delivered, even if the sender goes offline after sending.
Such delivery guarantees are routinely achieved in practice, for instance in gossip networks used in blockchain systems.\footnote{We underline that this assumption is not required for dynamic systems with evolving membership.
For instance, Carbon~\cite{DBLP:journals/tdsc/CamaioniGKMRVV25} avoids it and instead relies on a guaranteed message delivery only when both the sender and receiver remain online.
Similar techniques could be employed in our setting to relax the requirement of delivery despite the sender going offline.}
We stress that our model assumes only guaranteed eventual delivery of messages.
It does not preclude faulty processes from equivocating; that is, a faulty process may send different messages to different recipients.
A detailed formal description of our network model is provided in \Cref{subsection:partial_synchrony_environment_adversary}.

\smallskip
\noindent\textbf{Partial synchrony.}
This work focuses on the standard partially synchronous model~\cite{DLS88}.
In a nutshell, there exists an unknown timeslot GST such that (1) the system behaves asynchronously before GST, and (2) the system behaves synchronously after GST with the known upper-bound $\Delta$ on message delays.
Moreover, if any correct process $p$ is active at any timeslot $\tau \geq \text{GST}$, then $p$ is not waiting at timeslot $\tau$, i.e., no clock drift occurs after GST.
Lastly, each execution is associated with an unknown duration $\delta \leq \Delta$ that denotes the \emph{actual} bound on message delays.

\smallskip
\noindent\textbf{Logs \& stake.}
A \emph{log} is a non-empty ordered list of transactions.
Given any log $\mathcal{L}$, the following methods are defined:
\begin{compactitem}
    \item $\mathcal{L}.\mathsf{length}$: the number of transactions in $\mathcal{L}$.

    \item $\mathcal{L}[i]$, for any $i \in [1, \mathcal{L}.\mathsf{length}]$: the $i$-th transaction of $\mathcal{L}$.
\end{compactitem} 
Two logs $\mathcal{L}_1$ and $\mathcal{L}_2$ are \emph{consistent} if and only if $\mathcal{L}_1[i] = \mathcal{L}_2[i]$, for every $i$ with $1 \leq i \leq \min(\mathcal{L}_1.\mathsf{length}, \mathcal{L}_2.\mathsf{length})$.
Otherwise, the logs are \emph{inconsistent}.
Similarly, a log $\mathcal{L}_2$ \emph{extends} a log $\mathcal{L}_1$ if and only if (1) logs $\mathcal{L}_1$ and $\mathcal{L}_2$ are consistent, and (2) $\mathcal{L}_1.\mathsf{length} \leq \mathcal{L}_2.\mathsf{length}$.
If a transaction $\mathsf{tr}$ belongs to a log $\mathcal{L}$, we write ``$\mathsf{tr} \in \mathcal{L}$''.
We denote by $\mathsf{Logs}$ the set of all logs.

\mysubparagraphNoDot{Genesis \& local log.}
Each execution is associated with a unique \emph{genesis log} known to all processes.
The genesis log generalizes the concept of a ``genesis block'' and acts as the initial log from which all subsequent logs are built.

Each process maintains its \emph{local log}.
Formally, each process $p$ has a special log-register denoted by $\mathsf{log}(p)$.
For every correct process $p$, $\mathsf{log}(p) = \mathcal{L}_{\mathrm{g}}$ at timeslot $0$, where $\mathcal{L}_{\mathrm{g}}$ denotes the unique genesis log (of that specific execution).
Given any correct process $p$ and any timeslot $\tau$, $\mathsf{log}(p, \tau)$ denotes the value in the $\mathsf{log}(p)$ register at timeslot $\tau$.
If $\mathsf{log}(p, \tau) = \mathcal{L}$, we say that $p$ \emph{outputs} $\mathcal{L}$ at timeslot $\tau$.

\mysubparagraphNoDot{Stake.}
Every log defines its \emph{stake distribution}.
Formally, there exists a function $\mathsf{S}: \mathsf{Logs} \times \mathsf{IDs} \to \IN_{\geq 0}$.
Intuitively,  stake refers to each identifier's amount of on-chain resources.
For each log $\mathcal{L} \in \mathsf{Logs}$, we define its \emph{total stake}:
\begin{equation*}
    \mathcal{L}.\mathsf{total\_stake} = \sum_{\mathit{id} \in \mathsf{IDs}} \mathsf{S}(\mathcal{L}, \mathit{id}).
\end{equation*}
We assume that, for each log $\mathcal{L} \in \mathsf{Logs}$, $\mathcal{L}.\mathsf{total\_stake} > 0$.
Moreover, we set the following restriction on the considered stake function $\mathsf{S}(\cdot, \cdot)$:
\begin{equation*}
    \forall (\mathcal{L}_1, \mathcal{L}_2) \in \mathsf{Logs}^2: \mathcal{L}_1.\mathsf{total\_stake} = \mathcal{L}_2.\mathsf{total\_stake}.
\end{equation*}
Given this restriction, let $\mathbb{T}$ denote the total stake, i.e., $\mathbb{T} = \mathcal{L}.\mathsf{total\_stake}$, for every $\mathcal{L} \in \mathsf{Logs}$.
Importantly, we require protocols to be agnostic to the stake distribution function: given \emph{any} stake distribution function $\mathsf{S}(\cdot, \cdot)$ satisfying the conditions above, the protocol must meet its specification to be deemed correct.

\smallskip
\noindent\textbf{Permissioned setting.}
Here, the set of processes $\Pi$ is finite and known.
Additionally, $\Pi$'s cardinality $n$ is known.
Moreover, each process has a single identifier: $\forall p \in \Pi: \mathsf{id}(p) = \{ p\}$.
Finally, each process is active at every timeslot.

\mysubparagraphNoDot{Static $\rho$-bounded adversary.}
A static $\rho$-bounded adversary, for any $\rho \in [0, 1]$, corruptes at most $\rho \cdot n$ processes at the beginning of each execution.

\mysubparagraphNoDot{Known vs.\ unknown facts.}
The following facts are known to processes:
\begin{compactitem}
    \item the set of processes $\Pi$, its cardinality, and the identifier function $\mathsf{id}(\cdot)$;

    \item the bound on the power of the adversary $\rho$, the stake distribution function $\mathsf{S}(\cdot, \cdot)$, the genesis log, and the  upper-bound $\Delta$ on message delays;

    \item the process allocation function as every process is active at every timeslot.

\end{compactitem}
In contrast, the following facts are unknown to processes:
\begin{compactitem}
    \item the set of corrupted processes, its cardinality, and GST.

\end{compactitem}

\smallskip
\noindent\textbf{Quasi-permissionless setting.}
In the quasi-permissionless setting, the set of processes $\Pi$ is not necessarily finite.
Moreover, processes might have more than a single associated identifier.
In the quasi-permissionless setting, only processes with non-zero stake are guaranteed to be active.
Specifically, for any timeslot $\tau$ and any correct process $p$ for which there exist a $\tau$-active correct process $q$ and an identifier $\mathit{id}_p \in \mathsf{id}(p)$ with $\mathsf{S}( \mathsf{log}(q, \tau), \mathit{id}_p ) > 0$, process $p$ is active at $\tau$.

\mysubparagraphNoDot{Static $\rho$-bounded adversary.}
Intuitively, a static $\rho$-bounded adversary cannot control more than a $\rho$ fraction of the total stake.
Formally, for every correct process $p$ and every timeslot $\tau$, at most $\rho$ fraction of $\mathsf{log}(p, \tau)$'s total stake ($\mathsf{log}(p, \tau).\mathsf{total\_stake} = \mathbb{T}$) belongs to identifiers associated with faulty processes according to the stake distribution specified by $\mathsf{log}(p, \tau)$.

\mysubparagraphNoDot{Known vs. unknown facts.}
The following facts are known to processes:
\begin{compactitem}
    \item the bound on the power of the adversary $\rho$, the stake distribution function $\mathsf{S}(\cdot, \cdot)$, and the genesis log;

    \item the upper-bound on message delays $\Delta$.
\end{compactitem}
The following facts are unknown:
\begin{compactitem}
    \item the set of processes $\Pi$, its cardinality, and the identifier function $\mathsf{id}(\cdot)$;

    \item the set of corrupted processes, its cardinality, and GST;

    \item the process allocation function.

\end{compactitem}

%% file: sections/30_properties.tex
\section{Consensus Properties}
\label{section:composable_safety_log_specific_properties}

This section outlines the consensus properties we aim to translate from the permissioned to the quasi-permissionless setting.
These properties are divided into two categories: (1) core properties (\Cref{subsection:core_properties}), including consistency, liveness, optimistic responsiveness, and accountability, and (2) composable log-specific safety properties (\Cref{subsection:log_specific_properties}).

\subsection{Core Properties} \label{subsection:core_properties}

We begin by defining the \emph{core properties} of (permissioned or quasi-permissionless) consensus protocols, which are found in (almost) all of them. 

\smallskip
\noindent\textbf{Consistency.}
Intuitively, consistency guarantees that logs of correct processes never diverge.
\begin{definition}[Consistency]
A (permissioned or quasi-permissionless) protocol satisfies \emph{consistency} if and only if the following two conditions hold:
\begin{compactitem}
    \item No roll-backs: For every correct process $p \in \Pi$ and every two timeslots $\tau_1, \tau_2 \in \mathbb{N}_{\geq 0}$ with $\tau_1 < \tau_2$, $\mathsf{log}(p, \tau_2)$ extends $\mathsf{log}(p, \tau_1)$.

    \item No divergence: For every pair of correct processes $(p_1, p_2) \in \Pi^2$ and every timeslot $\tau \in \mathbb{N}_{\geq 0}$, logs $\mathsf{log}(p_1, \tau)$ and $\mathsf{log}(p_2, \tau)$ are consistent.
\end{compactitem}
\end{definition}
If a protocol satisfies the consistency property against a $\rho$-bounded static adversary, we say the protocol is \emph{$\rho$-consistent}.

\smallskip
\noindent\textbf{Liveness.}
The liveness property ensures that every transaction is finalized within a known time frame after GST.

\begin{definition}[$\ell$-Liveness] \label{definition:liveness}
A (permissioned or quasi-permissionless) protocol satisfies \emph{$\ell$-liveness} if and only if the following condition is satisfied for every timeslot $\tau \in \mathbb{N}_{\geq 1}$.
Suppose the following holds:
\begin{compactitem}
    \item Let $\tau^* = \max(\tau, \text{GST}) + \ell$.

    \item Let a transaction $\mathsf{tr}$ be received by a correct process from the environment at some timeslot $\leq \tau$.

    \item Let $p$ be any correct process active (and non-waiting) at a timeslot $\geq \tau^*$ and let $\tau_{\mathrm{a}}$ denote the first timeslot $\geq \tau^*$ at which $p$ is active and non-waiting.
\end{compactitem}
Then, $\mathsf{tr} \in \mathsf{log}(p, \tau_{\mathrm{a}})$.
\end{definition}
If a protocol satisfies the $\ell$-liveness property against a $\rho$-bounded static adversary, we say the protocol is \emph{$(\rho, \ell)$-live}.

\smallskip
\noindent\textbf{Optimistic responsiveness.}
Informally, the optimistic responsiveness property guarantees that transactions are finalized at network speed whenever all processes are correct.

\begin{definition}[$\ell_{\mathrm{or}}$-Responsiveness] \label{definition:responsiveness}
A (permissioned or quasi-permissionless) protocol satisfies $\ell_{\mathrm{or}}$-\emph{responsiveness}, where $\ell_{\mathrm{or}} \in O(\delta)$, 
if and only if the following condition is satisfied for every timeslot $\tau \in \mathbb{N}_{\geq 1}$ in every execution where all processes are correct.
Suppose the following holds:
\begin{compactitem}
    \item Let $\tau^* = \max(\tau, \text{GST}) + \ell_{\mathrm{or}}$.

    \item Let a transaction $\mathsf{tr}$ be received by a correct process at some timeslot $\leq \tau$.

    \item Let $p$ be any correct process active (and non-waiting) at a timeslot $\geq \tau^*$ and let $\tau_{\mathrm{a}}$ denote the first timeslot $\geq \tau^*$ at which $p$ is active and non-waiting.
\end{compactitem}
Then, $\mathsf{tr} \in \mathsf{log}(p, \tau_{\mathrm{a}})$.
\end{definition}
If a protocol satisfies the $\ell_{\mathrm{or}}$-optimistic responsiveness property against a $\rho$-bounded static adversary, we say that the protocol is \emph{$(\rho,\ell_{\mathrm{or}})$-responsive}.
Let us briefly compare our definition of $\ell_{\mathrm{or}}$-responsiveness (\Cref{definition:responsiveness}) and our definition of $\ell$-liveness (\Cref{definition:liveness}).
As shown, $\ell_{\mathrm{or}} \in O(\delta)$, where $\delta$ denotes the actual (and unknown) upper bound on message delays after GST (\cf \Cref{section:model_overview}), while $\ell$ may depend on the known upper bound $\Delta$ and does not need to reflect actual network delays.
Thus, while a responsive protocol can finalize transactions at the speed of the network, a live protocol may do so slower, depending on conservative bounds.
Importantly, we note that our transformation ensures responsiveness even in non-failure-free executions assuming that all processes do behave correctly after GST.

\smallskip
\noindent\textbf{Accountability.}
Accountability is a property ensuring that, if consistency is ever violated, a sufficient number of faulty processes are conclusively identified through undeniable proofs of guilt.
We begin by introducing the concept of a proof of guilt in quasi-permissionless consensus algorithms; this definition is inspired by that from~\cite{lewis2025beyond}.

\begin{definition}[Proof of guilt]
Let $\mathcal{P}$ be any quasi-permissionless protocol, and let $\mathit{id} \in \mathsf{IDs}$ be any identifier.
Consider a set of messages $\mathcal{M}$, each of which is signed by $\mathit{id}$ (i.e., using the corresponding private key).
The set $\mathcal{M}$ constitutes a \emph{proof of guilt} for $\mathit{id}$ with respect to $\mathcal{P}$ if and only if there exists no execution of $\mathcal{P}$ in which (1) $\mathit{id}$ (i.e., the corresponding process) sends all the messages from $\mathcal{M}$, and (2) $\mathit{id}$ (i.e., the corresponding process) is correct.
\end{definition}

We are ready to define the accountability property in quasi-permissionless algorithms.
Recall that our transformation guarantees accountability in the resulting quasi-permissionless protocol, regardless of whether the original permissioned protocol satisfies it.\footnote{That is why we define accountability solely with respect to quasi-permissionless protocols.}

\begin{definition}[$\rho_{\mathrm{a}}$-Accountability] \label{definition:accountability}
A quasi-permissionless protocol satisfies \emph{$\rho_{\mathrm{a}}$-accountability} if and only if the following condition holds in every execution where there exist correct processes $p$ and $q$, and timeslots $\tau_p \in \mathbb{N}_{\geq 1}$ and $\tau_q \in \mathbb{N}_{\geq 1}$ such that $\mathsf{log}(p, \tau_p)$ is inconsistent with $\mathsf{log}(q, \tau_q)$.
Let $\mathcal{M}_p$ (resp., $\mathcal{M}_q$) be the set of all messages received by process $p$ (resp., $q$) by timeslot $\tau_p$ (resp., $\tau_q$).
Let $\mathcal{F}$ be the set of identifiers for which a proof of guilt exists in $\mathcal{M}_p \cup \mathcal{M}_q$.
Then, there must exist a correct process $z$ and a timeslot $\tau_z \in \mathbb{N}_{\geq 0}$ such that the identifiers in $\mathcal{F}$ collectively hold at least a $\rho_{\mathrm{a}}$-fraction of the total stake recorded in $\mathsf{log}(z, \tau_z)$:
\begin{equation*}
    \sum\limits_{\mathit{id} \in \mathcal{F}} \mathsf{S}(\mathsf{log}(z, \tau_z), \mathit{id}) \geq \rho_{\mathrm{a}} \cdot \mathsf{log}(z, \tau_z).\mathsf{total\_stake} = \rho_{\mathrm{a}} \cdot \mathbb{T}.
\end{equation*}
\end{definition}
Let us analyze the definition of the $\rho_{\mathrm{a}}$-accountability property.
The property is ``triggered'' when a consistency violation occurs, meaning that two correct processes $p$ and $q$ output inconsistent logs.
In such a case, the definition ensures that $p$ and $q$ collectively hold enough information to correctly identify as faulty a set of participants (i.e., identifiers) whose stake represents at least a $\rho_{\mathrm{a}}$-fraction of the total stake.

\mysubparagraphNoDot{Accountability vs.\ consistency.}
Accountability and consistency are inherently at odds.  
A $\rho$-consistent protocol guarantees that consistency is preserved as long as the adversary controls at most a $\rho$-fraction of the total stake; consistency may be violated only if this threshold is exceeded.  
Ideally, such a protocol would also achieve $\rho_{\mathrm{a}}$-accountability for some $\rho_{\mathrm{a}} > \rho$, meaning that whenever consistency is violated, the protocol can identify a set of faulty processes holding at least a $\rho_{\mathrm{a}}$-fraction of the total stake.  
In other words, if consistency is violated, the adversary controls more than a $\rho$-fraction of the stake---hence, we seek to hold accountable a stake weight that exceeds the consistency threshold.

\subsection{Composable Log-Specific Safety Properties} \label{subsection:log_specific_properties}

In this subsection, we define \emph{composable log-specific safety properties}, a generic class of properties we translate from the permissioned to the quasi-permissionless setting.

\smallskip
\noindent\textbf{Definition.}
A \emph{log-specific property} $P$ is a function $P: \mathbb{P}(\mathsf{Logs}) \to \{ \mathit{true}, \mathit{false} \}$, where $\mathbb{P}$ denotes the power set and $\mathsf{Logs}$ denotes the set of all logs (\cf \Cref{section:model_overview}).
Next, we define log-specific safety properties.

\begin{definition}[Log-specific safety property]
\label{definition:safety-log-specific-property}
A \emph{log-specific safety property} $S$ is a log-specific property with the following constraint: 
\begin{compactitem}

    \item Let $\mathit{logs} \subseteq \mathsf{Logs}$ be any set of logs such that $S(\mathit{logs}) = \mathit{false}$.
    Then, there exists a finite subset $\mathit{logs}' \subseteq \mathit{logs}$ such that, 
    for every set $\mathit{logs}''$ with $\mathit{logs}' \subseteq \mathit{logs}''$, $S(\mathit{logs}'') = \mathit{false}$. 
\end{compactitem}
\end{definition}

Intuitively,
the constraint specifies that if a set of logs fails to satisfy $S$, there must exist a finite subset that also does not satisfy $S$, and none of its supersets satisfy $S$.
We underline that \Cref{definition:safety-log-specific-property} 
follows the spirit of safety properties as defined by Alpern and Schneider in their seminal work~\cite{DBLP:journals/ipl/AlpernS85}.  
We are now ready to define composable log-specific safety properties.

\begin{definition}[Composable log-specific safety property]
\label{definition:composable-log-specific}
A log-specific safety property $S$ is \emph{composable} if and only if the following constraint holds:
\begin{compactitem}
    \item Let $(\mathit{logs}_1, \mathit{logs}_2) \subseteq \mathsf{Logs}^2$ be any pair of sets of logs such that $S(\mathit{logs}_1) = S(\mathit{logs}_2) = \mathit{true}$.
    Then, for every set  $\mathit{logs} \subseteq \mathit{logs}_1 \cup \mathit{logs}_2$, $S(\mathit{logs}) = \mathit{true}$.
\end{compactitem}
\end{definition}

In essence, a composable log-specific safety property $S$ indicates that if two sets of logs satisfy $S$, then every subset of their union also satisfies $S$. 
Properties that require correct processes to output only valid logs (according to some predetermined validity condition) are composable log-specific safety properties.
These include, for example, the following properties: (1) no log contains futile transactions, and (2) no log contains transactions not signed by their issuers.

\smallskip
\noindent\textbf{Satisfying log-specific safety properties.}
Lastly, we define what it means for a protocol to satisfy a (composable or not) log-specific safety property.
Fix any (permissioned or quasi-permissionless) protocol $\mathcal{P}$.
Given any execution $\mathcal{E}$ of the protocol $\mathcal{P}$, let $\mathsf{logs}(\mathcal{E})$ denote the set of all logs held by correct processes in $\mathcal{E}$:
\begin{equation*}
    \mathsf{logs}(\mathcal{E}) \equiv \{ \mathcal{L} \in \mathsf{Logs} \,|\, \exists p \in \Pi: p \text{ is correct} \land p \text{ outputs $\mathcal{L}$ in $\mathcal{E}$} \}.
\end{equation*}
Finally, we present the definition.

\begin{definition}[Satisfying log-specific safety properties]
Let $S$ be any log-specific safety property and let $\mathcal{P}$ be any (permissioned or quasi-permissionless) protocol.
We say that $\mathcal{P}$ \emph{satisfies} $S$ if and only if, in every execution $\mathcal{E}$ with $S(\{ \mathcal{L}_{\mathrm{g}} \}) = \mathit{true}$, where $\mathcal{L}_{\mathrm{g}}$ denotes the genesis log in $\mathcal{E}$, $S\big( \mathsf{logs}(\mathcal{E}) \big) = \mathit{true}$.
\end{definition}

A protocol $\mathcal{P}$ satisfies a log-specific safety property $S$ if correct processes output only logs that are allowed by $S$.
Notably, we exclude executions in which the genesis log is not allowed by $S$; otherwise, no protocol could satisfy $S$ as it is initially violated.
We underline that time is irrelevant in the context of satisfying $S$: for any two (arbitrarily different) executions $\mathcal{E}_1$ and $\mathcal{E}_2$ that have the same set of output logs (i.e., $\mathsf{logs}(\mathcal{E}_1) = \mathsf{logs}(\mathcal{E}_2)$), the property $S$ is either satisfied in \emph{both} executions or in \emph{neither}.

%% file: sections/40_construction.tex
\section{Transformation}
\label{section:construction}

In this section, we show how to transform any permissioned protocol $\mathcal{P}$ into a quasi-permissionless PoS protocol $\mathcal{T}(\mathcal{P})$.
We begin by presenting an overview of our transformation (\Cref{subsection:transformation_overview}).
Then, we introduce its pseudocode (\Cref{subsection:transformation_pseudocode}).
Finally, we provide an informal analysis of $\mathcal{T}(\mathcal{P})$'s correctness (\Cref{subsection:correctness_analysis}).
A formal proof is relegated to \Cref{section:proof}.

\subsection{Overview} \label{subsection:transformation_overview}

We partition the execution of the quasi-permissionless PoS protocol $\mathcal{T}(\mathcal{P})$ into \emph{epochs} (\Cref{fig:highlevel-construction-overview}).
In each epoch, the permissioned protocol $\mathcal{P}$ is executed, meaning that every epoch is responsible for producing a particular segment of the ``global'' log of finalized transactions.
Each epoch runs for (at least) a predetermined duration greater than the liveness parameter of protocol $\mathcal{P}$.
This ensures that, following GST, each epoch runs for sufficiently long to finalize new transactions.
Once an epoch completes, the log produced up to (and including) that epoch is treated as the genesis log for the next epoch.
Moreover, the identifiers holding stake according to the produced log validate the next epoch; the genesis log $\mathcal{L}_{\mathrm{g}}$ determines the validators of the first epoch.
During the entire execution, protocol $\mathcal{T}(\mathcal{P})$ keeps progressing through epochs, with each subsequent epoch building on the output of the previous one, resulting in an ever-growing log of finalized transactions.

\input{figures/alg-highlevel-construction-overview}

\smallskip
\noindent\textbf{Guarantees of $\mathcal{P}$ in $\mathcal{T}(\mathcal{P})$'s epoch-based structure.}
The correctness of our quasi-permissionless PoS protocol $\mathcal{T}(\mathcal{P})$ crucially relies on the guarantees provided by the permissioned protocol $\mathcal{P}$.
However, we observe that, 
within the epoch-based structure of $\mathcal{T}(\mathcal{P})$, protocol $\mathcal{P}$ might display unexpected behavior as described below.
In the standard permissioned setting, correct processes run $\mathcal{P}$ forever, as this is a fundamental characteristic of the setting.
However, the epoch-based structure of $\mathcal{T}(\mathcal{P})$ introduces a different environment for $\mathcal{P}$: correct processes \emph{stop} executing $\mathcal{P}$ associated with epoch $e$ once they transition to epoch $e + 1$.
This minor change necessitates a re-evaluation of the guarantees offered by $\mathcal{P}$: $\mathcal{P}$'s guarantees
in the standard permissioned setting do \emph{not} directly extend to this new context. 

To analyze $\mathcal{P}$'s behavior within the epoch-based structure of $\mathcal{T}(\mathcal{P})$, one needs to (1) enrich $\mathcal{P}$'s interface by allowing correct processes to stop executing it, and (2) examine the guarantees provided by this enriched protocol.
We achieve this by introducing the concept of \emph{quit-enhanced} permissioned protocols (\cf \Cref{subsection:standard_quit_enhanced}): given any standard permissioned protocol $\mathcal{P}'$, where correct processes are expected to participate forever, we define $\mathit{quit}(\mathcal{P}')$ as the quit-enhanced version of $\mathcal{P}'$, where correct processes have the option to stop participating.
(We emphasize that $\mathit{quit}(\mathcal{P})$ merely extends the interface of the original permissioned protocol $\mathcal{P}$.
In particular, analyzing the behavior of $\mathit{quit}(\mathcal{P})$ relies exclusively on understanding the behavior of $\mathcal{P}$ itself---there is no need to examine the internal mechanisms of the protocol.
This makes our quit-based extension inherently ``closed-box'' in nature.)
Using our concept of quit-enhanced protocols, we prove that consistency, liveness, optimistic responsiveness, and all log-specific safety properties translate from $\mathcal{P}$ to $\mathit{quit}(\mathcal{P})$ (\cf \Cref{subsection:quit_preservation}), allowing us to build $\mathcal{T}(\mathcal{P})$ on top of $\mathcal{P}$ that satisfies these properties in the permissioned model.\footnote{Liveness and optimistic responsiveness are preserved in only some (and not all) executions of $\mathit{quit}(\mathcal{P})$. However, as we show in \Cref{section:proof}, only these executions are needed to prove $\mathcal{T}(\mathcal{P})$'s liveness and optimistic responsiveness.}

\input{figures/alg_simulation1.tex}

\subsection{Pseudocode} \label{subsection:transformation_pseudocode}

Transformation $\mathcal{T}$ takes a permissioned $\rho$-consistent protocol $\Pro$ satisfying liveness with parameter $\ell$ and a stake distribution function $\mathsf{S}(\cdot, \cdot)$ as input, and outputs a quasi-permissionless PoS protocol $\mathcal{T}(\Pro)$ (\cf \Cref{fig:construction-overview} and~\Cref{algorithm:pos-code_2}). 
Each epoch of $\mathcal{T}(\mathcal{P})$ is associated with a set of active identifiers, denoted by $\mathsf{validators}_e$.
At the beginning of each epoch $e$, these identifiers execute 
an instance of $\Pro$
(\Cref{line:start_simulation_2}, \Cref{line:start_pseudocode_simulation}-\Cref{line:final_line_pseudocode_simulation}).

\smallskip
\noindent\textbf{Executing $\mathcal{P}$ within an epoch $e$.}
As $\mathbb{T}$ denotes the total stake (\cf \Cref{section:model_overview}), identifiers utilized in (executed-in-an-epoch) permissioned protocol $\mathcal{P}$ are integers in the range $[1, \mathbb{T}]$; we refer to these identifiers as ``permissioned-ids''.
Similarly, we refer to identifiers from the set $\mathsf{IDs}$ (\cf \Cref{section:model_overview}) as ``PoS-ids''.

\input{figures/alg_simulation2.tex}
Consider a correct process $p$ executing $\mathcal{P}$ in epoch $e$.
To start executing $\mathcal{P}$ in epoch $e$, process $p$ invokes the $\mathsf{start\_simulation}(\cdot)$ function (\Cref{line:function_start_simulation}).\footnote{Throughout the remainder of this section and in the pseudocode, we say that processes simulate $\mathcal{P}$ to emphasize that this is the underlying permissioned protocol being executed, and that each PoS-id may be running $\mathcal{P}$'s state machine on behalf of multiple permissioned-ids.}
There, process $p$ begins by setting the current log of the PoS protocol $\mathcal{T}(\Pro)$, finalized at the end of the previous epoch $e-1$, as the genesis log $\mathcal{L}$ for epoch $e$ (\Cref{line:start_establishing_genesis}-\Cref{line:end_establishing_genesis}).
It then maps each permissioned-id (from the $[1, \mathbb{T}]$ range) into one PoS-id (from the $\mathsf{IDs}$ set) using the $\mathit{id\_map}_p = \mathsf{map\_stake}(\mathcal{L})$ map (\Cref{line:id_map}).
For instance, if $\mathcal{L}$ assigns 3 tokens to a PoS-id $\mathit{id}$, then there exist $x_1, x_2, x_3 \in [1, \mathbb{T}]$ such that $\mathit{id\_map}_p[x_1] = \mathit{id\_map}_p[x_2] = \mathit{id\_map}_p[x_3] = \mathit{id}$; intuitively, in epoch $e$, permissioned-ids $x_1$, $x_2$ and $x_3$ correspond to the PoS-id $\mathit{id}$, meaning that PoS-id $\mathit{id}$ is responsible for simulating $\mathcal{P}$'s state machines associated with permissioned-ids $x_1$, $x_2$ and $x_3$.
This is also crucial because each process must know which identifier to use when forwarding messages related to the permissioned protocol $\mathcal{P}$. Specifically, if a permissioned-id $i_1$ needs to communicate with another permissioned-id $i_2$ in $\mathcal{P}$ (associated with some epoch), the process $p$ ``responsible'' for the permissioned-id $i_1$ must know a PoS-id corresponding to $i_2$ in order to forward the message correctly (\Cref{line:forward_received_simulation}).
After establishing the aforementioned permissioned-ids to PoS-ids mapping, process $p$ verifies whether it is validating epoch $e$, i.e., whether any of its identifiers have been assigned a positive stake by $\mathcal{L}$.
If there exists a permissioned-id $x \in [1, \mathbb{T}]$ that maps into a $p$'s PoS-id $\mathit{id} \in \mathsf{id}(p)$, process $p$ instantiates $\mathcal{P}$'s state machine initialized with permissioned-id $x$ and genesis log $\mathcal{L}$ (\Cref{line:initialize_simulation_id,line:genesis_log}).
Then, $p$ updates $\mathit{simulation\_map}_p[x]$ to the initialized state machine (\Cref{line:simulation_map}).
Note that $\mathit{id\_map}_p$ associates each permissioned-id with its PoS-id counterpart, whereas $\mathit{simulation\_map}_p$ maps only $p$'s permissioned-ids into their respective state machines.
Finally, if $p$ is indeed validating epoch $e$,
it instructs its timer $\mathit{epoch\_timer}_p$ to measure $\ed = \ell + \Delta$ time (\Cref{line:epoch_timer_measure}).
Here, $\ed = \ell + \Delta$ is selected large enough so that each simulated instance $\Pro$ is run sufficiently long to allow new transactions to be finalized after $\GST$. 
Specifically, the time period $\ed = \ell + \Delta$ ensures that, after GST, all correct validators of some post-GST epoch $e$ overlap in their execution of $\Pro$ (associated with epoch $e$) for at least $\ell$ time. 
This overlap, together with the $\ell$-liveness property of $\Pro$, guarantees that new transactions are finalized.

\smallskip
\noindent\textbf{Outputting logs.}
Within every epoch $e$, processes output a log produced by $\Pro$ as the $\mathcal{T}(\mathcal{P})$ log if the log is signed by any quorum of the set $\mathsf{validators}_e$ (\cref{line:new_log_received_gossip} and \Cref{definition:fully-certified}).

\input{figures/alg-construction-overview}

\smallskip
\noindent\textbf{Epoch change.}
When $\mathit{epoch\_timer}_i$ expires, process $p$ gossips a special \emph{epoch-ending} transaction $(\textsc{finish}, e)$ on behalf of each of its validating PoS-ids (\Cref{line:epoch_ending_transaction}).
The purpose of these \textsc{finish} transactions is ensuring that epoch $e$ eventually concludes by producing a complete log (\cf \Cref{definition:epoch-completed}), i.e., an epoch-ending block.

Process $p$ stops simulating $\Pro$ upon observing 
epoch-ending transactions from a quorum of identifiers with sufficient stake, \ie, some quorum $I \subseteq \{\mathit{id} | \mathit{id} \in \mathsf{validators}_e\}$ such that $\sum_{\mathit{id} \in I} \mathsf{S}(\LOG, \mathit{id}) \geq (1-\rho) \mathbb{T}$ (\Cref{line:new_epoch} and~\Cref{definition:epoch-completed}).
At that point, $p$ stops each of its currently simulated state machines (\Cref{line:stop_executing}), cancels $\mathit{epoch\_timer}_p$ (\Cref{line:epoch_timer_cancel}), and resets the simulation-specific variables (\Cref{line:start_reseting_variables}-\Cref{line:stop_reseting_variables}).

\smallskip
\noindent\textbf{Determining the next set of validators.}
The set $\mathsf{validators}_{e+1}$ of validators is selected based on the stake distribution determined by the log produced in epoch $e$ (\Cref{line:id_map} and~\Cref{definition:identifiers}).
More specifically, any identifier that has positive stake according to the log $\mathcal{L}$ produced in epoch $e$ is eligible to participate in the permissioned protocol instance $\Pro$ associated with the next epoch $e + 1$.

\smallskip
\noindent\textbf{On leaving \& joining validators.}
How do we ensure that all correct validators of epoch $e$ eventually complete that epoch?
And how do we ensure that all correct validators of epoch $e+1$ learn the up-to-date state of the system---that is, all blocks finalized up to epoch $e$?
Both questions are resolved through the underlying 
communication model (\cf~\Cref{section:model_overview}).
Specifically, before any correct validator of epoch $e$ departs, it gossips its current log of finalized blocks, which includes all blocks from epoch $e$ (\Cref{line:gossip_local_log}).
This log is fully certified, allowing any recipient to locally verify its validity and commitment status (see the following paragraph for details).
This mechanism guarantees that every correct validator eventually receives the fully-certified log, finalizes the blocks of epoch $e$, and is thus permitted to exit.
Similarly, joining validators are ensured to eventually receive the finalized log up to epoch $e$, enabling them to reconstruct the current system state and safely begin executing epoch $e+1$.
Importantly, this mechanism allows us to avoid imposing any restrictions on validator departures: the sets of validators in epochs $e$ and $e+1$ may be entirely disjoint.

\smallskip
\noindent\textbf{Methods defined on logs.}
The pseudocode of our transformation $\mathcal{T}$ utilizes many methods defined on logs.
Their formal definition can be found in \Cref{subsection:omitted_definitions}, while an informal description is given below:
\begin{compactitem}
    \item $\mathcal{L}.\mathsf{current\_ids}$: the set of identifiers with positive stake according to $\mathcal{L}$.

    \item $\mathcal{L}.\mathsf{epoch}$: the epoch with which $\mathcal{L}$ is associated.

    \item $\mathcal{L}.\mathsf{validators}$: the set of identifiers validating the epoch of $\mathcal{L}$, i.e., the set of identifiers that ``produced'' $\mathcal{L}$.

    \item $\mathcal{L}.\mathsf{completed}$: $\mathit{true}$ if and only if $\mathcal{L}$ contains sufficiently many (\ie $(1 - \rho)\mathbb{T}$ worth of stake) epoch-ending transactions.

    \item $\mathcal{L}.\mathsf{ep\_prefix}\big(e \in [0, \mathcal{L}.\mathsf{epoch}) \big)$: the completed log $\mathcal{L}'$ of epoch $e$ such that $\mathcal{L}$ extends $\mathcal{L}'$.
\end{compactitem}
Moreover, we say that a log $\mathcal{L}$ is \emph{certified} at a process $p$ if and only if $p$ receives signatures on the log $\mathcal{L}$ from a quorum of members (\ie $(1 - \rho)\mathbb{T}$ worth of stake) of $\mathcal{L}.\mathsf{validators}$.
Finally, a certified log $\mathcal{L}$ is \emph{fully-certified} at a process $p$ if and only if, for every $e \in [0, \mathcal{L}.\mathsf{epoch})$, $\mathcal{L}.\mathsf{ep\_prefix}(e)$ is certified at process $p$.
Note that two fully-certified logs cannot be conflicting.
A log attains full certification only if it is approved by a quorum of validators for each epoch it covers.
Since any two quorums intersect in at least one correct validator, it is impossible for two conflicting logs to both become fully certified.
Consequently, upon receiving a fully-certified log, any correct validator can safely adopt it as its local log (\Cref{line:update_log_gossip}).

\subsection{Correctness}
\label{subsection:correctness_analysis}

In this subsection, we provide an informal analysis of the correctness of our transformation.
\smallskip
\noindent\textbf{Consistency.}
The following theorem proves that our transformation preserves the consistency property from the permissioned to the quasi-permissionless setting.

\begin{theorem}[Consistency]
\label{theorem:consistency}
Consider a permissioned protocol $\mathcal{P}$ that satisfies consistency against a $\rho$-bounded adversary.
Then, the quasi-permissionless protocol $\mathcal{T}(\mathcal{P})$ satisfies consistency against a $\rho$-bounded adversary.
\end{theorem}
\begin{proofsketch}
We show that the consistency property translates from $\mathcal{P}$ to $\mathcal{T}(\mathcal{P})$ by induction.
Initially, all correct processes have $\LOG_{\mathrm{g}}$ as the genesis log.
Given that the logs output by the first epoch are consistent (due to $\mathcal{P}$'s consistency), consistency is satisfied in the first epoch.
Moreover, all correct processes agree on the genesis log and the set of validators for the second epoch.
By inductively applying the same argument, we can show that the logs output by the correct processes remain consistent throughout the entire execution.
A formal proof of the theorem is given in~\Cref{section:proof-of-consistency}.
\end{proofsketch}

\smallskip
\noindent\textbf{Composable log-specific safety properties.}
The theorem below asserts that our transformation carries over any composable log-specific safety property from the permissioned setting to the quasi-permissionless setting.

\begin{theorem}[Composable log-specific safety property]
\label{theorem:log-specific-properties}
Consider a permissioned protocol $\Pro$ that satisfies consistency and some composable log-specific safety property $S$ against a $\rho$-bounded adversary.
Then, $\mathcal{T}(\Pro)$ satisfies $S$ against a $\rho$-bounded adversary.
\end{theorem}
\begin{proofsketch}
    To prove that any composable log-specific safety property $S$ translates from $\mathcal{P}$ to $\mathcal{T}(\mathcal{P})$, we again rely on induction.
As $\mathcal{P}$ satisfies $S$, the set of logs output by the first epoch adheres to $S$.
Similarly, the set of logs output by the second epoch adheres to $S$.
Hence, their union adheres to $S$ as $S$ is composable (\cf~\cref{definition:composable-log-specific}):
\begin{equation*}
    S( \{ \mathcal{L} \,|\, \mathcal{L}.\mathsf{epoch} \in \{ 1, 2\} \land \mathcal{L} \text{ is output by a correct process} \}  ) = \mathit{true}.
\end{equation*}
Again, the set of logs output by the third epoch adheres to $S$, which then implies:
\begin{equation*}
    S( \{ \mathcal{L} \,|\, \mathcal{L}.\mathsf{epoch} \in \{ 1, 2, 3\} \land \mathcal{L} \text{ is output by a correct process} \}  ) = \mathit{true}.
\end{equation*}
By inductively applying this argument, we can show that $\mathcal{T}(\mathcal{P})$ satisfies $S$.
A formal proof of the theorem is given in~\Cref{section:proof-of-log-specific}.
\end{proofsketch}

\smallskip
\noindent\textbf{Liveness \& optimistic responsiveness.}
The following theorem demonstrates the preservation of liveness and optimistic responsiveness by our transformation.

\begin{theorem}[Liveness \& optimistic responsiveness]
\label{theorem:liveness}
Consider a permissioned protocol $\Pro$ that satisfies consistency and $\ell$-liveness, for some $\ell < \infty$, against a $\rho$-bounded adversary.
Then, $\mathcal{T}(\Pro)$ satisfies $\ell^{\star}$-liveness, with $\ell^{\star} = 2\Delta + 2\ell$, against a $\rho$-bounded adversary.
Furthermore, if $\Pro$ additionally satisfies $\ell_{\mathrm{or}}$-responsiveness, for some $\ell_{\mathrm{or}} \in O(\delta)$, where $\delta$ denotes the actual bound on message delays after GST, against a $\rho$-bounded adversary, 
then $\mathcal{T}(\Pro)$ satisfies $\ell^{\star}_{\mathrm{or}}$-responsiveness, with $\ell^{\star}_{\mathrm{or}} = 2\delta + 2\ell_{\mathrm{or}}$, against a $\rho$-bounded adversary.
\end{theorem}
\begin{proofsketch}
 If the first correct process $p$ for which there exists an identifier $\mathit{id} \in \mathsf{id}(p) \cap \mathsf{validators}_e$ enters an epoch $e$ at some timeslot $\tau \geq \text{GST}$, then all other correct validators enter epoch $e$ within $\Delta$ timeslots after observing the log seen by $p$.
Moreover, no correct process sends an epoch-ending transaction until $\ed = \ell + \Delta$ timeslots into epoch $e$, which implies that epoch $e$ cannot be completed before timeslot $\tau + \ed = \tau +  \ell + \Delta$.
Hence, all correct validators stay in epoch $e$ together for at least $\ell$ timeslots.
As $\Pro$ satisfies $\ell$-liveness, this overlap is sufficient to finalize new transactions in epoch $e$.
To prove optimistic responsiveness, we follow the previous argument while factoring in that the actual message delay bound is $\delta$.
As a result, all correct validators join epoch $e$ within $\delta$ timeslots---rather than $\Delta > \delta$---after seeing the log observed by $p$.
A formal proof is relegated to~\Cref{section:proof-of-liveness}.
\end{proofsketch}

\smallskip
\noindent\textbf{Accountability.}
Finally, the following theorem demonstrates that our transformation ensures that the resulting quasi-permissionless protocol satisfies accountability.
Note that this holds even if the original permissioned protocol lacks accountability. 

\begin{theorem} [Accountability] \label{theorem:accountability}
Consider a permissioned protocol $\mathcal{P}$ that satisfies consistency against a $\rho$-bounded adversary.
Then, the quasi-permissionless protocol $\mathcal{T}(\mathcal{P})$ satisfies $(1 - 2\rho)$-accountability.
\end{theorem}
\begin{proofsketch}
Suppose two correct processes $p$ and $q$ output inconsistent logs $\mathcal{L}_p$ and $\mathcal{L}_q$, respectively.
Therefore, $\mathcal{L}_p$ (resp., $\mathcal{L}_q$) is fully-certified at $p$ (resp., $q$).
Hence, disseminating these two logs, along with their respective signatures, enables accountability: every correct process eventually receives these inconsistent logs, combines the received signatures, and obtains a set of identifiers $C$ such that, for each $\mathit{id} \in C$, $\mathit{id}$ signs two inconsistent logs associated with the same epoch, thus proving $\mathit{id}$'s culpability.
Finally, since each of the two logs is signed by a quorum of validators---i.e., those holding at least $(1 - \rho)\mathbb{T}$ stake---the identifiers in $C$ collectively represent at least $(1 - \rho)\mathbb{T} + (1 - \rho)\mathbb{T} - \mathbb{T} = (1 - 2\rho)\mathbb{T}$ stake.
A formal proof of the theorem can be found in \Cref{section:accountability}.
\end{proofsketch}

\smallskip
In terms of message complexity, the transformation $\mathcal{T}(\mathcal{P})$ introduces an additional quadratic term to that of $\mathcal{P}$ in order to satisfy the accountability property in the quasi-permissionless setting, stemming from the signatures that processes must exchange before outputting a log.
However, since any consensus protocol inherently requires a quadratic number of messages in the worst case~\cite{dolev1985bounds}, our transformation does not introduce any asymptotic worst-case overhead. 
A proof is deferred to \Cref{section:message-complexity}.

%% file: figures/alg-highlevel-construction-overview.tex
\begin{figure}[tbp]
    \centering
    \begin{tikzpicture}[
            x=2.5cm,
            y=1cm,
            MyAction/.style={
                -latex,
            },
            MyProtocol/.style={
                draw=none,
                minimum width=2.3cm,
                minimum height=0.8cm,
                rounded corners=2pt,
                fill=black!20,
                align=center,
            },
        ]
        \scriptsize

        \pgfdeclarelayer{background}
        \pgfsetlayers{background,main}

        \begin{scope}[yshift=-0.5cm]
            \draw [MyAction] (0,0) -- (4.25,0) node [pos=1.0,below,yshift=-1pt] {Time};
            \draw ([yshift=2pt] 0,0) -- ([yshift=-2pt] 0,0);
            \draw ([yshift=2pt] 1,0) -- ([yshift=-2pt] 1,0);
            \draw ([yshift=2pt] 2,0) -- ([yshift=-2pt] 2,0);
            \draw ([yshift=2pt] 3,0) -- ([yshift=-2pt] 3,0);
            \draw ([yshift=2pt] 4,0) -- ([yshift=-2pt] 4,0);

            \draw [decorate,decoration={brace,mirror,amplitude=3pt,raise=5pt},densely dotted] (0,0) -- (1,0) node [midway,anchor=north,yshift=-8pt,align=center] (epoch1label) {Epoch $1$};
            \draw [decorate,decoration={brace,mirror,amplitude=3pt,raise=5pt},densely dotted] (1,0) -- (2,0) node [midway,anchor=north,yshift=-8pt,align=center] (epoch2label) {Epoch $2$};
            \draw [decorate,decoration={brace,mirror,amplitude=3pt,raise=5pt},densely dotted] (2,0) -- (3,0) node [midway,anchor=north,yshift=-8pt,align=center] (epoch3label) {Epoch $3$};
            \draw [decorate,decoration={brace,mirror,amplitude=3pt,raise=5pt},densely dotted] (3,0) -- (4,0) node [midway,anchor=north,yshift=-8pt,align=center] (epoch4label) {Epoch $4$};
        \end{scope}

        \node [MyProtocol] (p1) at (0.5,0) {Permissioned $\CP$\\Instance $\CP_1$};
        \node [MyProtocol] (p2) at (1.5,0) {Permissioned $\CP$\\Instance $\CP_2$};
        \node [MyProtocol] (p3) at (2.5,0) {Permissioned $\CP$\\Instance $\CP_3$};
        \node [MyProtocol] (p4) at (3.5,0) {Permissioned $\CP$\\Instance $\CP_4$};
        \node at (4.15,0) {.....};

        \begin{scope}[yshift=1cm]
            \node [inner sep=1pt] (v11) at (0,0) {\notoemoji[height=1.2em]{desktop computer}};
            \node [inner sep=1pt,anchor=south west] (v12) at ([xshift=-0.5em] v11.north east) {\notoemoji[height=1.2em]{desktop computer}};
            \node [inner sep=1pt,anchor=south east] (v13) at ([xshift=0.5em] v11.north west) {\notoemoji[height=1.2em]{desktop computer}};
            \begin{pgfonlayer}{background}
                \node [circle,fill=black!10,inner sep=12pt] (committee1) at ([yshift=1pt] v11.north) {};
            \end{pgfonlayer}
            \node [anchor=south,align=center] (committeelabel1) at (committee1.north) {Committee\\$\mathsf{validators}_1$};
        \end{scope}

        \begin{scope}[yshift=1cm]
            \node [inner sep=1pt] (v21) at (1,0) {\notoemoji[height=1.2em]{desktop computer}};
            \node [inner sep=1pt,anchor=south west] (v22) at ([xshift=-0.5em] v21.north east) {\notoemoji[height=1.2em]{desktop computer}};
            \node [inner sep=1pt,anchor=south east] (v23) at ([xshift=0.5em] v21.north west) {\notoemoji[height=1.2em]{desktop computer}};
            \begin{pgfonlayer}{background}
                \node [circle,fill=black!10,inner sep=12pt] (committee2) at ([yshift=1pt] v21.north) {};
            \end{pgfonlayer}
            \node [anchor=south,align=center] (committeelabel2) at (committee2.north) {Committee\\$\mathsf{validators}_2$};
        \end{scope}

        \begin{scope}[yshift=1cm]
            \node [inner sep=1pt] (v31) at (2,0) {\notoemoji[height=1.2em]{desktop computer}};
            \node [inner sep=1pt,anchor=south west] (v32) at ([xshift=-0.5em] v31.north east) {\notoemoji[height=1.2em]{desktop computer}};
            \node [inner sep=1pt,anchor=south east] (v33) at ([xshift=0.5em] v31.north west) {\notoemoji[height=1.2em]{desktop computer}};
            \begin{pgfonlayer}{background}
                \node [circle,fill=black!10,inner sep=12pt] (committee3) at ([yshift=1pt] v31.north) {};
            \end{pgfonlayer}
            \node [anchor=south,align=center] (committeelabel3) at (committee3.north) {Committee\\$\mathsf{validators}_3$};
        \end{scope}

        \begin{scope}[yshift=1cm]
            \node [inner sep=1pt] (v41) at (3,0) {\notoemoji[height=1.2em]{desktop computer}};
            \node [inner sep=1pt,anchor=south west] (v42) at ([xshift=-0.5em] v41.north east) {\notoemoji[height=1.2em]{desktop computer}};
            \node [inner sep=1pt,anchor=south east] (v43) at ([xshift=0.5em] v41.north west) {\notoemoji[height=1.2em]{desktop computer}};
            \begin{pgfonlayer}{background}
                \node [circle,fill=black!10,inner sep=12pt] (committee4) at ([yshift=1pt] v41.north) {};
            \end{pgfonlayer}
            \node [anchor=south,align=center] (committeelabel4) at (committee4.north) {Committee\\$\mathsf{validators}_4$};
        \end{scope}

        \begin{scope}[yshift=1cm]
            \node [inner sep=1pt] (v51) at (4,0) {\notoemoji[height=1.2em]{desktop computer}};
            \node [inner sep=1pt,anchor=south west] (v52) at ([xshift=-0.5em] v51.north east) {\notoemoji[height=1.2em]{desktop computer}};
            \node [inner sep=1pt,anchor=south east] (v53) at ([xshift=0.5em] v51.north west) {\notoemoji[height=1.2em]{desktop computer}};
            \begin{pgfonlayer}{background}
                \node [circle,fill=black!10,inner sep=12pt] (committee5) at ([yshift=1pt] v51.north) {};
            \end{pgfonlayer}
            \node [anchor=south,align=center] (committeelabel5) at (committee5.north) {Committee\\$\mathsf{validators}_5$};
        \end{scope}

        \draw [MyAction] (committee1) -- (p1);
        \draw [MyAction] (p1) -- (committee2);
        \draw [MyAction] (committee2) -- (p2);
        \draw [MyAction] (p2) -- (committee3);
        \draw [MyAction] (committee3) -- (p3);
        \draw [MyAction] (p3) -- (committee4);
        \draw [MyAction] (committee4) -- (p4);
        \draw [MyAction] (p4) -- (committee5);
        \draw [MyAction,densely dotted,shorten >=0.7cm] (committee5) -- (4.5,0);

    \end{tikzpicture}
    \caption{%
        High-level overview of $\mathcal{T}(\mathcal{P})$'s epoch-based structure.%
    }
    \label{fig:highlevel-construction-overview}
\end{figure}
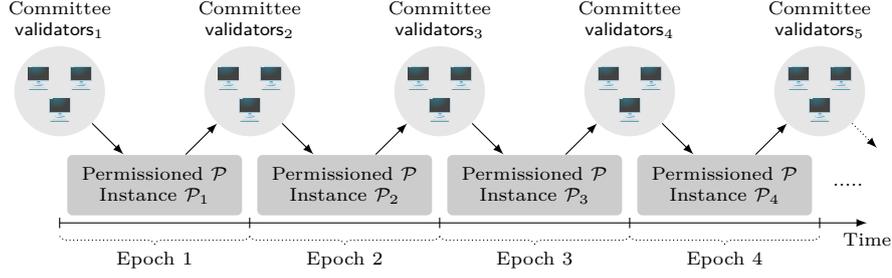

%% file: figures/alg_simulation1.tex
\begin{algorithm}[tbp]
    \caption{Permissioned to PoS transformation $\mathcal{T}$: Pseudocode for process $p$ [part 1/2]}
    \fontsize{9}{9.5}\selectfont
    \label{algorithm:pos-code_2}
    \begin{algorithmic}[1]

        \GenericBlock{\textbf{Inputs:}}
            \State Permissioned $\rho$-consistent protocol $\Pro$ with liveness parameter $\ell$
            \BlueComment{to be transformed}
        \EndGenericBlock

        \smallskip
        \GenericBlock{\textbf{Constants:}}
            \State $\mathsf{Log}$ $\CL_{\mathrm{g}}$
            \BlueComment{the genesis log; \cf \Cref{section:model_overview}}

            \State $\mathsf{Integer}$ $\mathbb{T}$ \BlueComment{the total stake; \cf \Cref{section:model_overview}}
        \EndGenericBlock

        \smallskip
        \GenericBlock{\textbf{Local variables:}}
            \State $\mathsf{Log}$ $\mathsf{log}(p) \gets \CL_{\mathrm{g}}$ \BlueComment{local log of $p$, initially set to genesis log}
            \State $\mathsf{Set}(\mathsf{Transactions})$ $\mathit{received\_txs}_p \gets \emptyset$
            \BlueComment{set of received transactions}

            \State $\mathsf{Epoch}$ $\mathit{epoch}_p \gets 0$ \BlueComment{the current epoch}

            \State $\mathsf{Set}(\mathsf{IDs})$ $\mathit{current\_validators}_p \gets \emptyset$ \BlueComment{set of $p$'s identifiers validating the current epoch}
            
            \State $\mathsf{Map}(\mathsf{Log} \to \mathsf{Boolean})$ $\mathit{signed}_p \gets \{ \mathit
        {false}$, for every log $\CL \in \mathsf{Logs} \}$
        \EndGenericBlock

        \medskip
        \At{\textbf{every timeslot $\tau$}}
            \State $\mathit{received\_txs}_p \gets$ the set of transactions received by timeslot $\tau$ (from the environment and \hphantom{||||l|}other processes)
            \State \textbf{invoke} $\mathsf{feed}(\mathit{received\_txs}_p)$ \BlueComment{forward the transactions to simulated $\Pro$}
            \LineComment{The next line can be optimized: $p$ can only gossip updates not previously gossiped.}
            \LineComment{For simplicity, we maintain the unoptimized pseudocode.}
            \State Gossip fully-certified $\mathsf{log}(p)$ and $\mathit{received\_txs}_p$ \label{line:gossip_local_log}
        \EndAt

        \smallskip
        \Upon{start}
            \State \textbf{invoke} $\mathsf{start\_simulation}\big( \mathsf{log}(p) \big)$ \label{line:start_simulation_1} \BlueComment{start simulating the first epoch; $\mathsf{log}(p) = \mathcal{L}_{\mathrm{g}}$ here}
        \EndUpon

        \smallskip
        \label{line:new_log_received}
        \Upon{receiving a fully-certified log $\CL$ such that $\CL$ extends $\mathsf{log}(p)$} \BlueComment{hence, $\mathcal{L}.\mathsf{epoch} \geq \mathit{epoch}_p$}
            \label{line:new_log_received_gossip}
            \State $\mathsf{log}(p) \gets \CL$ \BlueComment{update the local log}
            \label{line:update_log_gossip}
            \If{$\mathcal{L}.\mathsf{epoch} > \mathit{epoch}_p$ or $\mathcal{L}.\mathsf{completed} = \mathit{true}$} \BlueComment{check if new epoch should be started}
                \State \textbf{invoke} $\mathsf{stop\_simulation}$ \BlueComment{if so, stop simulation associated with the previous epoch}
                \label{line:new_epoch}
                \State \textbf{invoke} $\mathsf{start\_simulation}\big( \mathsf{log}(p) \big)$ \BlueComment{and start simulation associated with the new epoch}\label{line:start_simulation_2}
            \EndIf

        \EndUpon

        \smallskip
        \Upon{$\mathsf{obtain\_log}(\mathsf{Log} \text{ } \CL)$} \BlueComment{$\CL$ is obtained from simulated $\Pro$} \label{line:upon_obtain_log}
            
            \ForEach{$\mathit{id} \in \mathit{current\_validators}_p$}
                \ForEach{$\mathsf{Log}$ $\CL'$ with $\CL'.\mathsf{epoch} = \CL.\mathsf{epoch}$ and $\CL$ extends $\CL'$ and $\mathit{signed}_p[\CL'] = \mathit{false}$}
                    \If{no log inconsistent with $\mathcal{L}'$ has previously been signed} \label{line:check_inconsistent}
                        \State Sign $\CL'$ using the private key of $\mathit{id}$ and gossip $\CL'$ accompanied by the signature
                        \State $\mathit{signed}_p[\CL'] \gets \mathit{true}$
                        \EndIf
                \EndForEach
            \EndForEach
        \EndUpon

        \algstore{transformation}
    \end{algorithmic}
\end{algorithm}

%% file: figures/alg_simulation2.tex
\setcounter{algorithm}{0}

\begin{algorithm}[tbp]
    \caption{Permissioned to PoS transformation $\mathcal{T}$: Pseudocode for process $p$ [part 2/2]}
    \fontsize{9}{9.5}\selectfont
    \label{algorithm:simulation}
    \begin{algorithmic}[1]
    \algrestore{transformation}
        \LineComment{This section of the pseudocode is dedicated to simulating the permissioned protocol $\Pro$.} \label{line:start_pseudocode_simulation}
        
        \GenericBlock{\textbf{Local variables:}} \BlueComment{variables dedicated to simulating $\Pro$}
            \State $\mathsf{Map}([1, \mathbb{T}] \to \mathsf{IDs})$ $\mathit{id\_map}_p \gets $ empty map \BlueComment{permissioned-id to PoS-id}
            \State $\mathsf{Map}([1, \mathbb{T}] \to \mathsf{Simulation})$ $\mathit{simulation\_map}_p \gets$ empty map \BlueComment{permissioned-id to simulation}
            \State $\mathsf{Set}(\mathsf{Simulation})$ $\mathit{simulations}_p \gets \emptyset$ \BlueComment{set of $p$'s currently executed simulations}
            \State $\mathsf{Timer}$ $\mathit{epoch\_timer}_p$ \BlueComment{for measuring the duration of epochs}
        \EndGenericBlock

        \smallskip
        \GenericBlock{\textbf{Local functions:}}
            \Function{$\mathsf{Map}([1, \mathbb{T}] \to \mathsf{IDs})$}{$\mathsf{map\_stake}$}{$\mathsf{Logs} \text{ } \CL$} \label{line:map_function} \BlueComment{returns permissioned-id to PoS-id mapping}
                \State $\mathsf{Map}([1, \mathbb{T}] \to \mathsf{IDs})$ $\mathit{map} \gets $ empty map
                \State $\mathsf{List}(\mathsf{IDs}) \text{ } V \gets \CL.\mathsf{current\_ids}$ \BlueComment{find identifiers with positive stake according to $\CL$}
                \State Sort $V$ in lexicographical order
                \State $\mathsf{Integer} \text{ } \mathit{counter} \gets 1$
                \ForEach{$\mathit{id} \in V$} \BlueComment{iterate through $V$ in the ascending lexicographical order}
                    \ForEach{$j \in [1, \mathsf{S}(\CL, \mathit{id})]$} \BlueComment{associate a number of permissioned-ids equal to the stake}
                        \State $\mathit{map}[\mathit{counter}] \gets \mathit{id}$
                        \State $\mathit{counter} \gets \mathit{counter} + 1$
                    \EndForEach
                \EndForEach
                \State \textbf{return} $\mathit{map}$
            \EndFunction
        \EndGenericBlock

        \medskip
        \At{\textbf{every timeslot $\tau$}}
            \If{$\mathit{simulations}_p \neq \emptyset$} \BlueComment{check if $p$ is validating the current epoch}
                \ForEach{$S \in \mathit{simulations}_p$} 
                    \State Let $\mathcal{L} \gets \mathsf{log}(S.\mathsf{identifier}, \tau)$ \BlueComment{obtain the current log of the simulated instance $S$}
                    \If{$\mathcal{L}.\mathsf{epoch} > \mathit{epoch}_p$} \BlueComment{check if the current log is ``too long''}
                        \State $\mathcal{L} \gets \mathcal{L}.\mathsf{ep\_prefix}(\mathit{epoch}_p)$
                    \EndIf 

                    \State \textbf{trigger} $\mathsf{obtain\_log}( \mathcal{L})$ \BlueComment{report the current log}
                \EndForEach
            \EndIf
        \EndAt

        \smallskip
        \Upon{$\mathsf{start\_simulation}(\mathsf{Logs} \text{ } \CL)$} \label{line:function_start_simulation}          
            \If{$\mathcal{L}.\mathsf{completed} = \mathit{true}$} \label{line:start_establishing_genesis}
                \State $\mathit{epoch}_p \gets \CL.\mathsf{epoch} + 1$ \BlueComment{$\mathcal{L}$ is the genesis log for the epoch}
            \Else \BlueComment{$\mathcal{L}$ is a log from the ``middle'' of epoch}
                \State $\mathit{epoch}_p \gets \CL.\mathsf{epoch}$ \BlueComment{the current epoch is $\mathcal{L}$'s epoch}
                \State $\mathcal{L} \gets \mathcal{L}.\mathsf{ep\_prefix}(\mathit{epoch}_p - 1)$ \BlueComment{$(\mathit{epoch}_p - 1)$-prefix of $\mathcal{L}$ is the genesis log for the epoch} \label{line:end_establishing_genesis}
            \EndIf
            \State $\mathit{current\_validators}_p \gets \mathcal{L}.\mathsf{current\_ids} \cap \mathsf{id}(p)$ \BlueComment{set $p$'s identifiers validating the current epoch}
            \State $\mathit{id\_map}_p \gets \mathsf{map\_stake}(\CL)$ \label{line:id_map} \BlueComment{update the permissioned-id to PoS-id map}
            \ForEach{$i \in [1, \mathbb{T}]$ such that $\mathit{id\_map}_p[i] \in \mathsf{id}(p)$} \label{line:simulate_ids}
                \State Let $\mathit{simulation} \gets $ initialize $\Pro$ with permissioned-id $i$ \label{line:initialize_simulation_id}\BlueComment{instance of $\Pro$ with id $i$}
                \State Start $\mathit{simulation}$ with genesis log $\CL$ \label{line:genesis_log} \BlueComment{the genesis log for $\Pro$ with permissioned-id $i$ is $\CL$}
                \State $\mathit{simulations}_p \gets \mathit{simulations}_p \cup \{\mathit{simulation}\}$
                \State $\mathit{simulation\_map}_p[i] \gets \mathit{simulation}$ \label{line:simulation_map} 
            \EndForEach
            \If{$\mathit{current\_validators}_p \neq \emptyset$}
                \State \textbf{invoke} $\mathit{epoch\_timer}_p.\mathsf{measure}(\ed = \ell + \Delta)$ \BlueComment{if validating the epoch, start the timer} \label{line:epoch_timer_measure}
            \EndIf
        \EndUpon

        \smallskip
        \Upon{$\mathsf{stop\_simulation}$}
            \ForEach{$S \in \mathit{simulations}_p$}
                \State Stop executing $\mathcal{S}$ \label{line:stop_executing} \BlueComment{stop each instance $S$ of permissioned protocol $\Pro$}
            \EndForEach
            \LineComment{Reset the variables and cancel the timer}
            \State \textbf{invoke} $\mathit{epoch\_timer}_p.\mathsf{cancel()}$ \label{line:epoch_timer_cancel}
            \State $\mathit{id\_map}_p \gets $ empty map \label{line:start_reseting_variables}
            \State $\mathit{simulation\_map}_p \gets$ empty map
            \State $\mathit{simulations}_p \gets \emptyset$
            \State $\mathit{current\_validators}_p \gets \emptyset$ \label{line:stop_reseting_variables}
        \EndUpon

        \smallskip
        \Upon{$\mathit{epoch\_timer}_p$ expires} \BlueComment{the epoch should be completed, i.e., the timer has expired}
            \ForEach{$\mathit{id} \in \mathit{current\_validators}_p$}
                \State Gossip the $(\textsc{finish}, \mathit{epoch}_p)$ transaction issued by $\mathit{id}$ \label{line:epoch_ending_transaction}
            \EndForEach
        \EndUpon

        \smallskip
        \Upon{$S \in \mathit{simulations}_p$ sends a message $m$} \label{line:simulation_sends} \BlueComment{simulation instance $S$ sends $m$}
            \State $M \gets \langle \textsc{simulation}, \mathit{epoch}_p, m \rangle$ \BlueComment{tag $m$ with the current epoch}
            \LineComment{Send the simulation message to the PoS-id associated with the permissioned-id $m.\mathsf{receiver}$}
            \State Send $M$ to $\mathit{id\_map}_p[m.\mathsf{receiver}]$ \label{line:send_simulation_message}
        \EndUpon

        \smallskip
        \Upon{receiving a simulation message $M$ with $M.\mathsf{epoch} = \mathit{epoch}_p$ and $\mathit{simulations}_p \neq \emptyset$} \label{line:simulation_received}
            \State Forward $M.\mathsf{message}$ to $\mathit{simulation\_map}_p[M.\mathsf{message}.\mathsf{receiver}]$ \label{line:forward_received_simulation}
        \EndUpon

        \smallskip
        \Upon{$\mathsf{feed}\big( \mathsf{Set}(\mathsf{Transaction}) \text{ } \mathit{txs} \big)$}
            \ForEach{$S \in \mathit{simulations}_p$}
                \State Forward transactions $\mathit{txs}$ to $S$ \label{line:final_line_pseudocode_simulation} \BlueComment{simulation instance $S$ receives transactions $\mathit{txs}$} \label{line:forward_transactions}
            \EndForEach
        \EndUpon

    \end{algorithmic}
\end{algorithm}

%% file: figures/alg-construction-overview.tex
\begin{figure}[tbp]
    \centering
    \begin{tikzpicture}[
            x=1cm,
            y=1cm,
            MyAction/.style={
                -latex,
            }]
        \tiny

        \pgfdeclarelayer{background}
        \pgfsetlayers{background,main}

        \node [draw=none,fill=black!20,minimum width=1.25cm,rounded corners=2pt] (Lg1) at (0,0) {$\CL_{\mathrm{g}}$};
        \node [anchor=west,inner sep=0pt,xshift=0.5em] (tx11) at (Lg1.east) {$\tx_1$};
        \node [anchor=west,inner sep=0pt,xshift=2pt] (tx12) at (tx11.east) {$\tx_2$};
        \node [anchor=west,inner sep=0pt,xshift=1em] (tx13) at (tx12.east) {$...$};
        \node [anchor=west,inner sep=0pt,xshift=1em] (tx14) at (tx13.east) {$\tx_k$};
        \node [anchor=west,inner sep=0pt,xshift=2pt] (tx15) at (tx14.east) {$(\textsc{finish}, 1)_{\mathit{id}_1}$};
        \node [anchor=west,inner sep=0pt,xshift=2pt] (tx16) at (tx15.east) {$\tx_{k+1}$};
        \node [anchor=west,inner sep=0pt,xshift=2pt] (tx17) at (tx16.east) {$(\textsc{finish}, 1)_{\mathit{id}_2}$};
        \node [anchor=west,inner sep=0pt,xshift=2pt] (tx18) at (tx17.east) {$...$};
        \node [anchor=west,inner sep=0pt,xshift=2pt] (tx19) at (tx18.east) {$(\textsc{finish}, 1)_{\mathit{id}_{\frac{2\mathbb{T}}{3}}}$};

        \draw [decorate,decoration={brace,amplitude=3pt,raise=5pt,aspect=0.12},densely dotted] (tx11.north west |- tx19.north east) -- (tx19.north east) node [pos=0.12,anchor=south,yshift=7pt,align=center] (P11) {run $\CP_1$ with\\genesis log $\CL_{\mathrm{g}}$};

        \node [inner sep=1pt] (v11) at ([yshift=3em] P11.north) {\notoemoji[height=1.2em]{desktop computer}};
        \node [inner sep=1pt,anchor=south west] (v12) at ([xshift=-0.5em] v11.north east) {\notoemoji[height=1.2em]{desktop computer}};
        \node [inner sep=1pt,anchor=south east] (v13) at ([xshift=0.5em] v11.north west) {\notoemoji[height=1.2em]{desktop computer}};
        \begin{pgfonlayer}{background}
            \node [circle,fill=black!10,inner sep=12pt] (committee1) at ([yshift=1pt] v11.north) {};
        \end{pgfonlayer}
        \node [anchor=north east] (committeelabel1) at (committee1.west) {$\mathsf{validators}_1$};

        \draw [MyAction] (committee1) -- (P11);
        \draw [MyAction] (committee1) -| (tx15);
        \draw [MyAction] (committee1) -| (tx17);
        \draw [MyAction] (committee1) -| (tx19) node [pos=0.25,above,align=center] {send epoch ending transactions\\after $\ed = \ell + \Delta$ timeslots};
        \node [anchor=south,align=center] (genesislog1) at ([yshift=2pt] Lg1.north) {genesis\\log};
        \draw [MyAction] (genesislog1) |- (committee1) node [midway,above] {determines};

        \begin{pgfonlayer}{background}
            \draw [fill=black!10,draw=none,rounded corners=2pt] ([xshift=-2pt,yshift=2pt] Lg1.north west |- tx19.north east) rectangle ([xshift=2pt,yshift=-2pt] tx19.south east);
            \coordinate (overalllognorthwest) at ([xshift=-2pt,yshift=2pt] Lg1.north west |- tx19.north east);
            \coordinate (overalllogsouthwest) at ([xshift=-2pt,yshift=-2pt] Lg1.south west |- tx19.south east);
            \coordinate (overalllognortheast) at ([xshift=2pt,yshift=2pt] tx19.north east);
            \coordinate (overalllogsoutheast) at ([xshift=2pt,yshift=-2pt] tx19.south east);
        \end{pgfonlayer}

        \draw [decorate,decoration={brace,amplitude=3pt,raise=3pt}] (overalllogsouthwest) -- (overalllognorthwest) node [rotate=90,anchor=south,yshift=5pt,align=center] {log of PoS protocol};

        \draw [decorate,decoration={brace,mirror,amplitude=3pt,raise=3pt},densely dotted] (overalllogsouthwest) -- (overalllogsoutheast) node [midway,anchor=north,yshift=-6pt,align=center] (epoch1) {output log $\CL_1$ of epoch $1$ produced by $\CP_1$ with $\CL_{\mathrm{g}}$ prepended};

        \node [anchor=west,inner sep=0pt,xshift=1.5em] (tx21) at (tx19.east) {$\tx_{k+2}$};
        \node [anchor=west,inner sep=0pt,xshift=1em] (tx22) at (tx21.east) {$\tx_{k+3}$};
        \node [anchor=west,inner sep=0pt,xshift=1em] (tx23) at (tx22.east) {$... ...$};
        \node [anchor=west,inner sep=0pt,xshift=1em] (tx28) at (tx23.east) {$...$};
        \node [anchor=west,inner sep=0pt,xshift=2pt] (tx29) at (tx28.east) {$(\textsc{finish}, 1)_{\mathit{id}_{\frac{2\mathbb{T}}{3}}}$};

        \draw [decorate,decoration={brace,amplitude=3pt,raise=5pt,aspect=0.25},densely dotted] (tx21.north west |- tx29.north east) -- (tx29.north east) node [pos=0.25,anchor=south,yshift=7pt,align=center] (P21) {run $\CP_2$ with\\genesis log $\CL_1$};

        \node [inner sep=1pt] (v21) at ([yshift=3em] P21.north) {\notoemoji[height=1.2em]{desktop computer}};
        \node [inner sep=1pt,anchor=south west] (v22) at ([xshift=-0.5em] v21.north east) {\notoemoji[height=1.2em]{desktop computer}};
        \node [inner sep=1pt,anchor=south east] (v23) at ([xshift=0.5em] v21.north west) {\notoemoji[height=1.2em]{desktop computer}};
        \begin{pgfonlayer}{background}
            \node [circle,fill=black!10,inner sep=12pt] (committee2) at ([yshift=1pt] v21.north) {};
        \end{pgfonlayer}
        \node [anchor=north east] (committeelabel2) at (committee2.west) {$\mathsf{validators}_2$};

        \draw [MyAction] (committee2) -- (P21);
        \draw [MyAction] (committee2) -| (tx29);
        \draw [MyAction] (epoch1) -| ($(tx19.east)!0.5!(tx21.west)$) |- (committee2) node [midway,above] {determines};

        \coordinate (overalllogsoutheast2) at ([xshift=2pt,yshift=-2pt] tx29.south east);

        \draw [decorate,decoration={brace,mirror,amplitude=3pt,raise=18pt},densely dotted] (overalllogsouthwest) -- (overalllogsoutheast2) node [midway,anchor=north,yshift=-21pt,align=center] (epoch2) {output log $\CL_2$ of epoch $2$ produced by $\CP_2$ with $\CL_1$ prepended};

        \node [anchor=north,yshift=0.65em,xshift=2cm] (epoch3) at (epoch2.south) {$\ddots$};

    \end{tikzpicture}
    \caption{
    The quasi-permissionless PoS protocol $\mathcal{T}(\mathcal{P})$ proceeds in epochs (see \cref{fig:highlevel-construction-overview}). At each epoch $e$, a set of identifiers $\mathsf{validators}_e$ run a new instance $\Pro_e$ of the permissioned protocol $\Pro$; the genesis log for $\Pro_e$ is the log output in epoch $e - 1$.
    $\ed = \ell + \Delta$ timeslots into the execution of $\Pro_e$, the validators issue a special epoch-ending transaction $(\textsc{finish}, e)$. 
    Once there are two-thirds-stake worth of finalized epoch-ending transactions, each validator stops executing the protocol $\Pro_e$.
    The log at that point determines the validators for the epoch $e+1$, and is input as the genesis log to $\Pro_{e+1}$.%
    }
    \label{fig:construction-overview}
\end{figure}

%% file: sections/50_extensions.tex
\section{Extensions}
\label{section:extensions}

We now present some promising directions that can further improve our transformation.

\smallskip
\noindent\textbf{Beyond public key infrastructure.}
In this work, we assumed the use of digital signatures (\cf \Cref{section:model_overview}).
However, many permissioned protocols rely on ``heavier'' cryptographic primitives (e.g.,~\cite{hotstuff,lewis2022quadratic,civit2022byzantine}) such as threshold signatures.
Importantly, our transformation can easily be adapted for these protocols, provided that the cryptographic primitives necessary for the underlying permissioned protocol are established in each epoch.
Concretely, any setup procedure necessary for the permissioned protocol being transformed should be treated as an integral \emph{part} of the permissioned protocol itself.
There exists a body of work on proactive secret sharing across dynamic committees (e.g., \cite{MaramZWLZJS19,GoyalKMPS22,DBLP:conf/tcc/BenhamoudaG0HK020}), which can be useful for avoiding a fresh DKG (or trusted) setup and public parameters for each epoch. 
Finally, we note that, to defeat long range attacks~\cite{deirmentzoglou2019survey}, key-erasure techniques~\cite{algorandchen} can be employed on the PoS protocol, so that the processes that have become passive, even if corrupted in the future, cannot create a log for the past epochs in retrospect.

\smallskip
\noindent\textbf{On transforming weighted permissioned protocols.}
For simplicity, we have presented our transformation assuming that the underlying permissioned protocol assigns unit weight per validator---that is, it does not support heterogeneous weights.  
Then, each PoS validator 
must run one ``virtual'' validator in the permissioned protocol \emph{per unit of stake}.
This ``virtualization'' can be inefficient,
especially when large amounts of stake are concentrated among few stakeholders.
Note, however, that our transformation applies equally well to weighted permissioned protocols, where each PoS validator simply runs a single permissioned validator with weight proportional to its stake.
This avoids the overhead of ``virtualization''.
Notably, many widely used permissioned protocols---such as Tendermint~\cite{buchman2018latest}, PBFT~\cite{pbft}, and HotStuff~\cite{hotstuff}---offer weighted variants, allowing our transformation to pass through the stake weights directly, and for the resulting PoS protocol to operate more efficiently.

\smallskip
\noindent\textbf{On lotteries and incentive mechanisms.}
Our transformation is agnostic to the choice of the underlying permissioned protocol; as long as the protocol supports lotteries (e.g., probabilistic leader selection), the resulting PoS protocol naturally inherits this property. 
State updates that reward participating processes can be incorporated seamlessly, though such incentive functionality belongs to the ``virtual machine'' layer rather than the consensus mechanism itself.
Importantly, \Cref{theorem:accountability} ensures that the PoS protocol produced by our compiler satisfies accountability. 
This, in turn, enables the implementation of slashing mechanisms---an essential aspect of cryptoeconomic incentive design.

\smallskip
\noindent\textbf{On establishing the EAAC property.}
Our transformation can be extended to enable the EAAC (``expensive to attack in the absence of collapse'') property introduced in~\cite{DBLP:journals/corr/abs-2304-14701}.
Informally, the EAAC property captures a strong form of security in PoS protocols: it guarantees that launching a successful attack is prohibitively costly.
More precisely, if an adversary attempts to violate the protocol's guarantees, the property ensures that the faulty participants responsible for the attack will be penalized by having their stake slashed. 
At the same time, the property protects correct participants by ensuring they remain unharmed and retain their stake, even during adversarial conditions.
Given the technical complexity and detailed formalism of the EAAC property, we omit the full formal definition here and instead provide the aforementioned informal description. 
We encourage interested readers to consult~\cite{DBLP:journals/corr/abs-2304-14701} for the complete and rigorous treatment of the EAAC property.

\mysubparagraphNoDot{Obtaining EAAC using our transformation.}
It is established in~\cite{DBLP:journals/corr/abs-2304-14701} that no non-trivial EAAC guarantees can be provided by any quasi-permissionless protocol within the standard partially synchronous model: if the adversary is strong enough to cause consistency violations, it can also evade punishment.
However, the same work demonstrates that, under a stronger notion of partial synchrony---one that imposes a pessimistic upper bound on message delays $\Delta'$ (potentially orders of magnitude greater than $\Delta$) even before GST---it becomes possible to design a quasi-permissionless protocol that satisfies the EAAC property, assuming the adversary controls less than two-thirds of the total stake in every execution.
We follow this approach: our transformation, in addition to the properties already discussed, enables the resulting quasi-permissioned protocol to satisfy the EAAC property assuming that this pessimistic bound $\Delta'$ on message delays (even before GST) holds.
Importantly, the resulting protocol satisfies EAAC, regardless of whether the original permissioned protocol does.

Let us now provide modifications to our transformation $\mathcal{T}(\mathcal{P})$ sufficient for satisfying the EAAC property (assuming the pessimistic bound $\Delta'$ on message delays): 
\begin{compactenum}
     \item Each epoch $e$ is executed for a duration (at least) proportional to the pessimistic bound on message delays $\Delta'$.
     Concretely, each epoch is executed for more than $2\Delta'$ time.

    \item When a process $p$ validating epoch $e$ receives a fully-certified log $\mathcal{L}$, the process (1) signs a message $m = \langle \textsc{confirm}, \mathcal{L} \rangle$, (2) sends $m$ to all identifiers validating epoch $e$, and (3) gossips the received signatures (that fully-certified $\mathcal{L}$).
    We underline that modified $\mathcal{T}(\mathcal{P})$ includes two rounds of voting on a log: the first makes logs fully-certified (and is present in original, EAAC-less $\mathcal{T}(\mathcal{P})$), and the second revolves around the aforementioned \textsc{confirm} messages (unique to $\mathcal{T}(\mathcal{P})$ modified for EAAC). 

    \item When a process $p$ receives two-thirds-stake worth of \textsc{confirm} messages for some log $\mathcal{L}$, process $p$ ``packs together'' the received \textsc{confirm} signatures and gossips $\mathcal{L}$ along with the signatures.

    \item A process $p$ outputs a log $\mathcal{L}$ (i.e., sets $\mathsf{log}(p)$ to $\mathcal{L}$) only upon receiving two-thirds-stake worth of signatures on $\langle \textsc{confirm}, \mathcal{L} \rangle$.

    \item Finally, once a consistency violation occurs, the processes repeatedly initiate instances of the Dolev-Strong protocol~\cite{DBLP:journals/siamcomp/DolevS83}.
    This protocol is used to collectively agree on an updated genesis block, which incorporates slashing penalties for any processes identified as faulty during the violation.
    When this updated genesis block is established, the system can safely resume normal execution from this new agreed-upon state.
    This recovery procedure follows the approach detailed in~\cite[Algorithm 2]{DBLP:journals/corr/abs-2304-14701}.
\end{compactenum}

\mysubparagraphNoDot{Why do these modifications enable the EAAC property?}
If there is a consistency violation in an epoch $e$, there exists a correct process $p$ (resp., $q$) validating epoch $e$ that receives two-thirds-stake worth of signatures on some log $\mathcal{L}_p$ (resp., $\mathcal{L}_q$) \emph{while being in epoch $e$}; these are the signatures making inconsistent logs $\mathcal{L}_p$ and $\mathcal{L}_q$ fully-certified.
As (1) both $p$ and $q$ gossip the received signatures, (2) message delays are bounded by $\Delta'$ even before GST, and (3) epoch $e + 1$ is executed for a period of time proportional to $\Delta'$, all validators of epoch $e + 1$ receive the conflicting signatures while in epoch $e + 1$ and ensure the slashing of the responsible identifiers (via the Dolev-Strong recovery procedure).
Moreover, no correct process is ever slashed as no correct process (i.e., validator) ever signs conflicting logs, which means that no proof of guilt of a correct process could ever be produced.

%% file: sections/60_relatedwork.tex
\section{Related Work}
\label{section:relatedwork}

Due to space constraints, expansive comparison to related work,
including on \emph{group membership \& view synchronous communication}~\cite{DBLP:journals/tocs/BirmanJ87,cachin2011introduction,DBLP:journals/csur/ChocklerKV01,DBLP:journals/tocs/BirmanJ87,DBLP:conf/icdcs/SchiperS93,DBLP:journals/tdsc/AmirNST05,spread,DBLP:conf/icdcs/MoserAMA94,DBLP:conf/podc/Bracha84} and on \emph{deterministic reconfiguration in asynchrony}~\cite{fischer1985impossibility,DBLP:journals/jacm/AguileraKMS11,DBLP:conf/opodis/AlchieriBGF17,DBLP:conf/wdag/JehlVM15,DBLP:conf/opodis/KuznetsovRP19,DBLP:conf/wdag/GafniM15,DBLP:conf/wdag/SpiegelmanKM17,DBLP:conf/opodis/GuerraouiKKPST20,DBLP:journals/tdsc/CamaioniGKMRVV25,kuznetsov2020asynchronous},
is relegated to \cref{section:relatedwork-long}.

For \emph{PoS blockchain protocols}, Ethereum~\cite{ethereum_consensus_specs,casperffg,gasper} is the largest PoS blockchain by market cap, but
the idea of PoS traces back to the Bitcoin community~\cite{cryptocurrencywithoutproofofwork,bentov2014proof,cunicula2011proof,tacotime2013netcoin,quantummechanic2011pos,maxwell2014impossibility,peercoin,bitcoin}.
The first provably-secure PoS protocols include SnowWhite~\cite{snowwhite,DBLP:conf/asiacrypt/PassS17} and Ouroboros Praos~\cite{praos,ouroboros}.
Hybrid Consensus~\cite{pass2017hybrid,byzcoin}
proceeds in epochs, each of which has an associated permissioned consensus instance, with epoch transition and output log construction similar to that of our transformation.
Lewis--Pye and Roughgarden~\cite{DBLP:journals/corr/abs-2304-14701} 
and Budish et al.~\cite{DBLP:journals/corr/abs-2405-09173} 
transform HotStuff~\cite{hotstuff} and Tendermint~\cite{buchman2018latest}, respectively, from the permissioned to the PoS setting.
PaLa~\cite{pala} is a partially synchronous protocol
with built-in reconfiguration.
Sui Lutris~\cite{suilutris} features a unique reconfiguration mechanism for 
its combined partially-ordering/totally-ordering consensus mechanism~\cite{abc2019,guerraoui2019consensus,baudet2020fastpay}.
There is a substantial line of work on \emph{reconfiguration of replicated state-machines}~\cite{DBLP:journals/tocs/Lamport98,lamport2009vertical,lamport2010reconfiguring,lamport2009stoppable,abraham2016bvp}.
Systems implementing reconfiguration include BFT-SMaRt~\cite{bessani2014state} and Raft~\cite{DBLP:conf/usenix/OngaroO14}.
Duan and Zhang~\cite{dynamicbftfoundations} propose Dyno, a family of total-order broadcast protocols with reconfiguration carefully woven in in a bespoke manner.
To the best of our knowledge, no earlier work
describes a closed-box transformation from permissioned to PoS consensus
that preserves and provides the range of desirable properties studied in this paper.

%% file: _references_used.tex
\nocite{abc2019}
\nocite{abraham2016bvp}
\nocite{algorandchen}
\nocite{algorandsosp}
\nocite{baudet2020fastpay}
\nocite{bchain}
\nocite{bentov2014proof}
\nocite{bessani2014state}
\nocite{bitcoin}
\nocite{bitcoinbackbone}
\nocite{buchman2016tendermint}
\nocite{buchman2018latest}
\nocite{byzcoin}
\nocite{Cachin2001}
\nocite{cachin2011introduction}
\nocite{casperffg}
\nocite{CGG19}
\nocite{civit2022byzantine}
\nocite{cosmos2023staking}
\nocite{crime_punishment}
\nocite{cryptocurrencywithoutproofofwork}
\nocite{cunicula2011proof}
\nocite{DBLP:conf/asiacrypt/PassS17}
\nocite{DBLP:conf/ccs/MillerXCSS16}
\nocite{DBLP:conf/ccs/ShengWNKV21}
\nocite{DBLP:conf/ccs/SpiegelmanGSK22}
\nocite{DBLP:conf/eurosys/DanezisKSS22}
\nocite{DBLP:conf/fc/NeuTT22}
\nocite{DBLP:conf/fc/NeuTT24}
\nocite{DBLP:conf/icdcs/MoserAMA94}
\nocite{DBLP:conf/icdcs/SchiperS93}
\nocite{DBLP:conf/opodis/AlchieriBGF17}
\nocite{DBLP:conf/opodis/GuerraouiKKPST20}
\nocite{DBLP:conf/opodis/KuznetsovRP19}
\nocite{DBLP:conf/podc/Bracha84}
\nocite{DBLP:conf/podc/KeidarKNS21}
\nocite{DBLP:conf/sp/NeuTT21}
\nocite{DBLP:conf/tcc/BenhamoudaG0HK020}
\nocite{DBLP:conf/usenix/OngaroO14}
\nocite{DBLP:conf/wdag/GafniM15}
\nocite{DBLP:conf/wdag/JehlVM15}
\nocite{DBLP:conf/wdag/SpiegelmanKM17}
\nocite{DBLP:journals/corr/abs-2304-14701}
\nocite{DBLP:journals/corr/abs-2306-03058}
\nocite{DBLP:journals/corr/abs-2310-14821}
\nocite{DBLP:journals/corr/abs-2405-09173}
\nocite{DBLP:journals/corr/abs-2405-20488}
\nocite{DBLP:journals/csur/ChocklerKV01}
\nocite{DBLP:journals/dc/AlpernS87}
\nocite{DBLP:journals/ipl/AlpernS85}
\nocite{DBLP:journals/jacm/AguileraKMS11}
\nocite{DBLP:journals/siamcomp/DolevS83}
\nocite{DBLP:journals/tdsc/AmirNST05}
\nocite{DBLP:journals/tdsc/CamaioniGKMRVV25}
\nocite{DBLP:journals/tocs/BirmanJ87}
\nocite{DBLP:journals/tocs/Lamport98}
\nocite{deirmentzoglou2019survey}
\nocite{DLS88}
\nocite{dolev1985bounds}
\nocite{dynamicbftfoundations}
\nocite{easy-accountability}
\nocite{ethereum_consensus_specs}
\nocite{fischer1985impossibility}
\nocite{fullversionpreprint}
\nocite{gasper}
\nocite{GoyalKMPS22}
\nocite{guerraoui2019consensus}
\nocite{HKD07}
\nocite{hotstuff}
\nocite{KermarrecS07}
\nocite{kuznetsov2020asynchronous}
\nocite{kwon2014tendermint}
\nocite{lamport2009stoppable}
\nocite{lamport2009vertical}
\nocite{lamport2010reconfiguring}
\nocite{lewis2022quadratic}
\nocite{lewis2025beyond}
\nocite{MaramZWLZJS19}
\nocite{maxwell2014impossibility}
\nocite{ouroboros}
\nocite{pala}
\nocite{pass2017hybrid}
\nocite{pbft}
\nocite{peercoin}
\nocite{praos}
\nocite{quantummechanic2011pos}
\nocite{snowwhite}
\nocite{spread}
\nocite{suilutris}
\nocite{tacotime2013netcoin}
\nocite{Tang2024}
\nocite{thunderella}

%% file: appendix/20_modelprelims.tex
\section{System Model: Detailed Description}
\label{section:preliminaries}

This section describes our model in full detail.
(Recall that the overview is given in \Cref{section:model_overview}.)
We start by introducing processes and how they communicate (\Cref{subsection:processes_communication}).
Then, we define the standard partially synchronous model of communication, the environment that issues transactions to processes and the adversary (\Cref{subsection:partial_synchrony_environment_adversary}).
Finally, we define permissioned and quasi-permissionless protocols with a particular focus on the knowledge processes possess within these protocols (\Cref{subsection:permissioned_quasi_permissionless}).

\subsection{Processes \& Means of Communication} \label{subsection:processes_communication}

\mysubparagraphNoDot{Processes.} 
We consider a (potentially infinite) set of processes denoted by $\Pi$.
Each process $p \in \Pi$ is assigned a non-empty and potentially infinite set of \emph{identifiers}, denoted by $\mathsf{id}(p)$.
Intuitively, $\mathsf{id}(p)$ determines the set of public keys for which process $p$ knows the corresponding private key (\cf the ``Public key encryption scheme'' paragraph below); process $p$ can use its identifiers to create sybils.
Importantly, identifier sets are disjoint:
\begin{equation*}
    \forall (p_1, p_2) \in \Pi^2: p_1 \neq p_2 \implies \mathsf{id}(p_1) \cap \mathsf{id}(p_2) = \emptyset.
\end{equation*}
We denote by $\mathsf{IDs}$ the set of all identifiers.
We note that there may exist an identifier $\mathit{id} \in \mathsf{IDs}$ not allocated to any process, i.e., $\forall p \in \Pi: \mathit{id} \notin \mathsf{id}(p)$.

\mysubparagraphNoDot{Time \& process allocation.}
Time is divided into discrete timeslots whose range is $\IN_{\geq 0}$.
Each process may or may not be \emph{active} at each timeslot.
To accommodate for clock drifts, our model permits processes to be idle even at timeslots at which they are active.
Concretely, at each timeslot at which a process is active, the process can either be \emph{waiting} or \emph{not waiting}.
(The slower a process's clock, the more frequently the process waits at active timeslots.)
A \emph{process allocation} is a function specifying, for each process $p \in \Pi$, the timeslots at which process $p$ is active and, for each such timeslot, if the process is waiting or not.

\mysubparagraphNoDot{Inputs.}
Each process $p \in \Pi$ is given a finite set of \emph{inputs}, which captures its knowledge at the beginning of the execution.
If a variable is specified as one of $p$'s inputs, we refer to it as \emph{determined for $p$}; otherwise, the variable is \emph{undetermined for $p$}.
If a variable is determined (resp., undetermined) for every process $p \in \Pi$, we say that the variable is \emph{determined} (resp., \emph{undetermined}).

\mysubparagraphNoDot{Public key infrastructure.}
In this work, we focus on protocols assuming a public key infrastructure (PKI) that allows processes to sign their messages and verify messages received from other processes.
Concretely, each process $p$ associates each $\mathit{id} \in \mathsf{id}(p)$ with a pair of \emph{private} and \emph{public keys}; the public key is disseminated to other processes, whereas the private key is kept secret.
Note that processes associate public keys with identifiers, and not with processes as the identifier function $\mathsf{id}(\cdot)$ is unknown (\cf \Cref{subsection:permissioned_quasi_permissionless} for more details).
A process $p$ can send a message $m$ signed using the private key of an identifier $\mathit{id}$ if and only if (1) $\mathit{id} \in \mathsf{id}(p)$, or (2) $p$ has previously received $m$ and an $\mathit{id}$'s signature on $m$.

\mysubparagraphNoDot{Communication.}
At each timeslot, each non-waiting active process may \emph{send} a finite set of messages on behalf of (any) one of its identifiers to any other specific identifier.
Intuitively, we assume that each identifier is associated with its IP address, enabling messages to be sent to that identifier directly.
Similarly, at each timeslot, each non-waiting active process may receive (a possibly empty) set of messages sent to its identifiers.
Waiting active or inactive processes do not send or receive any messages.

We also assume that processes may send their messages to the entire universe of processes.
Specifically, at each timeslot, each non-waiting active process may \emph{gossip} a finite set of messages on behalf of (any) one of its identifiers to \emph{all} other identifiers.
In practice, this dissemination can be implemented by some version of a gossip protocol~\cite{KermarrecS07}.

\subsection{Partial Synchrony \& Adversary} \label{subsection:partial_synchrony_environment_adversary}

\mysubparagraphNoDot{Partial synchrony.}
In this work, we consider the standard partially synchronous model~\cite{DLS88}.
Informally, the partially synchronous model is a hybrid between full asynchrony and full synchrony: an unknown timeslot exists where the system switches from acting asynchronously to behaving synchronously.
Formally, each execution is associated with an unknown timeslot $\text{GST} \in \IN_{\geq 0}$ such that the following two conditions hold:
\begin{compactitem}
    \item If any correct process $p$ is active at any timeslot $\tau \geq \text{GST}$, then $p$ is not waiting at timeslot $\tau$.
    Intuitively, our model forbids local clocks from drifting after GST, while allowing them to drift before GST.
    (We emphasize that all our results can easily be extended to the model in which bounded clock drifts occur even after GST.)

    \item There exists known duration $\Delta \in \IN_{\geq 1}$ such that the following is satisfied:
    \begin{compactitem}
        \item If (1) a process sends a message at any timeslot $\tau$ to any identifier $\mathit{id}$, and (2) process $p$ with $\mathit{id} \in \mathsf{id}(p)$ is active and non-waiting
        at some timeslot $\tau' \geq \max(\tau, \text{GST}) + \Delta$, then $p$ receives the message at a timeslot $\leq \tau'$.

        \item If (1) a process gossips a message at any timeslot $\tau$, and (2) another process $p$ is active and non-waiting
        at some timeslot $\tau' \geq \max(\tau, \text{GST}) + \Delta$, then $p$ receives that dissemination at a timeslot $\leq \tau'$.
    \end{compactitem}
    That is, message delays are bounded by $\Delta$ after GST.
\end{compactitem}
Lastly, each execution is associated with an unknown duration $\delta$ that denotes the \emph{actual} bound on message delays.
Concretely, $\delta \in \IN_{\geq 1}$ is the smallest duration such that if (1) a process sends a message at any timeslot $\tau$ to any identifier $\mathit{id}$, and (2) process $p$ with $\mathit{id} \in \mathsf{id}(p)$ is active and non-waiting at some timeslot $\tau' \geq \max(\tau, \text{GST}) + \delta$, then $p$ receives the message at a timeslot $\leq \tau'$.
The same holds for disseminated messages.
Naturally, $\delta \leq \Delta$. 

\mysubparagraphNoDot{Adversary.}
We consider a static adversary that corrupts a fraction of all processes at the beginning of each execution.
A corrupted process is said to be \emph{faulty}, whereas a non-faulty process is said to be \emph{correct}.
Faulty processes may behave arbitrarily (i.e., we consider Byzantine failures), while correct processes adhere strictly to the prescribed protocol.

\subsection{Permissioned Setting vs. Quasi-Permissionless Setting} \label{subsection:permissioned_quasi_permissionless}

We now examine the additional constraints that arise in each of the two settings.

\medskip
\noindent\textbf{Permissioned setting.}
In the permissioned setting, the set of processes $\Pi$ is finite and known.
Concretely, $|\Pi| = n$, for some $n \in \IN_{\geq 1}$; the number of processes $n$ is also known.
Moreover, each process has a single identifier:
\begin{equation*}
    \forall p \in \Pi: \mathsf{id}(p) = \{ p \}.
\end{equation*}
Moreover, the identifier function $\mathsf{id}(p)$ is also known.
Lastly, each process is active at every timeslot.

\mysubparagraphNoDot{Defining static $\rho$-bounded adversaries.}
A static $\rho$-bounded adversary, for any $\rho \in [0, 1]$, corrupts at most $\rho \cdot n$ processes at the beginning of each execution.

\mysubparagraphNoDot{Determined vs. undetermined inputs.}
The following inputs are determined:
\begin{compactitem}
    \item the set of processes $\Pi$, its finite cardinality $n = |\Pi|$, and the identifier function $\mathsf{id}(\cdot)$;

    \item the bound on the power of the adversary $\rho$, the stake distribution function $\mathsf{S}(\cdot, \cdot)$, and the genesis log;

    \item the upper-bound on message delays $\Delta$.

\end{compactitem}
In contrast, the following inputs are undetermined:
\begin{compactitem}
    \item the set of corrupted processes and its cardinality;

    \item GST and the actual bound on message delays $\delta \leq \Delta$; 

    \item the process allocation function: any specific process does not know if any other specific process is waiting or not at any specific timeslot $\tau$ (as the process does not know if $\tau \geq \text{GST}$).
\end{compactitem}

\medskip
\noindent\textbf{Quasi-permissionless setting.}
In the quasi-permissionless setting, the set of process $\Pi$ is not necessarily finite.
Moreover, processes might have more than a single associated identifier.
In the quasi-permissionless setting, only processes with non-zero stake are guaranteed to be active.
Specifically, for any timeslot $\tau$ and any correct process $p$ for which there exist a $\tau$-active correct process $q$ and an identifier $\mathit{id}_p \in \mathsf{id}(p)$ with $\mathsf{S}( \mathsf{log}(q, \tau), \mathit{id}_p ) > 0$, process $p$ is active at $\tau$.

\mysubparagraphNoDot{Defining static $\rho$-bounded adversaries.}
For every correct process $p$ and every timeslot $\tau$, for any $\LOG$ that is a prefix of $\mathsf{log}(p, \tau)$, at most $\rho$ fraction of $\LOG$'s total stake (i.e., $\LOG.\mathsf{total\_stake} = \mathbb{T}$) belongs to identifiers associated with faulty processes according to the stake distribution specified by $\LOG$.

\mysubparagraphNoDot{Determined vs. undetermined inputs.}
In the quasi-permissionless setting, the following inputs are determined:
\begin{compactitem}
    \item the bound on the power of the adversary $\rho$, the stake distribution function $\mathsf{S}(\cdot, \cdot)$, and the genesis log;

    \item the upper-bound on message delays $\Delta$.
\end{compactitem}
The following inputs are undetermined:
\begin{compactitem}
    \item the set of processes $\Pi$, its cardinality, and the identifier function $\mathsf{id}(\cdot)$;

    \item the set of corrupted processes and its cardinality;

    \item GST and the actual bound on message delays $\delta \leq \Delta$;

    \item the process allocation function.
\end{compactitem}

%% file: appendix/10_quit_enhanced.tex
\section{Permissioned Blockchain Protocols} \label{section:permissioned_blockchain}

In this section, we present a formal analysis of permissioned blockchain protocols.
We begin by defining the computational model underlying permissioned blockchain protocols (\Cref{subsection:quit_computational model}).
Next, we formally model executions of these protocols (\Cref{subsection:behaviors_executions}).
We then introduce the definitions of both standard and quit-enhanced permissioned blockchain protocols (\Cref{subsection:standard_quit_enhanced}).
Finally, we prove that the quit-enhanced protocols maintain the key properties of their standard counterparts (\Cref{subsection:quit_preservation}).
For simplicity, any reference to a "blockchain protocol" in this section implies a permissioned blockchain protocol.

\subsection{Computational Model}\label{subsection:quit_computational model}

Formally, a blockchain protocol $\mathcal{P}$ is a tuple
\begin{equation*}
    (\Pi, \rho, \mathsf{States}, \{ s_p^0 \,|\, p \in \Pi \}, \mathsf{Messages}, \mathsf{Transactions}, \mathsf{Logs}, \mathsf{Transition}),
\end{equation*}
as we explain below.

\mysubparagraphNoDot{Processes \& resilience.}
We denote by $\Pi$ the set of processes executing the blockchain protocol $\mathcal{P}$.
Moreover, we denote by $\rho \in [0, 1]$ the resilience of $\mathcal{P}$; we say that $\mathcal{P}$ is a $\rho$-resilient blockchain protocol.

\mysubparagraphNoDot{States.}
We denote by $\mathsf{States}$ the set of states processes can take while executing $\mathcal{P}$.
Each state $s \in \mathsf{States}$ encodes the following information:
\begin{compactitem}
    \item the process associated with $s$, denoted by $s.\mathsf{process}$;

    \item the log associated with $s$ (might be $\bot$), denoted by $s.\mathsf{log}$;

    \item if $s$ is the special quitting state, denoted by $s.\mathsf{quit} \in \{ \mathit{true}, \mathit{false} \}$.
\end{compactitem}
For each process $p \in \Pi$, there exists the \emph{initial state} $s_p^0 \in \mathsf{States}$ such that (1) $s_p^0.\mathsf{process} = p$, (2) $s_p^0.\mathsf{log} = \bot$, and (3) $s_p^0.\mathsf{quit} = \mathit{false}$. 
For every non-initial state $s \in \mathsf{States}$, $s.\mathsf{log} \neq \bot$.

\mysubparagraphNoDot{Messages.}
We denote by $\mathsf{Messages}$ the set of messages processes can send and receive while executing $\mathcal{P}$.
Each message $m \in \mathsf{Messages}$ encodes the following information:
\begin{compactitem}
    \item the sender of $m$, denoted by $m.\mathsf{sender}$;

    \item the receiver of $m$, denoted by $m.\mathsf{receiver}$.
\end{compactitem}

\mysubparagraphNoDot{Transactions.}
We denote by $\mathsf{Transactions}$ the set of transactions processes can receive from the environment while executing $\mathcal{P}$.
Each transaction $\mathit{tx} \in \mathsf{Transactions}$ encodes the following information:
\begin{compactitem}
    \item the process that receives $\mathit{tx}$, denoted by $\mathit{tx}.\mathsf{process}$;

    \item if $\mathit{tx}$ is the special starting transaction, denoted by $\mathit{tx}.\mathsf{start} \in \{ \mathit{true}, \mathit{false} \}$;

    \item if $\mathit{tx}$ is the special quitting transaction, denoted by $\mathit{tx}.\mathsf{quit} \in \{ \mathit{true}, \mathit{false} \}$;

    \item the log associated with $\mathit{tx}$ (might be $\bot$), denoted by $\mathit{tx}.\mathsf{log}$.
\end{compactitem}
The following holds for each transaction $\mathit{tx} \in \mathsf{Transactions}$: $\mathit{tx}.\mathsf{log} \neq \bot$ if and only if $\mathit{tx}.\mathsf{start} = \mathit{true}$.
Moreover, there is no transaction $\mathit{tx} \in \mathsf{Transactions}$ such that $\mathit{tx}.\mathsf{start} = \mathit{tx}.\mathsf{quit} = \mathit{true}$.
Finally, for each process $p \in \Pi$ and each log $\mathcal{L} \in \mathsf{Logs}$, there exists a transaction $\mathit{tx} \in \mathsf{Transactions}$ such that (1) $\mathit{tx}.\mathsf{process} = p$, (2) $\mathit{tx}.\mathsf{start} = \mathit{true}$, (3) $\mathit{tx}.\mathsf{quit} = \mathit{false}$, and (4) $\mathit{tx}.\mathsf{log} = \mathcal{L}$.

\mysubparagraphNoDot{State-transition function.}
We denote by $\mathsf{Transition}$ the state-transition function of $\mathcal{P}$ that maps (1) a state, (2) a boolean, (3) a set of messages, and (4) a set of transactions into (a) a new state, and (b) a set of messages.
Formally, given (1) a state $s \in \mathsf{States}$, (2) a set of messages $M^R \subseteq \mathsf{Messages}$ such that, for every message $m \in M^R$, $m.\mathsf{receiver} = s.\mathsf{process}$, and (3) a set of transactions $T^R \subseteq \mathsf{Transactions}$ such that for every transaction $\mathit{tx} \in T^R$, $\mathit{tx}.\mathsf{process} = s.\mathsf{process}$,
$\mathsf{Transition}(s, \mathit{waiting}, M^R, T^R) = (s', M^S)$, for some $s'$ and $M^S$, where
\begin{compactitem}
    \item $s' \in \mathsf{States}$ is a state for which $s'.\mathsf{process} = s.\mathsf{process}$;

    \item $M^S \subsetneq \mathsf{Messages}$ is a set of messages such that, for every message $m \in M^S$, $m.\mathsf{sender} = s.\mathsf{process}$;

    \item if $\mathit{waiting} = \mathit{true}$, then $s' = s$ and $M^S = \emptyset$.
\end{compactitem}
Given $\mathsf{Transition}(s, \mathit{waiting}, M^R, T^R) = (s', M^S)$, the following holds:
\begin{compactitem}
    \item if $\mathit{waiting} = \mathit{true}$, then $s' = s$ and $M^S = \emptyset$;

    \item if $s.\mathsf{quit} = \mathit{false}$ and there does not exist a transaction $\mathit{tx} \in T^R$ with $\mathit{tx}.\mathsf{quit} = \mathit{true}$, then $s'.\mathsf{quit} = \mathit{false}$;

     \item if $s.\mathsf{quit} = \mathit{false}$ and there exists a transaction $\mathit{tx} \in T^R$ with $\mathit{tx}.\mathsf{quit} = \mathit{true}$, then $s'.\mathsf{quit} = \mathit{true}$, $s'.\mathsf{log} = s.\mathsf{log}$ and $M^S = \emptyset$.

    \item if $s.\mathsf{quit} = \mathit{true}$, then $s' = s$ and $M^S = \emptyset$;

    \item if $s = s_p^0$, for some process $p \in \Pi$, and there exists a transaction $\mathit{tx} \in T^R$ with $\mathit{tx}.\mathsf{start} = \mathit{true}$, then $s'.\mathsf{log} = \mathit{tx}.\mathsf{log}$;

    \item if $s = s_p^0$, for some process $p \in \Pi$, and there does not exist a transaction $\mathit{tx} \in T^R$ with $\mathit{tx}.\mathsf{start} = \mathit{true}$, then $s' = s$ and $M^S = \emptyset$.
\end{compactitem}

\subsection{Behaviors \& Executions}\label{subsection:behaviors_executions}

We now model executions of blockchain protocols.
Hence, fix any such protocol $\mathcal{P} = (\Pi, \rho, \mathsf{States}, \{ s_p^0 \,|\, p \in \Pi \}, \mathsf{Messages}, \mathsf{Transactions}, \mathsf{Logs}, \mathsf{Transition})$.

\mysubparagraphNoDot{Fragments.} 
A tuple $\mathcal{FR} = \big( s, \tau, \mathit{waiting}, M^R, T^R, M^S \big)$, where
\begin{compactitem}
    \item $s \in \mathsf{States}$ is a state;

    \item $\tau \in \mathbb{N}_{\geq 0} \cup \{ +\infty \}$ is a timeslot;

    \item $\mathit{waiting} \in \{ \mathit{true}, \mathit{false} \}$;

    \item $M^R \subsetneq \mathsf{Messages}$ is a set of messages;
    
    \item $T^R \subsetneq \mathsf{Transactions}$ is a set of transactions;

    \item $M^S \subsetneq \mathsf{Messages}$ is a set of messages, 
\end{compactitem}
is a \emph{$\tau$-fragment} of a process $p \in \Pi$ according to the protocol $\mathcal{P}$ if and only if:
\begin{compactenum}
    \item $s.\mathsf{process} = p$;

    \item for every message $m \in M^R$, $m.\mathsf{receiver} = p$;

    \item for every transaction $\mathit{tx} \in T^R$, $\mathit{tx}.\mathsf{process} = p$;

    \item for every message $m \in M^S$, $m.\mathsf{sender} = p$;

    \item if $\mathit{waiting} = \mathit{true}$, then $M^R \cup T^R \cup M^S = \emptyset$.

\end{compactenum}
Additionally, we say that $\mathcal{FR}$ is a \emph{valid} $\tau$-fragment of process $p$ according to the protocol $\mathcal{P}$ if and only if:
\begin{compactitem}
    \item if $\tau = 0$, then $s = s_p^0$;

    \item no two transactions $\mathit{tx}_1, \mathit{tx}_2 \in T^R$ exist such that $\mathit{tx}_1.\mathsf{start} = \mathit{true}$ and $\mathit{tx}_2.\mathsf{quit} = \mathit{true}$;

    \item if $s = s_p^0$, then $\mathit{waiting} = \mathit{false}$;

    \item if $s = s_p^0$, then $M^R \cup T^R = \emptyset$ or $M^R \cup T^R = \{ \mathit{tx} \}$, for some transaction $\mathit{tx} \in \mathsf{Transactions}$ with $\mathit{tx}.\mathsf{start} = \mathit{true}$;

    \item if $s = s_p^0$ and there does not exist a transaction $\mathit{tx} \in T^R$, then $M^S = \emptyset$;

    \item $\mathsf{Transition}(s, \mathit{waiting}, M^R, T^R) = (\cdot, M^S)$.
\end{compactitem}
Intuitively, a $\tau$-fragment of a process captures what occurs at the process during timeslot $\tau$ from the viewpoint of an omniscient external observer.

Given a $\tau$-fragment $\mathcal{FR} = \big( s, \tau, \mathit{waiting}, M^R, T^R, M^S \big)$ of a process $p \in \Pi$ according to the protocol $\mathcal{P}$, let:
\begin{compactitem}
    \item $\mathcal{FR}.\mathsf{state} \equiv s$;
    
    \item $\mathcal{FR}.\mathsf{waiting} \equiv \mathit{waiting}$;

    \item $\mathcal{FR}.\mathsf{received\_msgs} \equiv M^R$;

    \item $\mathcal{FR}.\mathsf{received\_txs} \equiv T^R$;

    \item $\mathcal{FR}.\mathsf{sent\_msgs} \equiv M^S$;

    \item $\mathcal{FR}.\mathsf{quit} \equiv s.\mathsf{quit}$.
\end{compactitem}

\mysubparagraphNoDot{Behaviors.}
A tuple $\mathcal{B} = \Big\langle \mathcal{FR}^0 = \big( s^0, 0, \mathit{waiting}^0, M^{R(0)}, T^{R(0)}, M^{S(0)} \big), ..., \mathcal{FR}^k = \big( s^k, k, \mathit{waiting}^k, M^{R(k)}, T^{R(k)}, M^{S(k)} \big) \Big \rangle$ is a \emph{($k \in \mathbb{N}_{\geq 0} \cup \{ +\infty \}$)-long behavior} of a process $p \in \Pi$ according to the protocol $\mathcal{P}$ if and only if:
\begin{compactitem}
    \item for every $i \in [0, k]$, $\mathcal{FR}^i$ is an $i$-fragment of process $p$ according to the protocol $\mathcal{P}$;

    \item there exists at most one transaction $\mathit{tx} \in \bigcup\limits_{i \in [0, k]} T^{R(i)}$ with $\mathit{tx}.\mathsf{start} = \mathit{true}$;

    \item there exists at most one transaction $\mathit{tx} \in \bigcup\limits_{i \in [0, k]} T^{R(i)}$ with $\mathit{tx}.\mathsf{quit} = \mathit{true}$.
\end{compactitem}
Moreover, we say that $\mathcal{B}$ is a \emph{valid} $k$-long behavior of process $p \in \Pi$ according to the protocol $\mathcal{P}$ if and only if:
\begin{compactitem}
    \item for every $i \in [0, k]$, $\mathcal{FR}^i$ is a valid $i$-fragment of process $p$;

    \item for every $i \in [0, k - 1]$, $\mathsf{Transition}(s^i, M^{R(i)}, T^{R(i)}) = (s^{i + 1}, M^{S(i)})$.
\end{compactitem}

Given a $k$-long behavior $\mathcal{B} = \Big\langle \mathcal{FR}^0, ..., \mathcal{FR}^k \Big\rangle$ of a process $p \in \Pi$ according to the protocol $\mathcal{P}$, let:
\begin{compactitem}
    \item for every $i \in [0, k]$, $\mathcal{B}.\mathsf{fragment}(i) \equiv \mathcal{FR}^i$;

    \item for every $i \in [0, k]$, $\mathcal{B}.\mathsf{state}(i) \equiv \mathcal{FR}^i.\mathsf{state}$;

    \item for every $i \in [0, k]$, $\mathcal{B}.\mathsf{log}(i) = s^i.\mathsf{log}$;

    \item for every $i \in [0, k]$, $\mathcal{B}.\mathsf{received\_msgs}(i) \equiv \mathcal{FR}^i.\mathsf{received\_msgs}$;

    \item for every $i \in [0, k]$, $\mathcal{B}.\mathsf{received\_txs}(i) \equiv \mathcal{FR}^i.\mathsf{received\_txs}$;
    
    \item for every $i \in [0, k]$, $\mathcal{B}.\mathsf{sent\_msgs}(i) \equiv \mathcal{FR}^i.\mathsf{sent\_msgs}$;

    \item for every $i \in [0, k]$, $\mathcal{B}.\mathsf{quit}(i) \equiv \mathcal{FR}^i.\mathsf{quit}$;

    \item for every $i \in [0, k]$, $\mathcal{B}.\mathsf{waiting}(i) \equiv \mathcal{FR}^i.\mathsf{waiting}$;

    \item for every $i \in [0, k]$, $\mathcal{B}.\mathsf{prefix}(i) \equiv \Big\langle \mathcal{FR}^0, ..., \mathcal{FR}^i \Big\rangle$.
\end{compactitem}

\mysubparagraphNoDot{Executions.}
A $k$-long execution $\mathcal{E}$ of the protocol $\mathcal{P}$, for any $k \in \mathbb{N}_{\geq 0} \cup \{ +\infty \}$, is a tuple $[\mathcal{F}, \text{GST} \in \mathsf{N}_{\geq 0}, \{ \mathcal{B}_p \,|\, p \in \Pi \}]$ such that the following guarantees hold:
\begin{compactitem}
    \item \emph{Faulty processes:} $\mathcal{F} \subseteq \Pi$ is a set of $|\mathcal{F}| \leq \rho \cdot n$ processes.

    \item \emph{Composition:} For every process $p \in \Pi$, $\mathcal{B}_p$ is a $k$-long behavior of $p$ according to the protocol $\mathcal{P}$.

    \item \emph{Validity:} For every process $p \in \Pi \setminus{\mathcal{F}}$, $\mathcal{B}_p$ is a valid $k$-long behavior of $p$ according to the protocol $\mathcal{P}$.

    \item \emph{GST-validity:} For every process $p \in \Pi \setminus{ \mathcal{F} }$ and every $i \in [\text{GST}, k]$, the following holds: $\mathcal{B}_p.\mathsf{waiting}(i) = \mathit{false}$.

    \item \emph{Receive-validity:} If there exists a message $m$, where $s = m.\mathsf{sender}$ and $r = m.\mathsf{receiver}$, such that $m \in \mathcal{B}_r.\mathsf{received\_msgs}(\tau_r)$, for some $\tau_r \in [0, k]$, then there exists $\tau_s \in [0, \tau_r - 1]$ such that $m \in \mathcal{B}_s.\mathsf{sent\_msgs}(\tau_s)$.

    \item \emph{Send-validity:} Let there exist a message $m$, where $s = m.\mathsf{sender}$ and $r = m.\mathsf{receiver}$, such that $m \in \mathcal{B}_s.\mathsf{sent\_msgs}(\tau_s)$, for some $\tau_s \in [0, k]$.
    If $k \geq \max(\tau_s, \text{GST}) + \delta$, then there exists $\tau_r \leq \max(\tau_s, \text{GST}) + \delta$ such that $m \in \mathcal{B}_r.\mathsf{received\_msgs}(\tau_r)$.
\end{compactitem}

\subsection{Standard vs. Quit-Enhanced Blockchain Protocols}\label{subsection:standard_quit_enhanced}

Finally, we are ready to formally define quit-enhanced blockchain protocols.
To do so, we first provide a definition of standard blockchain protocols.

\begin{definition} [Standard protocols]
Fix any blockchain protocol $\mathcal{P} = (\Pi, \rho,\\ \mathsf{States}, \{ s_p^0 \,|\, p \in \Pi \}, \mathsf{Messages}, \mathsf{Transactions}, \mathsf{Logs}, \mathsf{Transition})$.
We say that $\mathcal{P}$ is a \emph{standard} blockchain protocol if and only if the following holds:
\begin{compactitem}
    \item for every state $s \in \mathsf{States}$, $s.\mathsf{quit} = \mathit{false}$;

    \item for every transaction $\mathit{tx} \in \mathsf{Transactions}$, $\mathit{tx}.\mathsf{quit} = \mathit{false}$.
\end{compactitem}
\end{definition}

Intuitively, standard blockchain protocols do not have special quitting states and do not accept special quitting transactions from the environment.
The following definition introduces quit-enhanced blockchain protocols.

\begin{definition} [Quit-enhanced protocols]
Fix any standard blockchain protocol $\mathcal{P} = (\Pi, \rho, \mathsf{States}, \{ s_p^0 \,|\, p \in \Pi \}, 
\mathsf{Messages}, \mathsf{Transactions}, \mathsf{Logs}, \mathsf{Transition})$.
Let $\mathcal{P}' = (\Pi, \rho, \mathsf{States}', \{ s_p^{0} \,|\, p \in \Pi \}, \mathsf{Messages}, \mathsf{Transactions}', \mathsf{Logs}, \mathsf{Transition}')$ be another blockchain protocol.
We say that $\mathcal{P}'$
is the \emph{quit-enhanced} version of $\mathcal{P}$, denoted by $\mathit{quit}(\mathcal{P})$, if and only if the following holds:
\begin{compactitem}

    \item $\mathsf{States}' = \mathsf{States} \cup \{ Q \,|\, (p \in \Pi, \mathcal{L} \in \mathsf{Logs}): Q.\mathsf{process} = p \land Q.\mathsf{start} = \mathit{false} \land Q.\mathsf{quit} = \mathit{true} \land Q.\mathsf{log} = \mathcal{L}\}$;

    \item $\mathsf{Transactions}' = \mathsf{Transactions} \cup \{ q \,|\, p \in \Pi: q.\mathsf{process} = p \land q.\mathsf{start} = \mathit{false} \land q.\mathsf{quit} = \mathit{true} \land q.\mathsf{log} = \bot  \}$;
    \item for every state $s \in \mathsf{States}$ (thus, $s.\mathsf{quit} = \mathit{false}$), every boolean $\mathit{waiting}$, every set of messages $M^R \subseteq \mathsf{Messages}$ and every set of transactions $T^R \subseteq \mathsf{Transactions}$, the following holds:
    \begin{equation*}
        \mathsf{Transition}'(s, \mathit{waiting}, M^R, T^R) = \mathsf{Transition}(s, \mathit{waiting}, M^R, T^R);
    \end{equation*}

\end{compactitem}
\end{definition}

\subsection{Preservation of Properties}\label{subsection:quit_preservation}

Finally, this subsection proves that quit-enhanced blockchain protocols preserve key properties of their standard counterparts.
This is crucial for proving the correctness of our transformation (\cf \Cref{section:proof}).
We fix any standard blockchain protocol $\mathcal{P} = (\Pi, \rho, \{ s_p^0 \,|\, p \in \Pi \}, \mathsf{Messages}, \mathsf{Transactions}, \mathsf{Logs}, \mathsf{Transition} )$ and its quit-enhanced version $\mathcal{P}' = \mathit{quit}(\mathcal{P})$.
We start by proving a crucial proposition used in showing that consistency and log-specific safety properties are translated from $\mathcal{P}$ to $\mathcal{P}'$.

\begin{proposition}\label{propositon:quit_mapping}
Let $\mathcal{E}' = [\mathcal{F}', \text{GST}', \{ \mathcal{B}_p' \,|\, p \in \Pi] \}]$ be any $k$-long execution of $\mathcal{P}'$, for some $k \in \mathbb{N}_{\geq 0}$ (thus, $k \neq +\infty$).
Then, there exists a $k$-long execution $\mathcal{E} = [\mathcal{F}, \text{GST}, \{ \mathcal{B}_p \,|\, p \in \Pi \}]$ of $\mathcal{P}$ such that:
\begin{compactitem}
    \item $\mathcal{F} = \mathcal{F}'$; 

    \item for every process $p \in \mathcal{F}$ and every $i \in [0, k]$, the following holds: (1) $\mathcal{B}_p.\mathsf{state} \in \mathsf{States}$, (2) $\mathcal{B}_p.\mathsf{received\_msgs}(i) = \mathcal{B}_p'.\mathsf{received\_msgs}(i)$, (3) $\mathcal{B}_p.\mathsf{received\_txs}(i) = \mathcal{B}_p'.\mathsf{received\_txs}(i) \setminus{ (\mathsf{Transactions}' \setminus{\mathsf{Transactions}}) }$, and (4) $\mathcal{B}_p.\mathsf{sent\_msgs}(i) = \mathcal{B}_p'.\mathsf{sent\_msgs}(i)$.

    \item for every process $p \in \Pi \setminus{ \mathcal{F} }$ and every timeslot $\tau \in [0, k]$, $\mathcal{B}_p.\mathsf{log}(\tau) = \mathcal{B}_p'.\mathsf{log}(\tau)$.
\end{compactitem}
\end{proposition}
\begin{proof}
We divide all processes from the $\Pi \setminus{ \mathcal
{F'}} $ set into two disjoint groups:
\begin{compactitem}
    \item Let $Q' = \{ p \in \Pi \setminus{ \mathcal{F}' } \,|\, \text{$\exists i \in [0, k]$: $\mathcal{B}_p'.\mathsf{quit}(i) = \mathit{true}$} \}$.

    \item Let $NQ' = (\Pi \setminus{ \mathcal{F}' }) \setminus{ Q' }$.
\end{compactitem}

\smallskip
\noindent\emph{Step 1: Constructing a valid $k$-long behavior $\mathcal{B}_p$ for a process $p \in Q'$.}
\\ We construct $\mathcal{B}_p$ in the following way:
\begin{compactenum}
    \item Let $i \in [0, k]$ denote the smallest integer such that $\mathcal{B}_p'.\mathsf{quit}(i) = \mathit{true}$.

    \item For every $j \in [0, i - 2]$, we set $\mathcal{B}_p.\mathsf{fragment}(j)$ to $\mathcal{B}_p'.\mathsf{fragment}(j)$.

    \item Let $\mathcal{FR}^{i - 1} = \big( \mathcal{B}_p'.\mathsf{fragment}(i - 1).\mathsf{state}, i - 1, \mathit{true}, \emptyset, \emptyset, \emptyset \big)$.

    \item We set $\mathcal{B}_p.\mathsf{fragment}(i - 1)$ to $\mathcal{FR}^{i - 1}$.

    \item For every $j \in [i, k]$, we set $\mathcal{B}_p'.\mathsf{fragment}(j)$ to $\big( \mathcal{FR}^{i - 1}.\mathsf{state}, j, \mathit{true}, \emptyset, \emptyset, \emptyset \big)$.
\end{compactenum}
Indeed, the constructed $k$-long behavior $\mathcal{B}_p$ is valid according to $\mathcal{P}$:
\begin{compactitem}
    \item Until (and including) timeslot $i - 2$, $\mathcal{B}_p$ is valid as (1) it is identical to $\mathcal{B}_p'$ until (and including) timeslot $i - 2$, and (2) $\mathcal{B}_p'$ is valid.

    \item We have that $\mathcal{B}_p.\mathsf{fragment}(i - 1).\mathsf{state} = \mathcal{B}_p'.\mathsf{fragment}(i - 1).\mathsf{state}$ because of the construction of $\mathcal{B}_p.\mathsf{fragment}(i - 1)$.
    Hence, the transition from the $(i - 2)$-nd to $(i - 1)$-st fragment in $\mathcal{B}_p$ is valid according to the $\mathsf{Transition}$ function.

    \item For every timeslot $i \in [i - 1, k]$, process $p$ is waiting at timeslot $i$, thus not receiving any messages or transactions, not sending any messages, and not changing its state.
\end{compactitem}
Moreover, for every timeslot $\tau \in [0, k]$, $\mathcal{B}_p.\mathsf{log}(\tau) = \mathcal{B}_p'.\mathsf{log}(\tau)$ due to (1) the definition of the quit-enhanced protocols, (2) the assumed constraints on the state transition function, and (3) our construction.

\smallskip
\noindent\emph{Step 2: Constructing a valid $k$-long behavior $\mathcal{B}_p$ for a process $p \in NQ'$.}
\\ We set $\mathcal{B}_p$ to $\mathcal{B}_p'$.
Hence, $\mathcal{B}_p$ is trivially valid (as $\mathcal{B}_p'$ is) and, for every timeslot $\tau \in [0, k]$, $\mathcal{B}_p.\mathsf{log}(\tau) = \mathcal{B}_p'.\mathsf{log}(\tau)$.

\smallskip
\noindent\emph{Step 3: Constructing a $k$-long behavior $\mathcal{B}_p$ for a process $p \in \mathcal{F}'$.}
\\ We set $\mathcal{B}_p$ to $\mathcal{B}_p'$.
Then, if there is a fragment of $\mathcal{B}_p$ in which $p$ is in a state $s \in \mathsf{States}' \setminus{ \mathsf{States}}$, we modify the fragment so that $p$ is in any state from the $\mathsf{States}$ set.
Similarly, if there is a fragment in which $p$ receives any transaction $\mathit{tx} \in \mathsf{Transactions}' \setminus{ \mathsf{Transactions} }$, we modify the fragment so that $p$ does not receive this transaction.

\smallskip
\noindent\emph{Epilogue: Constructing $\mathcal{E}$.}
\\ Finally, let $\mathcal{E} = [\mathcal{F} = \mathcal{F}', k + 1, \{ \mathcal{B}_p \,|\, p \in \Pi \}]$.
Note that the statement of the lemma is satisfied for $\mathcal{E}$.
It is left to prove that $\mathcal{E}$ satisfies the guarantees of an execution:
\begin{compactitem}
    \item \emph{Faulty processes:} Indeed, $|\mathcal{F}| \leq \rho \cdot n$ as $\mathcal{F} = \mathcal{F}'$ and $|\mathcal{F}'| \leq \rho \cdot n$.

    \item \emph{Composition:} For every process $p \in \Pi$, $\mathcal{B}_p$ is a $k$-long behavior of $p$ according to $\mathcal{P}$ (due to the construction of $\mathcal{B}_p$).

    \item \emph{Validity:} For every process $p \in \Pi \setminus{ \mathcal{F} }$, $\mathcal{B}_p$ is a valid $k$-long behavior of $p$ (due to the construction of $\mathcal{B}_p$; see steps 1 and 2).

    \item \emph{GST-validity:} As $k + 1 > k$, the guarantee is trivially satisfied.

    \item \emph{Receive-validity:} First, note that if any process $p \in \Pi$ sends a message $m$ in $\mathcal{B}_p'$ (i.e., $m \in \mathcal{B}_p'.\mathsf{sent\_msgs}$), then $\mathcal{B}_p$ sends $m$ in $\mathcal{B}_p$ (i.e., $m \in \mathcal{B}_p.\mathsf{sent\_msgs}$).
    Let us distinguish three cases:
    \begin{compactitem}
        \item Let $p \in \mathcal{F}$.
        In this case, the statement is trivially true due to the construction of $\mathcal{B}_p$ (see step 3).

        \item Let $p \in NQ'$.
        Again, the statement holds as $\mathcal{B}_p = \mathcal{B}_p'$ (see step 2).

        \item Let $p \in Q'$.
        Let $m$ be sent in the $x$-th fragment of $\mathcal{B}_p'$.
        Let $i \in [0, k]$ denote the smallest integer such that $\mathcal{B}_p'.\mathsf{quit} = \mathit{true}$.
        Note that $x \in [0, i - 2]$ (due to the definition of the $\mathsf{Transition}'$ function).
        As $\mathcal{B}_p$ is identical to $\mathcal{B}_p'$ until (and including) timeslot $i - 2$, $m$ is sent in the $x$-th fragment of $\mathcal{B}_p$. 
    \end{compactitem}
    Now, consider any message $m$ received in $\mathcal{E}$.
    This implies that $m$ is received in $\mathcal{E}'$, which then means that $m$ is sent in $\mathcal{E}'$.
    Due to the argument above, $m$ is sent in $\mathcal{E}$ as well, which proves the receive-validity property.

    \item \emph{Send-validity:}
    As $k > k + 1$, the property trivially holds.
\end{compactitem}
As $\mathcal{E}$ is indeed an execution of $\mathcal{P}$, the proof is concluded.
\end{proof}

We prove that the consistency property is translated from $\mathcal{P}$ to $\mathcal{P}'$.

\begin{lemma} [Preservation of consistency] \label{lemma:consistency_preservation_quit}
If $\mathcal{P}$ satisfies the consistency property, then $\mathcal{P}'$ satisfies the consistency property.
\end{lemma}
\begin{proof}
By contradiction, suppose $\mathcal{P}'$ does not satisfy consistency.
Let $\mathcal{E}' = [\mathcal{F}', \text{GST}', \{ \mathcal{B}_p' \,|\, p \in \Pi \}]$ be any $k$-long execution of $\mathcal{P}'$, for some $k \in \mathbb{N}_{\geq 0} \cup \{ + \infty \}$, in which the consistency property is not satisfied.
We study two cases:
\begin{compactitem}
    \item Let $k \in \mathbb{N}_{\geq 0}$ (thus, $k \neq + \infty$).
    By \Cref{propositon:quit_mapping}, there exists a $k$-long execution $\mathcal{E} = [\mathcal{F} = \mathcal{F}', \text{GST}, \{ \mathcal{B}_p \,|\, p \in \Pi \}]$ of $\mathcal{P}$ such that, for every process $p \in \Pi \setminus { \mathcal{F} }$ and every timeslot $\tau \in [0, k]$, $\mathcal{B}_p.\mathsf{log}(\tau) = \mathcal{B}_p'.\mathsf{log}(\tau)$.
    Hence, the consistency property is violated in $\mathcal{E}$, which represents a contradiction with the fact that $\mathcal{P}$ satisfies consistency.

    \item Let $k = + \infty$.
    We now define a $\mathit{prefix}$-long execution $\mathcal{E}_{\mathit{prefix}}'$ of $\mathcal{P}'$, for some $\mathit{prefix} \in \mathbb{N}_{\geq 0}$, in the following manner:
    \begin{compactitem}
        \item If there exist a process $p \in \Pi \setminus { \mathcal{F}' }$ and two timeslots $\tau_1, \tau_2 \in \mathbb{N}_{\geq 0}$ with $\tau_1 < \tau_2$ such that $\mathcal{B}_p'.\mathsf{log}(\tau_2)$ does not extend $\mathcal{B}_p'.\mathsf{log}(\tau_1)$, then let $\mathcal{E}_{\mathit{prefix}}' = [\mathcal{F}', \text{GST}', \{ \mathcal{B}_p^{\mathit{prefix}} \,|\, p \in \Pi \}]$, where, for each process $p \in \Pi$, $\mathcal{B}_p^{\mathit{prefix}}$ denotes the $\tau_2$-long prefix of $\mathcal{B}_p'$ (i.e., $\mathcal{B}_p^{\mathit{prefix}} = \mathcal{B}_p'.\mathsf{prefix}(\tau_2)$).

        \item Otherwise, there must exist a timeslot $\tau$ and a pair of processes $(p_1, p_2) \in (\Pi \setminus { \mathcal{F}' })^2$ such that logs $\mathcal{B}_{p_1}'.\mathsf{log}(\tau)$ and $\mathcal{B}_{p_2}'.\mathsf{log}(\tau)$ are not consistent.
        In this case, let $\mathcal{E}_{\mathit{prefix}}' = [\mathcal{F}', \text{GST}', \{ \mathcal{B}_p^{\mathit{prefix}} \,|\, p \in \Pi \}]$, where, for each process $p \in \Pi$, $\mathcal{B}_p^{\mathit{prefix}}$ denotes the $\tau$-long prefix of $\mathcal{B}_p'$.
    \end{compactitem}
    Note that the consistency property is violated in $\mathcal{E}_{\mathit{prefix}}'$.
    Finally, according to \Cref{propositon:quit_mapping}, there exists an execution of $\mathcal{P}$ that violates consistency.
    Given that $\mathcal{P}$ satisfies the property, this is impossible, which completes a contradiction in this case.
\end{compactitem}
As the statement of the lemma holds in both cases, the proof is concluded.
\end{proof}

Next, we prove that log-specific safety properties are preserved in $\mathcal{P}'$.

\begin{lemma} [Preservation of log-specific safety properties]
\label{lemma:log-specific-quit-enhanced}
If $\mathcal{P}$ satisfies any log-specific safety property $S_L$, then $\mathcal{P}'$ satisfies $S_L$.
\end{lemma}
\begin{proof}
By contradiction, suppose $\mathcal{P}'$ does not satisfy the property $S_L$.
Let $\mathcal{E}' = [\mathcal{F}', \text{GST}', \{ \mathcal{B}_p' \,|\, p \in \Pi \}]$ be any $k$-long execution of $\mathcal{P}'$, for some $k \in \mathbb{N}_{\geq 0} \cup \{ + \infty \}$, in which $S_L$ is not satisfied.
Let $\mathbb{L} = \bigcup\limits_{i \in [0, k]} \bigcup\limits_{p \in \Pi} \mathcal{B}_p'.\mathsf{log}(i)$.
As $S_L$ is not satisfied in $\mathcal{E}'$, we have that $S_L(\mathbb{L}) = \mathit{false}$.

We now consider two cases:
\begin{compactitem}
    \item Let $k \in \mathbb{N}_{\geq 0}$ (thus, $k \neq + \infty$).
    In this case, \Cref{propositon:quit_mapping} proves that there exists an execution of $\mathcal{P}$ in which the property is not satisfied.
    As this is impossible, the statement of the lemma holds in this case.

    \item Let $k = + \infty$.
    As $S_L$ is a log-specific safety property and $S_L(\mathbb{L}) = \mathit{false}$, there exists a finite subset $\mathit{logs}_{\mathit{prefix}} \subseteq \mathbb{L}$ such that $S_L(\mathit{logs}_{\mathit{prefix}}) = \mathit{false}$.
    Let $\tau$ denote the smallest timeslot such that, for every $\mathcal{L} \in \mathit{logs}_{\mathit{prefix}}$, there exist a process $\Pi \setminus { \mathcal{F}' }$ and a timeslot $\tau' \in [0, \tau]$ with $\mathcal{B}_p'.\mathsf{log}(\tau') = \mathcal{L}$.
    Now, let $\mathcal{E}_{\mathit{prefix}}' = [\mathcal{F}', \text{GST}', \{ \mathcal{B}_p^{\mathit{prefix}} \,|\, p \in \Pi \}]$, where, for each process $p \in \Pi$, $\mathcal{B}_p^{\mathit{prefix}}$ denotes the $\tau$-long prefix of $\mathcal{B}_p'$.

    Let $\mathbb{L}' = \bigcup\limits_{i \in [0, \tau]} \bigcup\limits_{p \in \Pi} \mathcal{B}_p^{\mathit{prefix}}.\mathsf{log}(i)$.
    Note that $\mathit{logs}_{\mathit{prefix}} \subseteq \mathbb{L}'$.
    Hence, $S_L(\mathbb{L}') = \mathit{false}$ according to the definition of log-specific safety properties.
    Finally, \Cref{propositon:quit_mapping} proves that $\mathcal{P}$ violates $S_L$ as well, thus concluding the proof in this case.
\end{compactitem}
As the statement of the lemma holds in both cases, the proof is concluded.
\end{proof}

Finally, we prove a proposition required to prove liveness and optimistic responsiveness of our quasi-permissionless PoS protocol.

\begin{proposition}
\label{proposition:quit-enhanced-liveness}
Let $\mathcal{E}' = [\mathcal{F}', \text{GST}', \{ \mathcal{B}_p' \,|\, p \in \Pi \}]$ be any $k$-long execution of $\mathcal{P}'$, for some $k \in \mathbb{N}_{\geq 0}$ (thus, $k \neq + \infty$), such that, for every $p \in \Pi \setminus{ \mathcal{F}' }$ and every $i \in [0, k]$, $\mathcal{B}_p'.\mathsf{quit}(i) = \mathit{false}$ and $\mathcal{B}_p'.\mathsf{received\_txs}(i) \cap \mathsf{Transactions}' = \emptyset$.
Then, there exists a $k$-long execution $\mathcal{E} = [\mathcal{F}, \text{GST}, \{ \mathcal{B}_p \,|\, p \in \Pi \}]$ of $\mathcal{P}$ such that:
\begin{compactitem}
    \item $\mathcal{F} = \mathcal{F}'$;

    \item $\text{GST} = \text{GST}'$';

    \item for every process $p \in \mathcal{F}$ and every $i \in [0, k]$, the following holds: (1) $\mathcal{B}_p.\mathsf{state} \in \mathsf{States}$, (2) $\mathcal{B}_p.\mathsf{received\_msgs}(i) = \mathcal{B}_p'.\mathsf{received\_msgs}(i)$, (3) $\mathcal{B}_p.\mathsf{received\_txs}(i) = \mathcal{B}_p'.\mathsf{received\_txs}(i) \setminus{ (\mathsf{Transactions}' \setminus{\mathsf{Transactions}}) }$, and (4) $\mathcal{B}_p.\mathsf{sent\_msgs}(i) = \mathcal{B}_p'.\mathsf{sent\_msgs}(i)$.

    \item for every process $p \in \Pi \setminus{\mathcal{F}}$, $\mathcal{B}_p = \mathcal{B}_p'$.
\end{compactitem}
\end{proposition}
\begin{proof}
We construct $\mathcal{E}$ in the following manner.

\smallskip
\noindent\emph{Step 1: Constructing a $k$-long behavior $\mathcal{B}_p$ for a process $p \in \mathcal{F}'$.}
\\ We set $\mathcal{B}_p$ to $\mathcal{B}_p'$.
Then, if there is a fragment of $\mathcal{B}_p$ in which $p$ is in a state $s \in \mathsf{States}' \setminus{ \mathsf{States}}$, we modify the fragment so that $p$ is in any state from the $\mathsf{States}$ set.
Similarly, if there is a fragment in which $p$ receives any transaction $\mathit{tx} \in \mathsf{Transactions}' \setminus{ \mathsf{Transactions} }$, we modify the fragment so that $p$ does not receive this transaction.

\smallskip
\noindent\emph{Step 2: Constructing a valid $k$-long behavior $\mathcal{B}_p$ for a process $p \in \Pi \setminus{\mathcal{F}'}$.}
We set $\mathcal{B}_p$ to $\mathcal{B}_p'$.

\smallskip
\noindent\emph{Epilogue: Constructing $\mathcal{E}$.}
\\ Let $\mathcal{E} = [\mathcal{F} = \mathcal{F}', \text{GST} = \text{GST}', \{ \mathcal{B}_p \,|\, p \in \Pi \}]$.
The statement of the proposition is satisfied by $\mathcal{E}$.
It is left to prove that $\mathcal{E}$ is indeed an execution of $\mathcal{P}$:
\begin{compactitem}
    \item \emph{Faulty processes:} We have that $|\mathcal{F}| \leq \rho \cdot n$ as $\mathcal{F} = \mathcal{F}'$ and $|\mathcal{F}'| \leq \rho \cdot n$.

    \item \emph{Composition:} For every process $p \in \Pi$, $\mathcal{B}_p$ is a $k$-long behavior of $p$ according to $\mathcal{P}$.

    \item \emph{Validity:} For every process $p \in \Pi \setminus{\mathcal{F}}$, $\mathcal{B}_p$ is a valid $k$-long behavior of $p$ according to $\mathcal{P}$.

    \item \emph{GST-validity:} This property is satisfied as (1) it is satisfied in $\mathcal{E}'$, and (2) for every process $p \in \Pi \setminus{\mathcal{F}}$, $\mathcal{B}_p = \mathcal{B}_p'$.

    \item \emph{Receive-validity:} The set of messages sent (resp., received) in $\mathcal{E}$ is identical to the set of messages sent (resp., received) in $\mathcal{E}'$.
    Precisely, if a process $p$ sends (resp., receives) a message a timeslot $\tau$ in $\mathcal{E}'$, then $p$ sends (resp., receives) the message at timeslot $\tau$ in $\mathcal{E}$.
    Hence, the property holds in $\mathcal{E}$ as it holds in $\mathcal{E}'$.

    \item \emph{Send-validity:} If a process $p$ sends (resp., receives) a message a timeslot $\tau$ in $\mathcal{E}'$, then $p$ sends (resp., receives) the message at timeslot $\tau$ in $\mathcal{E}$.
    Hence, the property holds in $\mathcal{E}$ as it holds in $\mathcal{E}'$.
\end{compactitem}
Hence, $\mathcal{E}$ is an execution of $\mathcal{P}$, which concludes the proof.
\end{proof}

%% file: appendix/20_pos_proof.tex
\section{Transformation: Omitted Definitions \& Proof of Correctness} \label{section:proof}

Throughout the entire section, we fix a permissioned protocol $\mathcal{P}$.

\subsection{Omitted Definitions} \label{subsection:omitted_definitions}

In this subsection, we present definitions omitted from \Cref{subsection:transformation_pseudocode}.
We start by defining a method that returns the set of PoS-ids with positive stake.

\begin{definition} \label{definition:current_ids}
Given any log $\mathcal{L}$, let $\mathcal{L}.\mathsf{current\_ids} \equiv \{ \mathit{id} \in \mathsf{IDs} \,|\, \mathsf{S}(\mathcal{L}, \mathit{id}) > 0 \}$.
\end{definition}

Given a log $\mathcal{L} = \mathcal{L}_{\mathrm{g}} || [\mathsf{tr}_1, \mathsf{tr}_2, ..., \mathsf{tr}_x]$ with $x \geq 1$ transactions after the genesis log, let $\mathcal{L}.\mathsf{predecessor} = \mathcal{L}_{\mathrm{g}} || [\mathsf{tr}_{1}, ..., \mathsf{tr}_{x - 1}]$ be the prefix of $\mathcal{L}$ with $x - 1$ transactions.
If $x = 1$, then $\mathcal{L}.\mathsf{predecessor} = \mathcal{L}_{\mathrm{g}}$.
Next, we define the epoch of a log.
Importantly, the following definition relies on the $\mathsf{completed}$ method of a log; this method is defined later, in \Cref{definition:epoch-completed}.
However, we underline that our definitions are not circular.
Specifically, \Cref{definition:epoch,definition:identifiers,definition:epoch-prefix,definition:epoch-completed} should be considered one large recursive definition.

\begin{definition} \label{definition:epoch}
Given any log $\mathcal{L} \neq \mathcal{L}_{\mathrm{g}}$ that extends $\mathcal{L}_{\mathrm{g}}$, we define $\mathcal{L}.\mathsf{epoch}$ as:
\begin{equation*}
        \LOG.\mathsf{epoch} \equiv \begin{cases}
            \LOG.\mathsf{predecessor}.\mathsf{epoch}, &\text{if $\LOG.\mathsf{predecessor}.\mathsf{completed} = \mathit{false}$}\\
            \LOG.\mathsf{predecessor}.\mathsf{epoch} + 1, &\text{if $\LOG.\mathsf{predecessor}.\mathsf{completed} = \mathit{true}$}.
        \end{cases}
    \end{equation*}
Moreover, $\mathcal{L}_{\mathrm{g}}.\mathsf{epoch} \equiv 0$.
For any log $\mathcal{L}' \neq \mathcal{L}_{\mathrm{g}}$ that does not extend $\mathcal{L}_{\mathrm{g}}$, $\mathcal{L}'.\mathsf{epoch} \equiv \mathit{undefined}$.
\end{definition}
Intuitively, if $\mathcal{L}.\mathsf{predecessor}$ is not a completed log (\cf \Cref{definition:epoch-completed}), then $\mathcal{L}$'s epoch is identical to the epoch of $\mathcal{L}.\mathsf{predecessor}$.
Otherwise, $\mathcal{L}$ belongs to a new epoch.
Next, we define the set of validators.

\begin{definition} \label{definition:identifiers}
Given any log $\mathcal{L} \neq \mathcal{L}_{\mathrm{g}}$ that extends $\mathcal{L}_{\mathrm{g}}$, we define $\mathcal{L}.\mathsf{validators}$ as:
\begin{equation*}
        \LOG.\mathsf{validators} \equiv \begin{cases}
            \LOG.\mathsf{predecessor}.\mathsf{validators}, &\text{if $\LOG.\mathsf{predecessor}.\mathsf{completed} = \mathit{false}$}\\
            \LOG.\mathsf{predecessor}.\mathsf{current\_ids}, &\text{if $\LOG.\mathsf{predecessor}.\mathsf{completed} = \mathit{true}$}.
        \end{cases}
    \end{equation*}
For any log $\mathcal{L}'$ such that (1) $\mathcal{L}'$ does not extend $\mathcal{L}_{\mathrm{g}}$, or (2) $\mathcal{L}' = \mathcal{L}_{\mathrm{g}}$, $\mathcal{L}'.\mathsf{validators} \equiv \mathit{undefined}$.
\end{definition}

Intuitively, $\mathcal{L}.\mathsf{validators}$ denotes the set of PoS-ids that produced $\mathcal{L}$.
If $\mathcal{L}$ and $\mathcal{L}.\mathsf{predecessor}$ belong to the same epoch, their respective validator sets are identical.
Otherwise, $\mathcal{L}.\mathsf{validators} = \mathcal{L}.\mathsf{predecessor}.\mathsf{current\_ids}$.
Next, we define the concept of an epoch prefix of a log.

\begin{definition}
\label{definition:epoch-prefix}
Given any log $\mathcal{L}$ that extends $\mathcal{L}_{\mathrm{g}}$ and any $e \in [0, \mathcal{L}.\mathsf{epoch})$, let $\mathcal{L}.\mathsf{ep\_prefix}(e) \equiv \mathcal{L}'$, where (1) $\mathcal{L}$ extends $\mathcal{L}'$, (2) $\mathcal{L}'.\mathsf{epoch} = e$, and (3) $\mathcal{L}'.\mathsf{completed} = \mathit{true}$.
For any $e \notin [0, \mathcal{L}.\mathsf{epoch})$, $\mathcal{L}.\mathsf{ep\_prefix}(e) \equiv \mathit{undefined}$.
\end{definition}

Given a log $\mathcal{L} \neq \mathcal{L}_{\mathrm{g}}$ that extends $\mathcal{L}_{\mathrm{g}}$, we define the $\mathcal{L}.\mathsf{finishers}$ set as:
\begin{equation*}
    \mathcal{L}.\mathsf{finishers} \equiv \{ \mathit{id} \in \mathcal{L}.\mathsf{validators} \,|\, \exists (\textsc{finish}, \mathcal{L}.\mathsf{epoch}) \in \mathcal{L} \text{ issued by $\mathit{id}$} \}.
\end{equation*}
Moreover, given a log $\mathcal{L} \neq \mathcal{L}_{\mathrm{g}}$ that extends $\mathcal{L}_{\mathrm{g}}$, let us define $\mathcal{L}.\mathsf{can\_complete}$:
\begin{equation*}
        \LOG.\mathsf{can\_complete} \equiv \begin{cases}
            \mathit{true}, &\text{if $\sum\limits_{\mathit{id} \in \mathcal{L}.\mathsf{finishers}} \mathsf{S}(\mathcal{L}.\mathsf{ep\_prefix}(\mathcal{L}.\mathsf{epoch} - 1), \mathit{id}) > \rho \mathbb{T}$}\\
            \mathit{false}, &\text{otherwise}.
        \end{cases}
\end{equation*}
Now, we define what it means for a log to be completed.

\begin{definition}
\label{definition:epoch-completed}
Given any log $\mathcal{L} \neq \mathcal{L}_{\mathrm{g}}$ that extends $\mathcal{L}_\mathrm{g}$, $\mathcal{L}.\mathsf{completed} = \mathit{true}$ if and only if (1) $\mathcal{L}.\mathsf{can\_complete} = \mathit{true}$, and (2) there does not exist a log $\mathcal{L}'$ such that (a) $\mathcal{L}$ strictly extends $\mathcal{L}'$ ($\mathcal{L}.\mathsf{length} > \mathcal{L}'.\mathsf{length}$), (b) $\mathcal{L}'.\mathsf{epoch} = \mathcal{L}.\mathsf{epoch}$, and (c) $\mathcal{L}'.\mathsf{can\_complete} = \mathit{true}$.
Moreover, $\mathcal{L}_{\mathrm{g}}.\mathsf{completed} = \mathit{true}$.
For any log $\mathcal{L}' \neq \mathcal{L}_{\mathrm{g}}$ that does not extend $\mathcal{L}_{\mathrm{g}}$, $\mathcal{L}'.\mathsf{completed} \equiv \mathit{undefined}$.
\end{definition}

Next, we define certified logs.

\begin{definition}
\label{definition:certified}
A log $\LOG \neq \mathcal{L}_{\mathrm{g}}$ that extends $\mathcal{L}_{\mathrm{g}}$ is said to be \emph{certified} at a process $p$ if and only if $p$ receives signatures on the log $\mathcal{L}$ from the members of a set $I \subseteq \mathcal{L}.\mathsf{validators}$ such that $\sum\limits_{\mathit{id} \in I} \mathsf{S}(\LOG.\mathsf{ep\_prefix}(\mathcal{L}.\mathsf{epoch} - 1), \mathit{id}) \geq (1-\rho) \mathbb{T}$.
Moreover, $\mathcal{L}_{\mathrm{g}}$ is certified at every process $p$.
\end{definition}

Lastly, we define fully-certified logs.

\begin{definition}
\label{definition:fully-certified}
A certified log $\LOG \neq \mathcal{L}_{\mathrm{g}}$ is said to be \emph{fully-certified} at a process $p$ if and only if $\forall e \in [0, \LOG.\mathsf{epoch})$, $\LOG.\mathsf{ep\_prefix}(e)$ is certified at process $p_i$. Moreover, $\mathcal{L}_{\mathrm{g}}$ is fully-certified at every process $p$.
\end{definition}

\subsection{Intermediate Propositions}

In this section, we prove some intermediate propositions used throughout the entire section.
For every log $\mathcal{L}$, let $\mathsf{processes}(\mathcal{L})$ be defined in the following way:
\begin{equation*}
    \mathsf{processes}(\mathcal{L}) \equiv \{ p \in \Pi \,|\, \exists \mathit{id} \in \mathcal{L}.\mathsf{validators}: \mathit{id} \in \mathsf{id}(p) \}.
\end{equation*}
We underline that $\mathsf{processes}(\mathcal{L})$ is \emph{not} known to processes (i.e., it is undetermined) as the identifier function $\mathsf{id}(\cdot)$ is undetermined.
Next, we define \emph{uncorrupted} logs.

\begin{definition} [Uncorrupted logs]
A log $\mathcal{L}$ is \emph{uncorrupted} if and only if
\begin{equation*}
    \sum\limits_{p \in \mathcal{F} \cap \mathsf{processes}(\mathcal{L})} \sum\limits_{\mathit{id} \in \mathsf{id}(p) \cap \mathcal{L}.\mathsf{validators}} \mathsf{S}(\mathcal{L}, \mathit{id}) \leq \rho \cdot \mathbb{T}.
\end{equation*}
\end{definition}

We start by proving that, if $\mathsf{log}(p, \tau) = \LOG$, where $p$ is a correct process and $\tau \in \mathbb{N}_{\geq 0}$ is a timeslot, then $\LOG.\mathsf{validators}$ is not overly corrupted.

\begin{proposition}\label{proposition:not_overly_corrupted}
Consider any correct process $p$ and any timeslot $\tau \in \mathbb{N}_{\geq 0}$.
Then, every log $\mathcal{L}$ such that $\mathsf{log}(p, \tau)$ extends $\mathcal{L}$ is uncorrupted.
\end{proposition}
\begin{proof}
The proposition holds because of the definition of the quasi-permissionless model (see \Cref{subsection:permissioned_quasi_permissionless}). 
\end{proof}

Next, we prove that correct processes only adopt fully-certified logs.

\begin{proposition}\label{proposition:fully_certified}
Consider any correct process $p$ and any timeslot $\tau \in \mathbb{N}_{\geq 0}$.
Then, $\mathsf{log}(p, \tau)$ is fully-certified at process $p$.
\end{proposition}
\begin{proof}
The proposition trivially holds due to the rule at~\Cref{line:new_log_received_gossip}.
\end{proof}

Next, we prove that correct processes only adopt logs that extend the genesis log $\mathcal{L}_{\mathrm{g}}$.

\begin{proposition}\label{proposition:log_extends_genesis}
Consider any correct process $p$ and any timeslot $\tau \in \mathbb{N}_{\geq 0}$.
Then, $\mathsf{log}(p, \tau)$ extends the genesis log $\mathcal{L}_{\mathrm{g}}$.
\end{proposition}
\begin{proof}
The proposition trivially holds as (1) process $p$ only extends its local log throughout every execution (due to the rule at~\Cref{line:new_log_received_gossip}), and (2) $\mathsf{log}(p, 0) = \mathcal{L}_{\mathrm{g}}$.
\end{proof}

We say that a correct process $p$ obtains a log $\mathcal{L}$ at a timeslot $\tau \in \mathbb{N}_{\geq 0}$ if and only if $p$ receives an $\mathsf{obtain\_log}(\mathcal{L})$ event (see~\Cref{line:upon_obtain_log}) at timeslot $\tau$.

\begin{proposition} \label{proposition:log_obtaining}
If any correct process $p$ obtains a log $\mathcal{L}$ with $\mathcal{L}.\mathsf{epoch} = e > 0$, then the following holds for log $\mathcal{L}' = \mathcal{L}.\mathsf{ep\_prefix}(e - 1)$: (1) $p \in \mathsf{processes}(\mathcal{L}')$, and (2) there exists a timeslot $\tau \in \mathbb{N}_{\geq 0}$ such that $\mathsf{log}(p, \tau) = \mathcal{L}'$.
\end{proposition}
\begin{proof}
As process $p$ obtains a log $\mathcal{L}$ (\Cref{line:upon_obtain_log}), process $p$ previously starts simulation of the underlying permissioned protocol $\mathcal{P}$ (\Cref{line:start_simulation_1} or~\Cref{line:start_simulation_2}) with some log $\mathcal{L}'$ and updates its local variable $e_p$ to $\mathcal{L}'.\mathsf{epoch} + 1$.
As $p$ does obtain a log, that implies that $p \in \mathsf{processes}(\mathcal{L}')$.
Moreover, when $p$ starts simulation with $\mathcal{L}'$, it sets $\mathcal{L}'$ as the genesis log for the underlying permissioned protocol $\mathcal{P}$.
As $\mathcal{P}$ satisfies consistency, any obtained log extends $\mathcal{L}'$.
Finally, as $p$ is only obtaining logs whose epoch is $e_p = \mathcal{L}'.\mathsf{epoch} + 1$, the statement of the proposition is satisfied for log $\mathcal{L}'$.
\end{proof}

\subsection{Proof of Consistency}
\label{section:proof-of-consistency}

First, we prove that no correct process roll-backs its log.

\begin{lemma} [No roll-backs] \label{lemma:no_roll_back}
Consider any correct process $p$ and any two timeslots $\tau_1, \tau_2 \in \mathbb{N}_{\geq 0}$ with $\tau_1 < \tau_2$.
Then, $\mathsf{log}(p, \tau_2)$ extends $\mathsf{log}(p, \tau_1)$.
\end{lemma}
\begin{proof}
The lemma trivially holds as $p$ only updates its local log upon receiving a log that extends its current one (\Cref{line:new_log_received_gossip}).
\end{proof}

Next, we need to prove that the logs of correct processes never diverge.
To this end, we define the set $\mathsf{logs}(e)$, for every epoch $e \in \mathbb{N}_{\geq 0}$:
\begin{equation*}
    \mathsf{logs}(e) = \{ \mathcal{L} \,|\, \mathcal{L}.\mathsf{epoch} = e \land \exists (p \in \Pi, \tau \in \mathbb{N}_{\geq 0}): p \text{ is correct} \land \mathsf{log}(p, \tau) = \mathcal{L}\}.
\end{equation*}
Note that $\mathsf{logs}(0) = \{ \mathcal{L}_{\mathrm{g}} \}$.
Now, we prove that only one completed log can exist within $\mathsf{logs}(e)$, for every epoch $e$, assuming that all logs in $e$ are mutually consistent.

\begin{proposition}\label{proposition:one_completed}
Consider any epoch $e \in \mathbb{N}_{\geq 0}$.
Suppose that for every pair of logs $(\mathcal{L}_1, \mathcal{L}_2) \in \big( \mathsf{logs}(e) \big)^2$, logs $\mathcal{L}_1$ and $\mathcal{L}_2$ are consistent.
Then, there exists at most one log $\mathcal{L} \in \mathsf{logs}(e)$ such that $\mathcal{L}.\mathsf{completed} = \mathit{true}$.
\end{proposition}
\begin{proof}
By contradiction, suppose at least two logs $\mathcal{L}_1$ and $\mathcal{L}_2$ exist such that (1) $\mathcal{L}_1 \in \mathsf{logs}(e)$, (2) $\mathcal{L}_2.\mathsf{logs}(e)$, and (3) $\mathcal{L}_1.\mathsf{completed} = \mathcal{L}_2.\mathsf{completed} = \mathit{true}$.
As both logs belong to $\mathsf{logs}(e)$, they are consistent.
However, it is then impossible that $\mathcal{L}_1.\mathsf{completed} = \mathcal{L}_2.\mathsf{completed} = \mathit{true}$ (due to~\Cref{definition:epoch-completed}), which concludes the proof.
\end{proof}

Next, we prove some constraints on a process that obtains a specific log.

\begin{proposition}\label{proposition:obtained_implies_completed}
If a correct process obtains a log $\mathcal{L}$ with $\mathcal{L}.\mathsf{epoch} = e > 0$, then $\mathcal{L}.\mathsf{ep\_prefix}(e - 1) \in \mathsf{logs}(e - 1)$.
\end{proposition}
\begin{proof}
Recall that $\mathcal{L}.\mathsf{ep\_prefix}(e - 1).\mathsf{epoch} = e - 1$.
Hence, the proposition follows directly from \Cref{proposition:log_obtaining}.
\end{proof}

Next, we prove that every log $\mathcal{L} \in \mathsf{logs}(e)$ is obtained by a correct process.

\begin{proposition} \label{proposition:obtained}
Consider any epoch $e \in \mathbb{N}$.
For every log $\mathcal{L} \in \mathsf{logs}(e)$, $\mathcal{L}$ is obtained by a correct process.
\end{proposition}
\begin{proof}
As $\mathcal{L} \in \mathsf{logs}(e)$, there exist a correct process $p$ and a timeslot $\tau$ such that $\mathsf{log}(p, \tau) = \mathcal{L}$.
Hence, $\mathcal{L}$ is fully-certified at process $p$ (by \Cref{proposition:fully_certified}).
Moreover, $\mathcal{L}.\mathsf{ep\_prefix}(e - 1)$ is uncorrupted (by \Cref{proposition:not_overly_corrupted}).
Therefore, there exists a correct process $q \in \mathsf{processes}(\mathcal{L}.\mathsf{ep\_prefix}(e - 1))$ that obtains $\mathcal{L}$.
Hence, the proposition holds.
\end{proof}

Next, we prove a direct consequence of \Cref{proposition:obtained_implies_completed,proposition:obtained}.

\begin{proposition} \label{proposition:obtained_implies_completed_all}
If a correct process obtains a log $\mathcal{L}$ with $\mathcal{L}.\mathsf{epoch} = e > 0$, then, for every $e' \in [0, e - 1]$, $\mathcal{L}.\mathsf{ep\_prefix}(e') \in \mathsf{logs}(e')$.
\end{proposition}
\begin{proof}
The proposition is proven by (1) inductively applying \Cref{proposition:obtained_implies_completed} and then \Cref{proposition:obtained}, and (2) the fact, for every epoch $e \in \mathbb{N}$ and every log $\mathcal{L} \in \mathsf{logs}(e)$, $\mathcal{L}_{\mathrm{g}} = \mathcal{L}.\mathsf{ep\_prefix}(0)$.
\end{proof}

Finally, we prove that all logs within $\mathsf{logs}(e)$ are mutually consistent, for every epoch $e \in \mathbb{N}_{\geq 0}$.

\begin{proposition} \label{proposition:all_consistent}
Consider any epoch $e \in \mathbb{N}_{\geq 0}$ and any pair of logs $(\mathcal{L}_1, \mathcal{L}_2) \in \big( \mathsf{logs}(e) \big)^2$.
Then, logs $\mathcal{L}_1$ and $\mathcal{L}_2$ are consistent.
\end{proposition}
\begin{proof}
We prove the proposition by induction.

\smallskip
\noindent\emph{Base step: We prove the proposition holds for $e = 0$.}
\\ The proposition trivially holds for $e = 0$ as $\mathsf{logs}(0) = \{ \mathcal{L}_{\mathrm{g}} \}$.

\smallskip
\noindent\emph{Inductive step: The proposition holds for some $e \geq 0$. We prove the proposition for $e + 1$.
}
\\ Both logs $\mathcal{L}_1$ and $\mathcal{L}_2$ are obtained by correct processes (by \Cref{proposition:obtained}).
Hence, $\mathcal{L}_1.\mathsf{ep\_prefix}(e) \in \mathsf{logs}(e)$ and $\mathcal{L}_2.\mathsf{ep\_prefix}(e) \in \mathsf{logs}(e)$ (by \Cref{proposition:obtained_implies_completed}).
As $\mathcal{L}_1.\mathsf{ep\_prefix}(e).\mathsf{completed} = \mathcal{L}_2.\mathsf{ep\_prefix}(e).\mathsf{completed} = \mathit{true}$, $\mathcal{L}_1.\mathsf{ep\_prefix}(e) = \mathcal{L}_2.\mathsf{ep\_prefix}(e)$ (by \Cref{proposition:one_completed}).

Let $r_1$ be a correct process that obtains $\mathcal{L}_1$ and $r_2$ be a correct process that obtains $\mathcal{L}_2$.
By \Cref{proposition:log_obtaining}, $p, q \in \mathsf{processes}(\mathcal{L}_1.\mathsf{ep\_prefix}(e))$.
Therefore, logs $\mathcal{L}_1$ and $\mathcal{L}_2$ are indeed consistent as, by \Cref{lemma:consistency_preservation_quit}, the underlying protocol $\mathcal{P}$ satisfies consistency (even though correct processes might have stopped executing it, which goes against the standard permissioned model $\mathcal{P}$ is designed for).
\end{proof}

Finally, we prove that the logs of correct processes never diverge.

\begin{lemma}[No divergence] \label{lemma:no_fork}
Consider any pair of correct processes $(p, q) \in \Pi^2$ and any timeslot $\tau \in \mathbb{N}_{\geq 0}$.
Then, logs $\mathsf{log}(p, \tau)$ and $\mathsf{log}(q, \tau)$ are consistent.
\end{lemma}
\begin{proof}
Let $\mathcal{L}_p = \mathsf{log}(p, \tau)$ and $\mathcal{L}_q = \mathsf{log}(q, \tau)$.
Moreover, let $e_p = \mathcal{L}_p.\mathsf{epoch}$ and $e_q = \mathcal{L}_q.\mathsf{epoch}$.
Without loss of generality, assume $e_p \leq e_q$.
We separate two cases:
\begin{compactitem}
    \item Let $e_p = e_q$.
    In this case, the lemma follows from \Cref{proposition:all_consistent}.

    \item Let $e_p < e_q$.
    In this case, a correct process obtains $\mathcal{L}_q$ (by \Cref{proposition:obtained}).
    Hence, $\mathcal{L}_q.\mathsf{prefix}(e_p) \in \mathsf{logs}(e_p)$ (by \Cref{proposition:obtained_implies_completed_all}).
    Therefore, $\mathcal{L}_q.\mathsf{prefix}(e_p)$ and $\mathcal{L}_p$ are consistent (by \Cref{proposition:all_consistent}), which implies that $\mathcal{L}_p$ and $\mathcal{L}_q$ are consistent.
\end{compactitem}
The lemma holds as its statement is satisfied in both possible cases.
\end{proof}

\subsection{Proof of Composable Log-Specific Safety Properties}
\label{section:proof-of-log-specific}

\begin{proof}[Proof of~\Cref{theorem:log-specific-properties}]

By~\Cref{theorem:consistency}, we know that $\mathcal{T}(\Pro)$ satisfies consistency.
We prove the lemma by induction and contradiction.
Consider an execution $\mathcal{E}$ of $\mathcal{T}(\Pro)$ and the set $\mathsf{logs}:=\mathsf{logs}(\mathcal{E})$.

\smallskip
\noindent\emph{Base step:}
Given a $\rho$-bounded adversary, there exists an execution $\mathcal{E}$ of the quit-enhanced version $\Pro'$ of the protocol $\Pro$ such that $\mathsf{logs}(\mathcal{E}) = \mathsf{logs}(1)$.
Then, since $\Pro$ satisfies $P$ against a $\rho$-bounded static adversary, by~\Cref{lemma:log-specific-quit-enhanced}, so does $\Pro'$, implying that $P(\mathsf{logs}(1)) = P(\mathsf{logs}(\mathcal{E})) = \mathit{true}$.
Hence, by~\Cref{definition:safety-log-specific-property}, for all $\mathsf{logs}' \subseteq \mathsf{logs}(1)$, it holds that $P(\mathsf{logs}') = \mathit{true}$.

\smallskip
\noindent\emph{Inductive step:}
Let $\mathsf{logs}_e = \bigcup_{i = 1, \ldots, e} \mathsf{logs}(e)$, and for all $\mathsf{logs}' \subseteq \mathsf{logs}_e$, it holds that $P(\mathsf{logs}') = \mathit{true}$.
By the safety of $\mathcal{T}(\Pro)$, for any logs $\mathcal{L},\mathcal{L}' \in \mathsf{logs}(e+1)$, it follows that $\mathcal{L}.\mathsf{prefix}(e) = \mathcal{L}'.\mathsf{prefix}(e)$, denoted by some $\mathcal{L}_e$.
Thus, given a $\rho$-bounded adversary, there exists an execution $\mathcal{E}$ of the quit-enhanced version $\Pro'$ of the protocol $\Pro$ with the genesis log $\mathcal{L}_e$ such that $\mathsf{logs}(\mathcal{E}) = \mathsf{logs}(e+1)$.
Then, since $\Pro$ satisfies $P$ against a $\rho$-bounded static adversary, by~\Cref{lemma:log-specific-quit-enhanced}, so does $\Pro'$, implying that $P(\mathsf{logs}(e+1)) = P(\mathsf{logs}(\mathcal{E})) = \mathit{true}$.
Therefore, by~\Cref{definition:safety-log-specific-property}, for all $\mathsf{logs}'' \subseteq \mathsf{logs}(e+1)$, it holds that $P(\mathsf{logs}'') = \mathit{true}$.
By~\Cref{definition:composable-log-specific} 
(composability), this implies for any $\mathsf{logs}' \subseteq \mathsf{logs}_e$ and $\mathsf{logs}'' \subseteq \mathsf{logs}(e+1)$; for every $\mathsf{logs}^* \subseteq \mathsf{logs}' \cup \mathsf{logs}''$, $P(\mathsf{log}^*) = \mathit{true}$.
Therefore, for all $\mathsf{logs}^* \subseteq \mathsf{logs}_{e+1}$, it holds that $P(\mathsf{logs}') = \mathit{true}$.

We have shown that for any $e < \infty$ and all $\mathsf{logs}^* \subseteq \mathsf{logs}_{e+1}$, $P(\mathsf{logs}^*) = \mathit{true}$.
Finally, suppose $P(\mathsf{logs}) = \mathit{false}$.
Then, by~\Cref{definition:safety-log-specific-property}, there exists a finite subset $\mathsf{logs}^* \subseteq \mathsf{logs}$ such that $P(\mathsf{logs}^*) = \mathit{false}$.
Now, let $e_{\mathit{max}} < \infty$ denote the largest epoch among the logs in $\mathsf{logs}^*$, \ie $e_{\mathit{max}} = \max\{\mathcal{L}.\mathsf{epoch} | \mathcal{L} \in \mathsf{logs}^*\}$, which is well-defined since $\mathsf{logs}^*$ is finite.
However, then $\mathsf{logs}^* \subseteq \mathsf{logs}_{e_{\mathit{max}}}$ and for all $\mathsf{logs}^* \subseteq \mathsf{logs}_{e_{\mathit{max}}}$, we know that $P(\mathsf{logs}') = \mathit{true}$, as $e_{\mathit{max}} < \infty$.
This is a contradiction, which implies $P(\mathsf{logs}) = \mathit{true}$.
\end{proof}

\subsection{Proof of Liveness}
\label{section:proof-of-liveness}

Suppose the adversary is a $\rho$-bounded static adversary in the quasi-permissionless setting.
Recall that $\mathsf{logs}$ denote the set of logs, including their prefixes, 
output by correct processes.
By~\Cref{theorem:consistency}, we observe that the protocol $\mathcal{T}(\Pro)$ is $\rho$-consistent.

\begin{lemma}
\label{lemma:unique_log}
For every epoch $e \in \mathbb{N}$, there exists a unique log $\LOG^*(e)$ such that (1) $\LOG^*(e) \in \mathsf{logs}$, (2) $\LOG^*(e).\mathsf{epoch} = e$, and (3) $\LOG^*(e).\mathsf{completed} = \mathit{true}$.
Moreover, for any $e < e'$, $\LOG^*(e')$ extends $\LOG^*(e)$.
\end{lemma}

\begin{proof}
By the consistency of $\mathcal{T}(\mathcal{S})$, all logs obtained by the correct processes are consistent.
Therefore, for any two logs $\LOG$ and $\LOG'$ obtained by the correct processes, if $\LOG.\mathsf{ep\_prefix}(e)$ and $\LOG'.\mathsf{ep\_prefix}(e)$ are both defined, it holds that $\LOG.\mathsf{ep\_prefix}(e) = \LOG'.\mathsf{ep\_prefix}(e)$ by property (1), $\LOG.\mathsf{epoch} = \LOG'.\mathsf{epoch} = e$ by property (2), and $\LOG.\mathsf{completed} = \LOG'.\mathsf{completed} =\mathit{true}$ by property (3) in~\Cref{definition:epoch-prefix}, all of which imply the first statement of the lemma.
Moreover, by consistency and~\Cref{definition:epoch}, for any $e < e'$, $\LOG^*(e')$ extends $\LOG^*(e)$.
\end{proof}

We say that a process $p$ is an $e$-process if there exists an identifier $\mathit{id} \in \mathsf{id}(p)$ such that $\mathit{id} \in \LOG.\mathsf{validators}$, where the $\LOG \in \mathsf{logs}(e)$. 
For each epoch $e \in \mathbb{N}_{\geq 1}$, let $\tau(e)$ denote the timeslot at which the first correct $e$-process enters epoch $e$ 
at~\Cref{line:start_simulation_1} or~\Cref{line:start_simulation_2}.
Formally, $\tau(e)$ is the first timeslot a correct process holds a log $\LOG$ such that $\LOG.\mathsf{epoch} = e-1$ and $\LOG.\mathsf{completed} = \mathit{true}$.
Lastly, we say that an epoch $e \in \mathbb{N}_{\geq 1}$ is a \emph{post-GST epoch} if $\tau(e) \geq \GST$.
Define $\ed = \ell + \Delta < \infty$.

\begin{lemma} 
\label{lemma:end_time}
Let $e \in \mathbb{N}_{\geq 1}$ and $e' \in \mathbb{N}_{\geq 1}$ be any two post-GST epochs such that $e < e'$.
Then, $\tau(e') \geq \tau(e) + \ed$. 
\end{lemma}
\begin{proof}
By~\Cref{lemma:unique_log}, $\LOG^{\star}(e')$ extends $\LOG^{\star}(e)$.
By the fact that $\LOG^{\star}(e) \in \mathsf{logs}$ and~\Cref{proposition:fully_certified}, the log $\LOG^{\star}(e)$ is fully-certified (\Cref{line:new_log_received}).
Moreover, for any $\tilde{e} \in [0, e']$, by~\Cref{proposition:not_overly_corrupted},
\begin{equation*}
\sum_{\text{corrupt }\mathit{id} \in \LOG^{\star}(\tilde{e}).\mathsf{validators}} \mathsf{S}(\LOG^{\star}(\tilde{e}), \mathit{id}) \leq \rho\cdot \LOG^{\star}(\tilde{e}).\mathsf{total\_stake}
\end{equation*}

A correct $e'$-process $p_i$ receives a fully-certified log $\LOG^{\star}(e'-1)$ (\Cref{line:new_log_received}) before entering epoch $e'$ (\Cref{line:start_simulation_1} and~\Cref{line:start_simulation_2}).
Therefore, as $1-\rho > \rho$, there exists a correct $(e'-1)$-process $p_j$ that signed the log $\LOG^{\star}(e'-1)$ by timeslot $\tau(e')$, and by~\Cref{definition:epoch-completed}, there is a transaction $(\textsc{finish}, e)$ on $\LOG^{\star}(e'-1)$ that was sent by a correct $(e'-1)$-process.
Since no correct $(e'-1)$-process sends such a transaction before timeslot $\tau(e' - 1) + \ed$ and as $\tau(e_1) \geq \tau(e_2)$ for any $e_1 \geq e_2$ by~\Cref{algorithm:pos-code_2}, it holds that no correct $(e'-1)$-process sends such a transaction before timeslot $\tau(e) + \ed$, implying that $\tau(e') \geq \tau(e) + \ed$.
\end{proof}

Next, we prove that, given any post-GST epoch $e$, all correct $e$-processs enter epoch $e$ by timeslot $\tau(e) + \delta$, where $\delta$ is the actual message delay.

\begin{lemma} \label{lemma:start_time}
Let $e \in \mathbb{N}_{\geq 1}$ be any post-GST epoch.
Then, all correct $e$-processs enter epoch $e$ by timeslot $\tau(e) + \delta$.
\end{lemma}
\begin{proof}
If $e = 1$, the lemma trivially holds as all correct $1$-processes start executing the protocol at time $0$, and $\GST = 0$.
Hence, let $e > 1$.

The first correct $e$-process enters epoch $e$ at timeslot $\tau(e)$ upon receiving a fully-certified log $\LOG^{\star}(e - 1)$.
As $\tau(e) \geq \GST$, all correct $e$-processs receive $\LOG^{\star}(e-1)$ by timeslot $\tau(e) + \delta$.
Let $\LOG'$ denote the log of a process $p_j$ right before the timeslot it received $\LOG^{\star}(e-1)$.
Either $\LOG'.\mathsf{epoch} \leq e$, which implies $p_j$ enters epoch $e$ by timeslot $\tau(e) + \delta$, or $\LOG'.\mathsf{epoch} = e' > e$, which implies $\tau(e') \leq \tau(e) + \delta$ and contradicts with~\Cref{lemma:end_time}
Therefore, all correct $e$-processes enter epoch $e$ by timeslot $\tau(e) + \delta$.
\end{proof}

\begin{proof}[Proof of~\Cref{theorem:liveness}]
Let $\tx$ be a transaction received by a correct process at some timeslot $\tau$ for the first time.
Let $e_{\mathit{max}}$ denote the largest epoch entered by an honest process by timeslot $\max(\tau, \GST) + \delta$, \ie,
\begin{equation*}
e_{\mathit{max}} = \max_{\LOG \in \mathsf{logs} \text{ at} \max(\tau, \GST) + \delta} \LOG.\mathsf{epoch}    
\end{equation*}
Importantly, every correct $e_{\mathit{max}}$-process $p_i$ obtains $\tx$ by timeslot $\max(\tau, \GST) + \delta$.
We distinguish the following two cases:
\begin{compactitem}
    \item Suppose $\tau(e_{\mathit{max}} + 1) > \max(\tau, \GST) + \delta + \ell$.
    By~\Cref{lemma:start_time}, all correct $e_{\mathit{max}}$-processes enter epoch $e_{\mathit{max}}$ by timeslot $\max(\tau,\GST) + \delta \geq \tau(e_{\mathit{max}}) + \delta$, by which they receive $\tx$, and after timeslot $\max(\tau,\GST) + \delta$, they all stay in epoch $e_{\mathit{max}}$ for at least $\ell$ more timeslots (all after $\GST)$.
    Therefore, by~\Cref{proposition:quit-enhanced-liveness} and the $(\rho, \ell)$-liveness of the protocol $\Pro$, every correct $e_{\mathit{max}}$-process obtains some log $\LOG$ with $\LOG.\mathsf{epoch} = e_{\mathit{max}}$ containing $\tx$ by timeslot $\max(\tau, \GST) + \delta + \ell$\footnote{Even though the correct processes stop executing the protocol at the end of the epoch, which goes against the standard permissioned model $\mathcal{P}$ is designed for, by~\Cref{proposition:quit-enhanced-liveness}, we know that the liveness of $\mathcal{P}$ carries over to the liveness of its quit-enhanced version.}.
    Since for all logs $\LOG$, $\LOG.\mathsf{epoch} > e_{\mathit{max}}$, obtained by a correct process it holds that $\LOG.\mathsf{ep\_prefix}(e_{\mathit{max}}) \in \mathsf{logs}(e_{\mathit{max}})$ by~\Cref{proposition:obtained_implies_completed}, for any correct process that is active (and non-waiting) at a timeslot $\tau_{\mathrm{a}} \geq \max(\tau, \GST) + \delta + \ell$, \ie, for any epoch $e$-process, $e \geq e_{\mathit{max}}$, it holds that $\mathsf{tr} \in \mathsf{log}(p, \tau_{\mathrm{a}})$.

    \smallskip
    Moreover, if $\Pro$ is also $(\rho,\ell_{\mathrm{or}})$-responsive, every correct $e_{\mathit{max}}$-process obtains some log $\LOG$ with $\LOG.\mathsf{epoch} = e_{\mathit{max}}$ containing $\tx$ by timeslot $\max(\tau, \GST) + \delta + \ell_{\mathrm{or}}$, where $\ell_{\mathrm{or}} \in O(\delta)$.
    Since for all logs $\LOG$, $\LOG.\mathsf{epoch} > e_{\mathit{max}}$, obtained by a correct process it holds that $\LOG.\mathsf{ep\_prefix}(e_{\mathit{max}}) \in \mathsf{logs}(e_{\mathit{max}})$ by~\Cref{proposition:obtained_implies_completed}, for any correct process that is active (and non-waiting) at a timeslot $\tau_{\mathrm{a}} \geq \max(\tau, \GST) + \delta + \ell_{\mathrm{or}}$, \ie, for any epoch $e$-process, $e \geq e_{\mathit{max}}$, it holds that $\mathsf{tr} \in \mathsf{log}(p, \tau_{\mathrm{a}})$.
    \smallskip
    
    \item Let $\tau(e_{\mathit{max}} + 1) \leq \max(\tau, \GST) + \delta + \ell$.
    In this case, by~\Cref{lemma:start_time}, every correct $(e_{\mathit{max}} + 1)$-process receives $\tx$ and enters epoch $\tau(e_{\mathit{max}} + 1)$ by timeslot $\tau(e_{\mathit{max}} + 1) + \delta > \GST$, and by~\Cref{lemma:end_time}, after timeslot $\tau(e_{\mathit{max}} + 1) + \delta$, they all stay in epoch $e_{\mathit{max}}+1$ for at least $\ell$ more timeslots as $\ed = \ell + \Delta > \ell + \delta$ (all after $\GST)$.
    Therefore, by~\Cref{proposition:quit-enhanced-liveness} and the $(\rho, \ell)$-liveness of the protocol $\Pro$, every correct $(e_{\mathit{max}}+1)$-process obtains a log $\LOG$ with $\LOG.\mathsf{epoch} = e_{\mathit{max}}+1$ containing $\tx$ by timeslot $\tau(e_{\mathit{max}} + 1) + \ell + \delta \leq \max(\tau, \GST) + 2\delta + 2\ell$.
    Since for all logs $\LOG$, $\LOG.\mathsf{epoch} > e_{\mathit{max}}+1$, obtained by a correct process it holds that $\LOG.\mathsf{ep\_prefix}(e_{\mathit{max}}+1) \in \mathsf{logs}(e_{\mathit{max}}+1)$ by~\Cref{proposition:obtained_implies_completed}, for any correct process that is active (and non-waiting) at a timeslot $\tau_{\mathrm{a}} \geq \max(\tau, \GST) + 2\delta + 2\ell$, \ie, for any epoch $e$-process, $e \geq e_{\mathit{max}}+1$, it holds that $\mathsf{tr} \in \mathsf{log}(p, \tau_{\mathrm{a}})$.

    \smallskip
    Now, suppose $\Pro$ is also $(\rho,\ell_{\mathrm{or}})$-responsive, and consider the following two cases:
    \begin{itemize}
        \item Suppose $\tau(e_{\mathit{max}} + 1) > \max(\tau, \GST) + \delta + \ell_{\mathrm{or}}$.
        Then, by~\Cref{lemma:start_time}, all correct $e_{\mathit{max}}$-processes enter epoch $e_{\mathit{max}}$ by timeslot $\max(\tau,\GST) + \delta \geq \tau(e_{\mathit{max}}) + \delta$, by which they receive $\tx$, and after timeslot $\max(\tau,\GST) + \delta$, they all stay in epoch $e_{\mathit{max}}$ for at least $\ell_{\mathrm{or}}$ more timeslots (all after $\GST)$.
        Therefore, by the $(\rho,\ell_{\mathrm{or}})$-responsiveness of the protocol $\Pro$, every correct $e_{\mathit{max}}$-process obtains some log $\LOG$ with $\LOG.\mathsf{epoch} = e_{\mathit{max}}$ containing $\tx$ by timeslot $\max(\tau, \GST) + \delta + \ell_{\mathrm{or}}$.
        Since for all logs $\LOG$, $\LOG.\mathsf{epoch} > e_{\mathit{max}}$, obtained by a correct process it holds that $\LOG.\mathsf{ep\_prefix}(e_{\mathit{max}}) \in \mathsf{logs}(e_{\mathit{max}})$ by~\Cref{proposition:obtained_implies_completed}, for any correct process that is active (and non-waiting) at a timeslot $\tau_{\mathrm{a}} \geq \max(\tau, \GST) + \delta + \ell_{\mathrm{or}}$, \ie, for any epoch $e$-process, $e \geq e_{\mathit{max}}$, it holds that $\mathsf{tr} \in \mathsf{log}(p, \tau_{\mathrm{a}})$.
        \item Suppose $\tau(e_{\mathit{max}} + 1) \leq \max(\tau, \GST) + \delta + \ell_{\mathrm{or}}$.
        In this case, by~\Cref{lemma:start_time}, every correct $(e_{\mathit{max}} + 1)$-process receives $\tx$ and enters epoch $\tau(e_{\mathit{max}} + 1)$ by timeslot $\tau(e_{\mathit{max}} + 1) + \delta > \GST$, and by~\Cref{lemma:end_time}, after timeslot $\tau(e_{\mathit{max}} + 1) + \delta$, they all stay in epoch $e_{\mathit{max}}+1$ for at least $\ell$ more timeslots as $\ed = \ell + \Delta > \ell + \delta$ (all after $\GST)$.
        Therefore, by the $(\rho,\ell_{\mathrm{or}})$-responsiveness of the protocol $\Pro$, every correct $(e_{\mathit{max}}+1)$-process obtains a log $\LOG$ with $\LOG.\mathsf{epoch} = e_{\mathit{max}}+1$ containing $\tx$ by timeslot $\tau(e_{\mathit{max}} + 1) + \ell_{\mathrm{or}} + \delta \leq \max(\tau, \GST) + 2\delta + \ell_{\mathrm{or}}$.
        Since for all logs $\LOG$, $\LOG.\mathsf{epoch} > e_{\mathit{max}}+1$, obtained by a correct process it holds that $\LOG.\mathsf{ep\_prefix}(e_{\mathit{max}}+1) \in \mathsf{logs}(e_{\mathit{max}}+1)$ by~\Cref{proposition:obtained_implies_completed}, for any correct process that is active (and non-waiting) at a timeslot $\tau_{\mathrm{a}} \geq \max(\tau, \GST) + 2\delta + 2\ell_{\mathrm{or}}$, \ie, for any epoch $e$-process, $e \geq e_{\mathit{max}}+1$, it holds that $\mathsf{tr} \in \mathsf{log}(p, \tau_{\mathrm{a}})$.
    \end{itemize}
\end{compactitem}
This concludes the theorem.
\end{proof}

\subsection{Proof of Accountability}\label{section:accountability}

In this subsection, we prove \Cref{theorem:accountability}.

\begin{proof}[Proof of~\Cref{theorem:accountability}]
Suppose there exist correct processes $p \in \Pi$ and $q \in \Pi$, and timeslots $\tau_p \in \mathbb{N}_{\geq 1}$ and $\tau_q \in \mathbb{N}_{\geq 1}$ such that $\mathsf{log}(p, \tau_p)$ is inconsistent with $\mathsf{log}(q, \tau_q)$.
Let $L$ denote the longest log (i.e., the log with the most transactions) such that (1) $\mathsf{log}(p, \tau_p)$ extends $L$, and (2) $\mathsf{log}(q, \tau_q)$ extends $L$.
Note that such log $L$ exists as $p$ and $q$ share the same genesis log $\mathcal{L}_{\mathrm{g}}$.
Moreover, note that $L.\mathsf{length} < \mathsf{log}(p, \tau_p).\mathsf{length}$ and $L.\mathsf{length} < \mathsf{log}(q, \tau_q).\mathsf{length}$.

Let $L_p$ denote the log such that (1) $L_p.\mathsf{predecessor} = L$, and (2) $\mathsf{log}(p, \tau_p)$ extends $L_p$.
Similarly, let $L_q$ denote the log such that (1) $L_q.\mathsf{predecessor} = L$, and (2) $\mathsf{log}(q, \tau_q)$ extends $L_q$.
By the definition of $L$, $L_p$, and $L_q$, logs $L_p$ and $L_q$ are inconsistent.
Crucially, $L_p.\mathsf{epoch} = L_q.\mathsf{epoch}$ and $L_p.\mathsf{validators} = L_q.\mathsf{validators}$ by \Cref{definition:identifiers},
since both logs share the same predecessor log $L$, i.e., $L_p.\mathsf{predecessor} = L_q.\mathsf{predecessor} = L$.

Moreover, let us define the log $L_p^{\star}$ in the following way:
\begin{equation*}
        L_p^{\star} \equiv \begin{cases}
            \mathsf{log}(p, \tau_p), &\text{if $\mathsf{log}(p, \tau_p).\mathsf{epoch} = L_p.\mathsf{epoch}$}\\
            \mathsf{log}(p, \tau_p).\mathsf{ep\_prefix}(L_p.\mathsf{epoch}), &\text{otherwise}.
        \end{cases}
\end{equation*}
We define the log $L_q^{\star}$ in the same way:
\begin{equation*}
        L_q^{\star} \equiv \begin{cases}
            \mathsf{log}(q, \tau_q), &\text{if $\mathsf{log}(q, \tau_q).\mathsf{epoch} = L_q.\mathsf{epoch}$}\\
            \mathsf{log}(q, \tau_q).\mathsf{ep\_prefix}(L_q.\mathsf{epoch}), &\text{otherwise}.
        \end{cases}
\end{equation*}
Note that (1) $L_p^{\star}.\mathsf{epoch} = L_q^{\star}.\mathsf{epoch} = L_p.\mathsf{epoch} = L_q.\mathsf{epoch}$, and (2) $L_p^{\star}.\mathsf{validators} = L_q^{\star}.\mathsf{validators} = L_p.\mathsf{validators} = L_q.\mathsf{validators}$.
Let $\mathbb{V} = L_p^{\star}.\mathsf{validators} = L_q^{\star}.\mathsf{validators}$.
Moreover, $L_p^{\star}$ extends $L_p$ and $L_q^{\star}$ extends $L_q$, which implies that $L_p^{\star}$ and $L_q^{\star}$ are inconsistent.

Finally, observe that since processes $p$ and $q$ have received fully certified logs $\mathsf{log}(p, \tau_p)$ and $\mathsf{log}(q, \tau_q)$, it follows that:
(1) by timeslot $\tau_p$, process $p$ obtained a quorum (with respect to $\mathbb{V}$) of signatures on $L_p^{\star}$, and
(2) by timeslot $\tau_q$, process $q$ obtained a quorum (with respect to $\mathbb{V}$) of signatures on $L_q^{\star}$.
Therefore, the messages received by $p$ and $q$ when collecting signatures on logs $L_p^{\star}$ and $L_q^{\star}$, respectively, contain proof of guilt of at least $(1 - 2\rho)\mathbb{T}$-worth of identifiers in $\mathbb{V}$.
This follows from the fact that each process received signatures from $(1 - \rho)\mathbb{T}$-worth of stake.
Thus, the quasi-permissionless protocol $\mathcal{T}(\mathcal{P})$ satisfies $(1 - 2\rho)$-accountability.
\end{proof}

\subsection{Proof of Message Complexity}
\label{section:message-complexity}

In this subsection, we study the message complexity of a quasi-permissionless PoS protocol $\mathcal{T}(\mathcal{P})$ obtained by transforming a permissioned protocol $\mathcal{P}$.
When studying message complexity, one can focus on two distinct aspects:
\begin{compactitem}
    \item %
    Messages exchanged among validators.

    \item 
    Messages exchanged between validators and non-validator processes.
\end{compactitem}
When it comes to messages among validators,
our transformation adds an additional ``signature-round'' to ensure accountability.
Hence, our transformation adds a quadratic additive factor to the number of the 
messages exchanged among the validators in the permissioned protocol $\mathcal{P}$.

Regarding the messages exchanged between validators and non-validator processes,
in our transformation, validators continue to disseminate their logs to all processes, not just validators. 
This is essential because all processes, including non-validators, must finalize new transactions. 
However, as even permissioned protocols must account for this type of dissemination (assuming they serve an undetermined universe of ``clients''), our definition of message complexity focuses exclusively on the messages exchanged among validators. 

Formally, consider a standard blockchain protocol $\mathcal{P}$
\begin{equation*}
    (\Pi, \rho, \mathsf{States}, \{ s_p^0 \,|\, p \in \Pi \}, \mathsf{Messages}, \mathsf{Transactions}, \mathsf{Logs}, \mathsf{Transition}),
\end{equation*}
with the set of processes $\Pi$ such that $|\Pi| = n$.
Let the tuple 
$\mathcal{B} = \Big\langle \mathcal{FR}^0_{\mathcal{E},p}, \ldots, \mathcal{FR}^k_{\mathcal{E},p} = \big( s^k_{\mathcal{E},p}, k, \mathit{waiting}^k_{\mathcal{E},p}, M^{R(k)}_{\mathcal{E},p}, T^{R(k)}_{\mathcal{E},p}, M^{S(k)}_{\mathcal{E},p} \big) \Big \rangle$ be a $k$-long behavior of a process $p \in \Pi$ according to an execution $\mathcal{E}$ of the protocol $\mathcal{P}$, where $T^{R(k)}_{\mathcal{E},p}$, $M^{S(k)}_{\mathcal{E},p}$ and $M^{R(k)}_{\mathcal{E},p}$ respectively denote the transactions received, and the messages sent and received by $p$ at timeslot $k$.
By $|T^{R(k)}_{\mathcal{E},p}|$, we denote the sums of the lengths of the transactions measured in bits within the set $T^{R(k)}_{\mathcal{E},p}$, and $|M^{S(k)}_{\mathcal{E},p}|$ and $|M^{R(k)}_{\mathcal{E},p}|$ are defined similarly.

\begin{definition}
We define the \emph{average message complexity} of $\mathcal{P}$ under $\rho$-bounded static adversaries, denoted by $\mathsf{AC}$, as follows:
\begin{equation*}
    \mathsf{AC} = \lim_{j \to \infty} \sup_{\mathcal{E} \in E} \frac{1}{j} \sum_{p \in \Pi \setminus \mathcal{F}} \sum_{i=0}^j (|T^{R(i)}_{\mathcal{E},p}|+|M^{S(i)}_{\mathcal{E},p}|+|M^{R(i)}_{\mathcal{E},p}|),
\end{equation*}
where $E$ denotes the set of executions of the protocol $\mathcal{P}$ under all $\rho$-bounded static adversaries. 
\end{definition} 

\begin{definition}
We define the \emph{peak message complexity} of $\mathcal{P}$, denoted by $\mathsf{PC}$ as follows:
\begin{equation*}
    \mathsf{PC} = \sup_{\mathcal{E} \in E} \sup_{i \in [0,\infty), p \in \Pi \setminus \mathcal{F}} (|T^{R(i)}_{\mathcal{E},p}|+|M^{S(i)}_{\mathcal{E},p}|+|M^{R(i)}_{\mathcal{E},p}|),
\end{equation*}
where $E$ denotes the set of executions of the protocol $\mathcal{P}$ under all $\rho$-bounded static adversary.
\end{definition}
We assume that all expressions within the limits above are continuous functions of $n$ and converge uniformly to a function of $n$, and $\mathsf{AC}$, $\mathsf{PC}$ are finite.

Given these definitions, we can state our theorem on the message complexity of $\mathcal{T}(\mathcal{P})$.
We assume that $\GST = 0$ in the following analysis to capture the communication complexity of the protocols while they finalize new transactions.
\begin{theorem}
\label{theorem:message-complexity}
Consider a permissioned protocol $\mathcal{P}$ that has finite average and peak message complexity of $\mathsf{AC} = \Omega(n^2)$ and $\mathsf{PC} = \Omega(n^2)$ against $\rho$-bounded static adversaries.
Then, there exist a parameter $\ell$ such that $\mathcal{T}(\mathcal{P})$ with parameter $\ell$ has average and peak message complexity of $\mathsf{AC}' = O(\mathsf{AC})$ and $\mathsf{PC}' = O(\mathsf{PC})$ against $\rho$-bounded static adversaries.
\end{theorem}
Intuitively,~\Cref{theorem:message-complexity} states that the message complexities are asymptotically the same for both the permissioned and PoS protocols.
It follows from the fact that the PoS protocol simply runs multiple iterations of the permissioned protocol stitched together, with quadratic overhead in the number of processes (\eg, epoch-ending messages), which is absorbed by the super-quadratic asymptotics of the permissioned protocol.
\begin{proof}
Let $\tilde{T}^{R(i)}_{\mathcal{E},p}, \tilde{M}^{S(i)}_{\mathcal{E},p}, \tilde{M}^{R(i)}_{\mathcal{E},p}$, $\tilde{\mathsf{AC}}$, $\tilde{\Pi}$ and $\tilde{E}$ denote the same definitions on the PoS protocol $\mathcal{T}(\mathcal{P})$, and suppose both protocols have infinite running times.
By the definition of $\mathsf{AC}$, 
for any positive $\epsilon < 1$, there exists an $\ell>0$ such that 
\begin{equation*}
    \sup_{\mathcal{E} \in E} \frac{1}{2\ell} \sum_{p \in \Pi \setminus \mathcal{F}} \sum_{i=0}^{2\ell} (|T^{R(i)}_{\mathcal{E},p}|+|M^{S(i)}_{\mathcal{E},p}|+|M^{R(i)}_{\mathcal{E},p}|) \leq (1+\epsilon) \mathsf{AC}
\end{equation*}
Consider $\mathcal{T}(\mathcal{P})$ instantiated with this parameter $\ell$. 
Then, counting the epoch-ending transactions and the transactions 
$\tx = \{ q \,|\, p \in \Pi: q.\mathsf{process} = p \land q.\mathsf{start} = \mathit{false} \land q.\mathsf{quit} = \mathit{true} \land q.\mathsf{log} = \bot  \}$, both of finite size,
we can state that 
for $\mathcal{T}(\mathcal{P})$ and $j = r\ell$; 
\begin{align*}
    & \sup_{\tilde{\mathcal{E}} \in \tilde{E}} \frac{1}{j} \sum_{p \in \tilde{\Pi} \setminus \mathcal{F}} \sum_{i=0}^j (|\tilde{T}^{R(i)}_{\tilde{\mathcal{E}},p}|+|\tilde{M}^{S(i)}_{\tilde{\mathcal{E}},p}|+|\tilde{M}^{R(i)}_{\tilde{\mathcal{E}},p}|) \\
    \leq& \sup_{\mathcal{E}_1, \ldots \mathcal{E}_r \in E} \frac{2}{r} \sum_{i=1}^r \frac{1}{2\ell} \sum_{p \in \Pi \setminus \mathcal{F}} \sum_{i=0}^{2\ell} (|T^{R(i)}_{\mathcal{E}_i,p}|+|M^{S(i)}_{\mathcal{E}_i,p}|+|M^{R(i)}_{\mathcal{E}_i,p}|) \\
    \leq& \frac{2}{r} \sum_{i=1}^r \sup_{\mathcal{E}_i \in E} \frac{1}{2\ell} \sum_{p \in \Pi \setminus \mathcal{F}} \sum_{i=0}^{2\ell} (|T^{R(i)}_{\mathcal{E}_i,p}|+|M^{S(i)}_{\mathcal{E}_i,p}|+|M^{R(i)}_{\mathcal{E}_i,p}|) \\
    \leq& n^2 C + \frac{2}{r} \sum_{i=1}^r (1+\epsilon) \mathsf{AC},
\end{align*}
for some executions $\mathcal{E}_i$ of $\mathcal{P}$ under a $\rho$-bounded adversary.
Here, $C$ is some constant finite value capturing the size of the epoch-ending transactions and the transactions $\tx$ mentioned above.
There are $n^2$ number of them; since $n$ identifiers send these transactions once, and they are re-broadcast to $n$ identifiers from $n$ identifiers once more as part of the fully-certified logs.
Moreover, note that we consider a duration $2\ell$ execution of the permissioned protocol in our bound; since no correct process will stay in an epoch longer than $2\ell$ (in fact, $\ell + O(\Delta)$) timeslots for a sufficiently long $\ell$.
Finally, for $\mathcal{T}(\mathcal{P})$, recalling $j = r \ell$, we can write;
\begin{align*}
    \tilde{\mathsf{AC}} =& \lim_{j \to \infty} \sup_{\tilde{\mathcal{E}} \in E} \frac{1}{j} \sum_{p \in \tilde{\Pi}} \sum_{i=0}^j (|\tilde{T}^{R(i)}_{\tilde{\mathcal{E}},p}|+|\tilde{M}^{S(i)}_{\tilde{\mathcal{E}},p}|+|\tilde{M}^{R(i)}_{\tilde{\mathcal{E}},p}|) \\
    =& \lim_{r \to \infty} \sup_{\tilde{\mathcal{E}} \in \tilde{E}} \frac{1}{j} \sum_{p \in \tilde{\Pi}} \sum_{i=0}^j (|\tilde{T}^{R(i)}_{\tilde{\mathcal{E}},p}|+|\tilde{M}^{S(i)}_{\tilde{\mathcal{E}},p}|+|\tilde{M}^{R(i)}_{\tilde{\mathcal{E}},p}|) \\
    =& \lim_{r \to \infty} n^2 C + \frac{2}{r} \sum_{i=1}^r (1+\epsilon) \mathsf{AC} = O(\mathsf{AC}),
\end{align*}
as $\mathsf{AC} = \Omega(n^2)$. 

Finally, for $\tilde{\mathsf{PC}}$, we note that
\begin{align*}
    \tilde{\mathsf{PC}} =& \sup_{\tilde{\mathcal{E}} \in E} \sup_{i \in [0,\infty), p \in \Pi \setminus \mathcal{F}} (|\tilde{T}^{R(i)}_{\tilde{\mathcal{E}},p}|+|\tilde{M}^{S(i)}_{\tilde{\mathcal{E}},p}|+|\tilde{M}^{R(i)}_{\tilde{\mathcal{E}},p}|) \\
    \leq& \sup_{\mathcal{E}_r \in E, r = 1, \ldots} \sup_{i \in [0,\ell+\Delta], p \in \Pi \setminus \mathcal{F}} (|T^{R(i)}_{\mathcal{E}_r,p}|+|M^{S(i)}_{\mathcal{E}_r,p}|+|M^{R(i)}_{\mathcal{E}_r,p}| + n^2 C) \\
    \leq& n^2 C + \mathsf{PC} = O(\mathsf{PC}),
\end{align*}
since $\mathsf{PC} = \Omega(n^2)$.
\end{proof}

%% file: appendix/60_relatedwork.tex
\section{Additional Related Work}
\label{section:relatedwork-long}

To the best of our knowledge, no earlier work
describes a closed-box transformation from permissioned to PoS consensus,
that preserves and provides the range of desirable properties studied in this paper.

\mysubparagraph{PoS blockchain protocols}
The first traces~\cite{cryptocurrencywithoutproofofwork,bentov2014proof,cunicula2011proof,tacotime2013netcoin,quantummechanic2011pos,maxwell2014impossibility,peercoin} of the idea of PoS blockchains can be found in the Bitcoin community~\cite{bitcoin}.
Today, Ethereum~\cite{ethereum_consensus_specs,casperffg,gasper} is the largest PoS blockchain by market capitalization.
The first protocols with a rigorous security analysis included SnowWhite~\cite{snowwhite} and Ouroboros Praos~\cite{praos}, PoS variants of Sleepy~\cite{DBLP:conf/asiacrypt/PassS17} and Ouroboros~\cite{ouroboros}, respectively. 
Sleepy and Ouroboros 
are adaptations of Nakamoto's longest-chain protocol~\cite{bitcoin,bitcoinbackbone}
to the classical permissioned 
setting.
They replace Nakamoto's proof-of-work with a lottery based on pseudo-random functions.
The constructions to handle stake shift in SnowWhite and Ouroboros Praos are not closed-box, 
but ``fold'' reconfiguration into the protocol 
in an inductive way such that the overall structure of Nakamoto consensus with a single block-tree per execution
is preserved.
To this end, time is divided into epochs,
and, conceptually, the stabilized prefix of the longest-chain 
as of the end of the current epoch 
governs the rights to block production in the epoch thereafter.
Other early PoS protocol include Algorand~\cite{algorandchen,algorandsosp}
and Tendermint~\cite{kwon2014tendermint,buchman2016tendermint,buchman2018latest,cosmos2023staking},
which implement state-machine replication multi-shot consensus through repeated single-shot Byzantine agreement, without pipelining.
A consequence is that these protocols don't show forking, and the 
stake distribution decided by one single-shot 
instance can immediately govern the next single-shot 
instance with ease.

ByzCoin~\cite{byzcoin} and Hybrid Consensus~\cite{pass2017hybrid}
use proof-of-work to sample a committee among the miners that then run a permissioned consensus protocol with low confirmation latency.
Specifically, Hybrid Consensus proceeds in epochs (called ``days''), each of which has an associated permissioned consensus instance, with epoch transition and output log construction similar to that of our transformation.
Since in Hybrid Consensus an epoch's committee is determined external to the permissioned/PoS protocol, rather than
based on the consensus of the epoch prior, consensus instances can overlap around the epoch boundary.
Thunderella~\cite{thunderella} builds on the techniques of Hybrid consensus to equip high-latency Nakamoto-style consensus with an optimistically responsive fast path.
Lewis--Pye and Roughgarden~\cite{DBLP:journals/corr/abs-2304-14701} 
and Budish et al.~\cite{DBLP:journals/corr/abs-2405-09173} 
provide an epoch-based transformation, similar to ours, from permissioned to PoS for HotStuff
and Tendermint, respectively.
PaLa~\cite{pala} is a partially synchronous PBFT-style~\cite{pbft} consensus protocol
with
a builtin reconfiguration mechanism where during reconfiguration both the outgoing and the incoming processes
vote for newly produced blocks.
Sui Lutris~\cite{suilutris} features a hybrid architecture integrating
a partially-ordering with a totally-ordering consensus mechanism~\cite{abc2019,guerraoui2019consensus,baudet2020fastpay},
and provides an epoch-based joint reconfiguration mechanism where the partially-ordering component is effectively paused during epoch transition.

\mysubparagraph{Reconfiguration for state-machine replication}
There is a substantial line of work on reconfiguration of replicated state-machines in the traditional distributed-computing literature.
Lamport already touched upon reconfiguration of replicated state-machines in the Paxos paper~\cite{DBLP:journals/tocs/Lamport98}
and refined the treatment in subsequent works~\cite{lamport2009vertical,lamport2010reconfiguring,lamport2009stoppable}.
The two-mode technique of~\cite{lamport2009vertical} was subsequently extended to Byzantine faults in~\cite{abraham2016bvp}.
BFT-SMaRt~\cite{bessani2014state} is among the first publicly available libraries that implement
a BFT protocol with support for reconfiguration.
The popular Raft algorithm~\cite{DBLP:conf/usenix/OngaroO14} also provides a built-in reconfiguration mechanism, but only in a crash-fault model.
BChain~\cite{bchain} describes a reconfiguration mechanism that is needed to remove faulty processes and restore liveness.
Duan and Zhang~\cite{dynamicbftfoundations} provide a recent formal treatment of Byzantine state-machine replication with reconfiguration.
Specifically, they provide Dyno, a family of total-order broadcast protocols with reconfiguration carefully woven in a bespoke manner.
Tang et al.~\cite{Tang2024} build on that work.

\mysubparagraph{Group membership \& view synchronous communication}
The following two classical distributed computing primitives accommodate dynamic changes in participation: group membership and view synchronous communication~\cite{DBLP:journals/tocs/BirmanJ87,cachin2011introduction,DBLP:journals/csur/ChocklerKV01}.
These two primitives were first introduced by Birman and Joseph~\cite{DBLP:journals/tocs/BirmanJ87}.
Group membership facilitates synchronization among processes regarding the current participants in the system. 
Specifically, processes advance through an agreed-upon sequence of views, with each view representing the system's membership at a given (logical) time.
View synchronous communication~\cite{DBLP:conf/icdcs/SchiperS93,DBLP:journals/tdsc/AmirNST05,spread,DBLP:conf/icdcs/MoserAMA94} combines the group membership primitive with reliable broadcast~\cite{DBLP:conf/podc/Bracha84}: guarantees of the reliable broadcast primitive are ensured within a view.

\mysubparagraph{Deterministic reconfiguration in asynchrony}
An intriguing and challenging research direction involves the development of deterministic reconfiguration protocols in a fully asynchronous setting.
The difficulty of this problem arises from the fact that such protocols cannot depend on consensus, as the seminal FLP impossibility result~\cite{fischer1985impossibility} demonstrates that consensus cannot be deterministically achieved in asynchronous systems. Consequently, agreement on the system's membership can never be guaranteed: reconfiguration must be achieved despite the fact that processes might have different views of the current system membership.
This problem was initially explored in the crash failure model~\cite{DBLP:journals/jacm/AguileraKMS11,DBLP:conf/opodis/AlchieriBGF17,DBLP:conf/wdag/JehlVM15,DBLP:conf/opodis/KuznetsovRP19,DBLP:conf/wdag/GafniM15,DBLP:conf/wdag/SpiegelmanKM17}.
Subsequently, reconfigurable Byzantine-resilient protocols were developed for various applications, including reliable broadcast~\cite{DBLP:conf/opodis/GuerraouiKKPST20}, payment systems~\cite{DBLP:journals/tdsc/CamaioniGKMRVV25}, and more general purposes~\cite{kuznetsov2020asynchronous}.

%% file: main.bbl
\begin{thebibliography}{10}

\bibitem{abraham2016bvp}
Ittai Abraham and Dahlia Malkhi.
\newblock {BVP}: Byzantine vertical {Paxos}.
\newblock In {\em Distributed Cryptocurrencies and Consensus Ledgers (DCCL)},
  2016.

\bibitem{DBLP:journals/jacm/AguileraKMS11}
Marcos~Kawazoe Aguilera, Idit Keidar, Dahlia Malkhi, and Alexander Shraer.
\newblock Dynamic atomic storage without consensus.
\newblock {\em J. {ACM}}, 58(2):7:1--7:32, 2011.

\bibitem{DBLP:conf/opodis/AlchieriBGF17}
Eduardo Alchieri, Alysson Bessani, Fab{\'{\i}}ola Greve, and Joni
  da~Silva~Fraga.
\newblock Efficient and modular consensus-free reconfiguration for
  fault-tolerant storage.
\newblock In {\em {OPODIS}}, volume~95 of {\em LIPIcs}, pages 26:1--26:17.
  Schloss Dagstuhl - Leibniz-Zentrum f{\"{u}}r Informatik, 2017.

\bibitem{DBLP:journals/ipl/AlpernS85}
Bowen Alpern and Fred~B. Schneider.
\newblock Defining liveness.
\newblock {\em Inf. Process. Lett.}, 21(4):181--185, 1985.

\bibitem{DBLP:journals/dc/AlpernS87}
Bowen Alpern and Fred~B. Schneider.
\newblock Recognizing safety and liveness.
\newblock {\em Distributed Comput.}, 2(3):117--126, 1987.

\bibitem{DBLP:journals/tdsc/AmirNST05}
Yair Amir, Cristina Nita{-}Rotaru, Jonathan~Robert Stanton, and Gene Tsudik.
\newblock Secure {Spread}: An integrated architecture for secure group
  communication.
\newblock {\em {IEEE} Trans. Dependable Secur. Comput.}, 2(3):248--261, 2005.

\bibitem{spread}
Yair Amir and Jonathan Stanton.
\newblock The {Spread} wide area group communication system.
\newblock Technical Report CNDS-98-4, The Center for Networking and Distributed
  Systems, The Johns Hopkins University, 1998.

\bibitem{DBLP:journals/corr/abs-2405-20488}
Balaji Arun, Zekun Li, Florian Suri{-}Payer, Sourav Das, and Alexander
  Spiegelman.
\newblock Shoal++: High throughput {DAG} {BFT} can be fast and robust!
\newblock In {\em {NSDI}}, pages 813--826. {USENIX} Association, 2025.

\bibitem{DBLP:journals/corr/abs-2310-14821}
Kushal Babel, Andrey Chursin, George Danezis, Anastasios Kichidis, Lefteris
  Kokoris-Kogias, Arun Koshy, Alberto Sonnino, and Mingwei Tian.
\newblock Mysticeti: Reaching the limits of latency with uncertified dags.
\newblock arXiv:2310.14821v4 [cs.DC], 2023.
\newblock URL: \url{http://arxiv.org/abs/2310.14821v4}, \href
  {https://arxiv.org/abs/2310.14821v4} {\path{arXiv:2310.14821v4}}.

\bibitem{baudet2020fastpay}
Mathieu Baudet, George Danezis, and Alberto Sonnino.
\newblock Fastpay: High-performance byzantine fault tolerant settlement.
\newblock In {\em {AFT}}, pages 163--177. {ACM}, 2020.

\bibitem{DBLP:conf/tcc/BenhamoudaG0HK020}
Fabrice Benhamouda, Craig Gentry, Sergey Gorbunov, Shai Halevi, Hugo Krawczyk,
  Chengyu Lin, Tal Rabin, and Leonid Reyzin.
\newblock Can a public blockchain keep a secret?
\newblock In {\em {TCC} {(1)}}, volume 12550 of {\em Lecture Notes in Computer
  Science}, pages 260--290. Springer, 2020.

\bibitem{cryptocurrencywithoutproofofwork}
Iddo Bentov, Ariel Gabizon, and Alex Mizrahi.
\newblock Cryptocurrencies without proof of work.
\newblock arXiv:1406.5694v9 [cs.CR], 2014.
\newblock URL: \url{http://arxiv.org/abs/1406.5694v9}, \href
  {https://arxiv.org/abs/1406.5694v9} {\path{arXiv:1406.5694v9}}.

\bibitem{bentov2014proof}
Iddo Bentov, Charles Lee, Alex Mizrahi, and Meni Rosenfeld.
\newblock Proof of activity: Extending {Bitcoin}'s proof of work via proof of
  stake [extended abstract].
\newblock {\em {SIGMETRICS} Perform. Evaluation Rev.}, 42(3):34--37, 2014.

\bibitem{bessani2014state}
Alysson~Neves Bessani, João Sousa, and Eduardo Adílio~Pelinson Alchieri.
\newblock State machine replication for the masses with {BFT-SMART}.
\newblock In {\em {DSN}}, pages 355--362. {IEEE} Computer Society, 2014.

\bibitem{DBLP:journals/tocs/BirmanJ87}
Kenneth~P. Birman and Thomas~A. Joseph.
\newblock Reliable communication in the presence of failures.
\newblock {\em {ACM} Trans. Comput. Syst.}, 5(1):47--76, 1987.

\bibitem{suilutris}
Sam Blackshear, Andrey Chursin, George Danezis, Anastasios Kichidis, Lefteris
  Kokoris{-}Kogias, Xun Li, Mark Logan, Ashok Menon, Todd Nowacki, Alberto
  Sonnino, Brandon Williams, and Lu~Zhang.
\newblock Sui lutris: {A} blockchain combining broadcast and consensus.
\newblock In {\em {CCS}}, pages 2606--2620. {ACM}, 2024.

\bibitem{DBLP:conf/podc/Bracha84}
Gabriel Bracha.
\newblock An asynchronous [(n-1)/3]-resilient consensus protocol.
\newblock In {\em {PODC}}, pages 154--162. {ACM}, 1984.

\bibitem{buchman2016tendermint}
Ethan Buchman.
\newblock {Tendermint}: Byzantine fault tolerance in the age of blockchains.
\newblock Master's thesis, University of Guelph,
  \url{https://allquantor.at/blockchainbib/pdf/buchman2016tendermint.pdf},
  2016.

\bibitem{buchman2018latest}
Ethan Buchman, Jae Kwon, and Zarko Milosevic.
\newblock The latest gossip on {BFT} consensus.
\newblock arXiv:1807.04938v3 [cs.DC], 2018.
\newblock URL: \url{http://arxiv.org/abs/1807.04938v3}, \href
  {https://arxiv.org/abs/1807.04938v3} {\path{arXiv:1807.04938v3}}.

\bibitem{DBLP:journals/corr/abs-2405-09173}
Eric Budish, Andrew Lewis{-}Pye, and Tim Roughgarden.
\newblock The economic limits of permissionless consensus.
\newblock In {\em {EC}}, pages 704--731. {ACM}, 2024.

\bibitem{casperffg}
Vitalik Buterin and Virgil Griffith.
\newblock Casper the friendly finality gadget.
\newblock arXiv:1710.09437v4 [cs.CR], 2017.
\newblock URL: \url{http://arxiv.org/abs/1710.09437v4}, \href
  {https://arxiv.org/abs/1710.09437v4} {\path{arXiv:1710.09437v4}}.

\bibitem{gasper}
Vitalik Buterin, Diego Hernandez, Thor Kamphefner, Khiem Pham, Zhi Qiao, Danny
  Ryan, Juhyeok Sin, Ying Wang, and Yan~X Zhang.
\newblock Combining ghost and casper.
\newblock arXiv:2003.03052v3 [cs.CR], 2020.
\newblock URL: \url{http://arxiv.org/abs/2003.03052v3}, \href
  {https://arxiv.org/abs/2003.03052v3} {\path{arXiv:2003.03052v3}}.

\bibitem{cachin2011introduction}
Christian Cachin, Rachid Guerraoui, and Lu{\'{\i}}s E.~T. Rodrigues.
\newblock {\em Introduction to Reliable and Secure Distributed Programming
  {(2.} ed.)}.
\newblock Springer, 2011.

\bibitem{Cachin2001}
Christian Cachin, Klaus Kursawe, Frank Petzold, and Victor Shoup.
\newblock Secure and efficient asynchronous broadcast protocols.
\newblock In {\em {CRYPTO}}, volume 2139 of {\em Lecture Notes in Computer
  Science}, pages 524--541. Springer, 2001.

\bibitem{DBLP:journals/tdsc/CamaioniGKMRVV25}
Martina Camaioni, Rachid Guerraoui, Jovan Komatovic, Matteo Monti,
  Pierre{-}Louis Roman, Manuel Vidigueira, and Gauthier Voron.
\newblock Carbon: Scaling trusted payments with untrusted machines.
\newblock {\em {IEEE} Trans. Dependable Secur. Comput.}, 22(2):1168--1180,
  2025.

\bibitem{pbft}
Miguel Castro and Barbara Liskov.
\newblock Practical byzantine fault tolerance and proactive recovery.
\newblock {\em {ACM} Trans. Comput. Syst.}, 20(4):398--461, 2002.

\bibitem{pala}
T-H.~Hubert Chan, Rafael Pass, and Elaine Shi.
\newblock {PaLa}: A simple partially synchronous blockchain.
\newblock Cryptology {ePrint} Archive, Paper 2018/981, 2018.
\newblock URL: \url{https://eprint.iacr.org/2018/981}.

\bibitem{algorandchen}
Jing Chen and Silvio Micali.
\newblock Algorand: {A} secure and efficient distributed ledger.
\newblock {\em Theor. Comput. Sci.}, 777:155--183, 2019.

\bibitem{DBLP:journals/csur/ChocklerKV01}
Gregory~V. Chockler, Idit Keidar, and Roman Vitenberg.
\newblock Group communication specifications: a comprehensive study.
\newblock {\em {ACM} Comput. Surv.}, 33(4):427--469, 2001.

\bibitem{civit2022byzantine}
Pierre Civit, Muhammad~Ayaz Dzulfikar, Seth Gilbert, Vincent Gramoli, Rachid
  Guerraoui, Jovan Komatovic, and Manuel Vidigueira.
\newblock Byzantine consensus is {$\Theta(n^2)$}: the {Dolev}-{Reischuk} bound
  is tight even in partial synchrony!
\newblock {\em Distributed Comput.}, 37(2):89--119, 2024.

\bibitem{CGG19}
Pierre Civit, Seth Gilbert, and Vincent Gramoli.
\newblock Polygraph: Accountable byzantine agreement.
\newblock In {\em {ICDCS}}, pages 403--413. {IEEE}, 2021.

\bibitem{easy-accountability}
Pierre Civit, Seth Gilbert, Vincent Gramoli, Rachid Guerraoui, and Jovan
  Komatovic.
\newblock As easy as {ABC:} optimal (a)ccountable (b)yzantine (c)onsensus is
  easy!
\newblock {\em J. Parallel Distributed Comput.}, 181:104743, 2023.

\bibitem{crime_punishment}
Pierre Civit, Seth Gilbert, Vincent Gramoli, Rachid Guerraoui, Jovan Komatovic,
  Zarko Milosevic, and Adi Seredinschi.
\newblock Crime and punishment in distributed byzantine decision tasks.
\newblock In {\em {ICDCS}}, pages 34--44. {IEEE}, 2022.

\bibitem{cosmos2023staking}
{Cosmos Network}.
\newblock Staking module: End-block.
\newblock
  \url{https://docs.cosmos.network/v0.46/modules/staking/05\_end\_block.html},
  2023.

\bibitem{snowwhite}
Phil Daian, Rafael Pass, and Elaine Shi.
\newblock {Snow White}: Robustly reconfigurable consensus and applications to
  provably secure proof of stake.
\newblock In {\em Financial Cryptography}, volume 11598 of {\em Lecture Notes
  in Computer Science}, pages 23--41. Springer, 2019.

\bibitem{DBLP:conf/eurosys/DanezisKSS22}
George Danezis, Lefteris Kokoris{-}Kogias, Alberto Sonnino, and Alexander
  Spiegelman.
\newblock {Narwhal and Tusk}: a {DAG}-based mempool and efficient {BFT}
  consensus.
\newblock In {\em EuroSys}, pages 34--50. {ACM}, 2022.

\bibitem{praos}
Bernardo David, Peter Gazi, Aggelos Kiayias, and Alexander Russell.
\newblock {Ouroboros Praos}: An adaptively-secure, semi-synchronous
  proof-of-stake blockchain.
\newblock In {\em {EUROCRYPT} {(2)}}, volume 10821 of {\em Lecture Notes in
  Computer Science}, pages 66--98. Springer, 2018.

\bibitem{deirmentzoglou2019survey}
Evangelos Deirmentzoglou, Georgios Papakyriakopoulos, and Constantinos
  Patsakis.
\newblock A survey on long-range attacks for proof of stake protocols.
\newblock {\em {IEEE} Access}, 7:28712--28725, 2019.

\bibitem{dolev1985bounds}
Danny Dolev and R{\"{u}}diger Reischuk.
\newblock Bounds on information exchange for byzantine agreement.
\newblock {\em J. {ACM}}, 32(1):191--204, 1985.

\bibitem{DBLP:journals/siamcomp/DolevS83}
Danny Dolev and H.~Raymond Strong.
\newblock Authenticated algorithms for byzantine agreement.
\newblock {\em {SIAM} J. Comput.}, 12(4):656--666, 1983.

\bibitem{bchain}
Sisi Duan, Hein Meling, Sean Peisert, and Haibin Zhang.
\newblock Bchain: Byzantine replication with high throughput and embedded
  reconfiguration.
\newblock In {\em {OPODIS}}, volume 8878 of {\em Lecture Notes in Computer
  Science}, pages 91--106. Springer, 2014.

\bibitem{dynamicbftfoundations}
Sisi Duan and Haibin Zhang.
\newblock Foundations of dynamic {BFT}.
\newblock In {\em {SP}}, pages 1317--1334. {IEEE}, 2022.

\bibitem{DLS88}
Cynthia Dwork, Nancy~A. Lynch, and Larry~J. Stockmeyer.
\newblock Consensus in the presence of partial synchrony.
\newblock {\em J. {ACM}}, 35(2):288--323, 1988.

\bibitem{ethereum_consensus_specs}
{Ethereum Foundation}.
\newblock Ethereum consensus specifications.
\newblock \url{https://github.com/ethereum/consensus-specs}, 2023.
\newblock Accessed: 2023-12-14.

\bibitem{fischer1985impossibility}
Michael~J. Fischer, Nancy~A. Lynch, and Mike Paterson.
\newblock Impossibility of distributed consensus with one faulty process.
\newblock {\em J. {ACM}}, 32(2):374--382, 1985.

\bibitem{DBLP:conf/wdag/GafniM15}
Eli Gafni and Dahlia Malkhi.
\newblock Elastic configuration maintenance via a parsimonious speculating
  snapshot solution.
\newblock In {\em {DISC}}, volume 9363 of {\em Lecture Notes in Computer
  Science}, pages 140--153. Springer, 2015.

\bibitem{bitcoinbackbone}
Juan~A. Garay, Aggelos Kiayias, and Nikos Leonardos.
\newblock The {Bitcoin} backbone protocol: Analysis and applications.
\newblock {\em J. {ACM}}, 71(4):25:1--25:49, 2024.

\bibitem{algorandsosp}
Yossi Gilad, Rotem Hemo, Silvio Micali, Georgios Vlachos, and Nickolai
  Zeldovich.
\newblock Algorand: Scaling byzantine agreements for cryptocurrencies.
\newblock In {\em {SOSP}}, pages 51--68. {ACM}, 2017.

\bibitem{GoyalKMPS22}
Vipul Goyal, Abhiram Kothapalli, Elisaweta Masserova, Bryan Parno, and Yifan
  Song.
\newblock Storing and retrieving secrets on a blockchain.
\newblock In {\em Public Key Cryptography {(1)}}, volume 13177 of {\em Lecture
  Notes in Computer Science}, pages 252--282. Springer, 2022.

\bibitem{DBLP:conf/opodis/GuerraouiKKPST20}
Rachid Guerraoui, Jovan Komatovic, Petr Kuznetsov, Yvonne{-}Anne Pignolet,
  Dragos{-}Adrian Seredinschi, and Andrei Tonkikh.
\newblock Dynamic byzantine reliable broadcast.
\newblock In {\em {OPODIS}}, volume 184 of {\em LIPIcs}, pages 23:1--23:18.
  Schloss Dagstuhl - Leibniz-Zentrum f{\"{u}}r Informatik, 2020.

\bibitem{guerraoui2019consensus}
Rachid Guerraoui, Petr Kuznetsov, Matteo Monti, Matej Pavlovic, and
  Dragos{-}Adrian Seredinschi.
\newblock The consensus number of a cryptocurrency.
\newblock {\em Distributed Comput.}, 35(1):1--15, 2022.

\bibitem{HKD07}
Andreas Haeberlen, Petr Kouznetsov, and Peter Druschel.
\newblock Peerreview: practical accountability for distributed systems.
\newblock In {\em {SOSP}}, pages 175--188. {ACM}, 2007.

\bibitem{DBLP:conf/wdag/JehlVM15}
Leander Jehl, Roman Vitenberg, and Hein Meling.
\newblock Smartmerge: {A} new approach to reconfiguration for atomic storage.
\newblock In {\em {DISC}}, volume 9363 of {\em Lecture Notes in Computer
  Science}, pages 154--169. Springer, 2015.

\bibitem{DBLP:conf/podc/KeidarKNS21}
Idit Keidar, Eleftherios Kokoris{-}Kogias, Oded Naor, and Alexander Spiegelman.
\newblock All you need is {DAG}.
\newblock In {\em {PODC}}, pages 165--175. {ACM}, 2021.

\bibitem{KermarrecS07}
Anne{-}Marie Kermarrec and Maarten van Steen.
\newblock Gossiping in distributed systems.
\newblock {\em {ACM} {SIGOPS} Oper. Syst. Rev.}, 41(5):2--7, 2007.

\bibitem{ouroboros}
Aggelos Kiayias, Alexander Russell, Bernardo David, and Roman Oliynykov.
\newblock Ouroboros: {A} provably secure proof-of-stake blockchain protocol.
\newblock In {\em {CRYPTO} {(1)}}, volume 10401 of {\em Lecture Notes in
  Computer Science}, pages 357--388. Springer, 2017.

\bibitem{peercoin}
Sunny King and Scott Nadal.
\newblock {PPCoin}: Peer-to-peer crypto-currency with proof-of-stake.
\newblock \url{https://peercoin.net/assets/paper/peercoin-paper.pdf}, 2012.

\bibitem{byzcoin}
Eleftherios Kokoris{-}Kogias, Philipp Jovanovic, Nicolas Gailly, Ismail Khoffi,
  Linus Gasser, and Bryan Ford.
\newblock Enhancing bitcoin security and performance with strong consistency
  via collective signing.
\newblock In {\em {USENIX} Security Symposium}, pages 279--296. {USENIX}
  Association, 2016.

\bibitem{fullversionpreprint}
Jovan Komatovic, Andrew Lewis-Pye, Joachim Neu, Tim Roughgarden, and
  Ertem~Nusret Tas.
\newblock From permissioned to proof-of-stake consensus.
\newblock Cryptology {ePrint} Archive, Paper 2025/1139, 2025.
\newblock URL: \url{https://eprint.iacr.org/2025/1139}.

\bibitem{DBLP:conf/opodis/KuznetsovRP19}
Petr Kuznetsov, Thibault Rieutord, and Sara {Tucci Piergiovanni}.
\newblock Reconfigurable lattice agreement and applications.
\newblock In {\em {OPODIS}}, volume 153 of {\em LIPIcs}, pages 31:1--31:17.
  Schloss Dagstuhl - Leibniz-Zentrum f{\"{u}}r Informatik, 2019.

\bibitem{kuznetsov2020asynchronous}
Petr Kuznetsov and Andrei Tonkikh.
\newblock Asynchronous reconfiguration with byzantine failures.
\newblock {\em Distributed Comput.}, 35(6):477--502, 2022.

\bibitem{kwon2014tendermint}
Jae Kwon.
\newblock Tendermint: Consensus without mining.
\newblock \url{https://tendermint.com/static/docs/tendermint.pdf}, 2014.

\bibitem{DBLP:journals/tocs/Lamport98}
Leslie Lamport.
\newblock The part-time parliament.
\newblock {\em {ACM} Trans. Comput. Syst.}, 16(2):133--169, 1998.

\bibitem{lamport2009stoppable}
Leslie Lamport, Dahlia Malkhi, and Lidong Zhou.
\newblock Stoppable {Paxos}.
\newblock Unpublished manuscript,
  \url{https://lamport.azurewebsites.net/pubs/stoppable.pdf}, 2009.

\bibitem{lamport2009vertical}
Leslie Lamport, Dahlia Malkhi, and Lidong Zhou.
\newblock Vertical {Paxos} and primary-backup replication.
\newblock In {\em {PODC}}, pages 312--313. {ACM}, 2009.

\bibitem{lamport2010reconfiguring}
Leslie Lamport, Dahlia Malkhi, and Lidong Zhou.
\newblock Reconfiguring a state machine.
\newblock {\em {SIGACT} News}, 41(1):63--73, 2010.

\bibitem{lewis2022quadratic}
Andrew Lewis-Pye.
\newblock Quadratic worst-case message complexity for state machine replication
  in the partial synchrony model.
\newblock arXiv:2201.01107v1 [cs.DC], 2022.
\newblock URL: \url{http://arxiv.org/abs/2201.01107v1}, \href
  {https://arxiv.org/abs/2201.01107v1} {\path{arXiv:2201.01107v1}}.

\bibitem{DBLP:journals/corr/abs-2304-14701}
Andrew Lewis-Pye and Tim Roughgarden.
\newblock Permissionless consensus.
\newblock arXiv:2304.14701v5 [cs.DC], 2023.
\newblock URL: \url{http://arxiv.org/abs/2304.14701v5}, \href
  {https://arxiv.org/abs/2304.14701v5} {\path{arXiv:2304.14701v5}}.

\bibitem{lewis2025beyond}
Andrew Lewis-Pye and Tim Roughgarden.
\newblock Beyond optimal fault tolerance.
\newblock arXiv:2501.06044v7 [cs.DC], 2025.
\newblock URL: \url{http://arxiv.org/abs/2501.06044v7}, \href
  {https://arxiv.org/abs/2501.06044v7} {\path{arXiv:2501.06044v7}}.

\bibitem{MaramZWLZJS19}
Sai Krishna~Deepak Maram, Fan Zhang, Lun Wang, Andrew Low, Yupeng Zhang, Ari
  Juels, and Dawn Song.
\newblock {CHURP:} dynamic-committee proactive secret sharing.
\newblock In {\em {CCS}}, pages 2369--2386. {ACM}, 2019.

\bibitem{maxwell2014impossibility}
Gregory Maxwell and Andrew Poelstra.
\newblock Distributed consensus from proof of stake is impossible.
\newblock \url{https://download.wpsoftware.net/bitcoin/pos.pdf}, 2014.

\bibitem{DBLP:conf/ccs/MillerXCSS16}
Andrew Miller, Yu~Xia, Kyle Croman, Elaine Shi, and Dawn Song.
\newblock The honey badger of {BFT} protocols.
\newblock In {\em {CCS}}, pages 31--42. {ACM}, 2016.

\bibitem{DBLP:conf/icdcs/MoserAMA94}
Louise~E. Moser, Yair Amir, P.~M. Melliar{-}Smith, and Deborah~A. Agarwal.
\newblock Extended virtual synchrony.
\newblock In {\em {ICDCS}}, pages 56--65. {IEEE} Computer Society, 1994.

\bibitem{bitcoin}
Satoshi Nakamoto.
\newblock Bitcoin: A peer-to-peer electronic cash system.
\newblock \url{https://bitcoin.org/bitcoin.pdf}, 2008.

\bibitem{DBLP:conf/sp/NeuTT21}
Joachim Neu, Ertem~Nusret Tas, and David Tse.
\newblock Ebb-and-flow protocols: {A} resolution of the availability-finality
  dilemma.
\newblock In {\em {SP}}, pages 446--465. {IEEE}, 2021.

\bibitem{DBLP:conf/fc/NeuTT22}
Joachim Neu, Ertem~Nusret Tas, and David Tse.
\newblock The availability-accountability dilemma and its resolution via
  accountability gadgets.
\newblock In {\em Financial Cryptography}, volume 13411 of {\em Lecture Notes
  in Computer Science}, pages 541--559. Springer, 2022.

\bibitem{DBLP:conf/fc/NeuTT24}
Joachim Neu, Ertem~Nusret Tas, and David Tse.
\newblock Short paper: Accountable safety implies finality.
\newblock In {\em {FC} {(1)}}, volume 14744 of {\em Lecture Notes in Computer
  Science}, pages 41--50. Springer, 2024.

\bibitem{DBLP:conf/usenix/OngaroO14}
Diego Ongaro and John~K. Ousterhout.
\newblock In search of an understandable consensus algorithm.
\newblock In {\em {USENIX} {ATC}}, pages 305--319. {USENIX} Association, 2014.

\bibitem{pass2017hybrid}
Rafael Pass and Elaine Shi.
\newblock Hybrid consensus: Efficient consensus in the permissionless model.
\newblock In {\em {DISC}}, volume~91 of {\em LIPIcs}, pages 39:1--39:16.
  Schloss Dagstuhl - Leibniz-Zentrum f{\"{u}}r Informatik, 2017.

\bibitem{DBLP:conf/asiacrypt/PassS17}
Rafael Pass and Elaine Shi.
\newblock The sleepy model of consensus.
\newblock In {\em {ASIACRYPT} {(2)}}, volume 10625 of {\em Lecture Notes in
  Computer Science}, pages 380--409. Springer, 2017.

\bibitem{thunderella}
Rafael Pass and Elaine Shi.
\newblock Thunderella: Blockchains with optimistic instant confirmation.
\newblock In {\em {EUROCRYPT} {(2)}}, volume 10821 of {\em Lecture Notes in
  Computer Science}, pages 3--33. Springer, 2018.

\bibitem{DBLP:conf/icdcs/SchiperS93}
Andr{\'{e}} Schiper and Alain Sandoz.
\newblock Uniform reliable multicast in a virtually synchronous environment.
\newblock In {\em {ICDCS}}, pages 561--568. {IEEE} Computer Society, 1993.

\bibitem{DBLP:conf/ccs/ShengWNKV21}
Peiyao Sheng, Gerui Wang, Kartik Nayak, Sreeram Kannan, and Pramod Viswanath.
\newblock {BFT} protocol forensics.
\newblock In {\em {CCS}}, pages 1722--1743. {ACM}, 2021.

\bibitem{abc2019}
Jakub Sliwinski and Roger Wattenhofer.
\newblock Abc: Proof-of-stake without consensus.
\newblock arXiv:1909.10926v3 [cs.CR], 2019.
\newblock URL: \url{http://arxiv.org/abs/1909.10926v3}, \href
  {https://arxiv.org/abs/1909.10926v3} {\path{arXiv:1909.10926v3}}.

\bibitem{DBLP:journals/corr/abs-2306-03058}
Alexander Spiegelman, Balaji Arun, Rati Gelashvili, and Zekun Li.
\newblock Shoal: Improving {DAG-BFT} latency and robustness.
\newblock In {\em {FC} {(1)}}, volume 14744 of {\em Lecture Notes in Computer
  Science}, pages 92--109. Springer, 2024.

\bibitem{DBLP:conf/ccs/SpiegelmanGSK22}
Alexander Spiegelman, Neil Giridharan, Alberto Sonnino, and Lefteris
  Kokoris{-}Kogias.
\newblock Bullshark: {DAG} {BFT} protocols made practical.
\newblock In {\em {CCS}}, pages 2705--2718. {ACM}, 2022.

\bibitem{DBLP:conf/wdag/SpiegelmanKM17}
Alexander Spiegelman, Idit Keidar, and Dahlia Malkhi.
\newblock Dynamic reconfiguration: Abstraction and optimal asynchronous
  solution.
\newblock In {\em {DISC}}, volume~91 of {\em LIPIcs}, pages 40:1--40:15.
  Schloss Dagstuhl - Leibniz-Zentrum f{\"{u}}r Informatik, 2017.

\bibitem{Tang2024}
Fei Tang, Jinlan Peng, Ping Wang, Huihui Zhu, and Tingxian Xu.
\newblock Improved dynamic byzantine fault tolerant consensus mechanism.
\newblock {\em Computer Communications}, 2024.
\newblock \url{https://doi.org/10.1016/j.comcom.2024.08.004}.
\newblock \href {https://doi.org/10.1016/j.comcom.2024.08.004}
  {\path{doi:10.1016/j.comcom.2024.08.004}}.

\bibitem{cunicula2011proof}
{User cunicula} and M.~Rosenfeld.
\newblock Proof of stake brainstorming.
\newblock \url{https://bitcointalk.org/index.php?topic=37194.0}, 2011.

\bibitem{quantummechanic2011pos}
{User QuantumMechanic}.
\newblock Proof of stake instead of proof of work.
\newblock \url{https://bitcointalk.org/index.php?topic=27787.0}, 2011.

\bibitem{tacotime2013netcoin}
{User tacotime}.
\newblock Netcoin proof-of-work and proof-of-stake hybrid design.
\newblock
  \url{https://web.archive.org/web/20131213085759/http://www.netcoin.io/wiki/Netcoin\_Proof-of-Work\_and\_Proof-of-Stake\_Hybrid\_Design},
  2013.

\bibitem{hotstuff}
Maofan Yin, Dahlia Malkhi, Michael~K. Reiter, Guy Golan{-}Gueta, and Ittai
  Abraham.
\newblock {HotStuff}: {BFT} consensus with linearity and responsiveness.
\newblock In {\em {PODC}}, pages 347--356. {ACM}, 2019.

\end{thebibliography}
